# Frontiers, challenges, and solutions in modeling of swift heavy ion effects in materials

N. Medvedev[1,2,a], A.E. Volkov[3,4,b], R. Rymzhanov[5,6], F. Akhmetov[7], S. Gorbunov[3], R. Voronkov[3], P. Babaev[3]

1) Institute of Physics, Czech Academy of Sciences, Na Slovance 1999/2, 182 21 Prague 8, Czech Republic
2) Institute of Plasma Physics, Czech Academy of Sciences, Za Slovankou 3, 182 00 Prague 8, Czech Republic
3) P.N. Lebedev Physical Institute of the Russian Academy of Sciences, Leninskij pr., 53,119991 Moscow, Russia
4) National Research Centre 'Kurchatov Institute', Kurchatov Sq. 1, 123182 Moscow, Russia
5) Joint Institute for Nuclear Research, Joliot-Curie 6, 141980, Dubna, Moscow Region, Russia
6) The Institute of Nuclear Physics, Ibragimov St. 1, 050032 Almaty, Kazakhstan
7) Industrial Focus Group XUV Optics, MESA+ Institute for Nanotechnology, University of Twente, Drienerlolaan 5, 7522 NB Enschede, The Netherlands

## Abstract

Since a few breakthroughs in the fundamental understanding of the effects of swift heavy ions (SHI) decelerating in the electronic stopping regime in the matter have been achieved in the last decade, it motivated us to review the state-of-the-art approaches in the modeling of SHI effects. The SHI track kinetics occurs *via* several well-separated stages and spans many orders of magnitude in time: from attoseconds in ion-impact ionization depositing an extreme amount of energy in a target, to femtoseconds of electrons transport and holes cascades, to picoseconds of lattice excitation and response, to nanoseconds of atomic relaxation, and even longer times of final macroscopic reaction. Each stage requires its own approaches for quantitative description. We discuss that understanding the links between the stages makes it possible to describe the entire track kinetics within a hybrid multiscale model without fitting procedures. The review focuses on the underlying physical mechanisms of each process, the dominant effects they produce, and the limitations of the existing approaches as well as various numerical techniques implementing these models. It provides an overview of *ab-initio*-based modeling of the evolution of the electronic properties; Monte Carlo simulations of nonequilibrium electronic transport; molecular dynamics modeling of atomic reaction including phase transformations and damage on the surface and in the bulk; kinetic Mote Carlo of atomic defect kinetics; finite-difference methods of tracks interaction with chemical solvents describing etching kinetics. We outline the modern methods that couple these approaches into multiscale and combined multidisciplinary models and point to their bottlenecks, strengths, and weaknesses. The analysis is accompanied by examples of important results improving the understanding of track formation in various materials. Summarizing the most recent advances in the field of the track formation process, the review delivers a comprehensive picture and detailed understanding of the phenomena. Important future directions of research and model development are also outlined.

[a] Corresponding author: ORCID: 0000-0003-0491-1090; Email: nikita.medvedev@fzu.cz
[b] Corresponding author: a.e.volkov@list.ru

# Contents

# I. Introduction

Since the discovery of radioactivity, it has been known that fission products modify the properties of exposed materials in a radical way [1]. At typical energies of fission fragments, the nuclear stopping mode of the ion energy loss dominates in damage formation [2]. In this case, a projectile knocks a target atom out of its equilibrium lattice position, providing the atom with a sufficiently large kinetic energy in a head-on collision [1,3]. These accelerated ions, in turn, may knock out secondary ions and so on, forming collision cascades.

At higher ion energies, e.g., in cosmic rays, a different type of interaction becomes dominant: energy transfer into the electronic subsystem of a target. This energy loss channel is called inelastic energy loss [2]. It dominates at heavy ion energies above some ~MeV energy, see Figure 1.

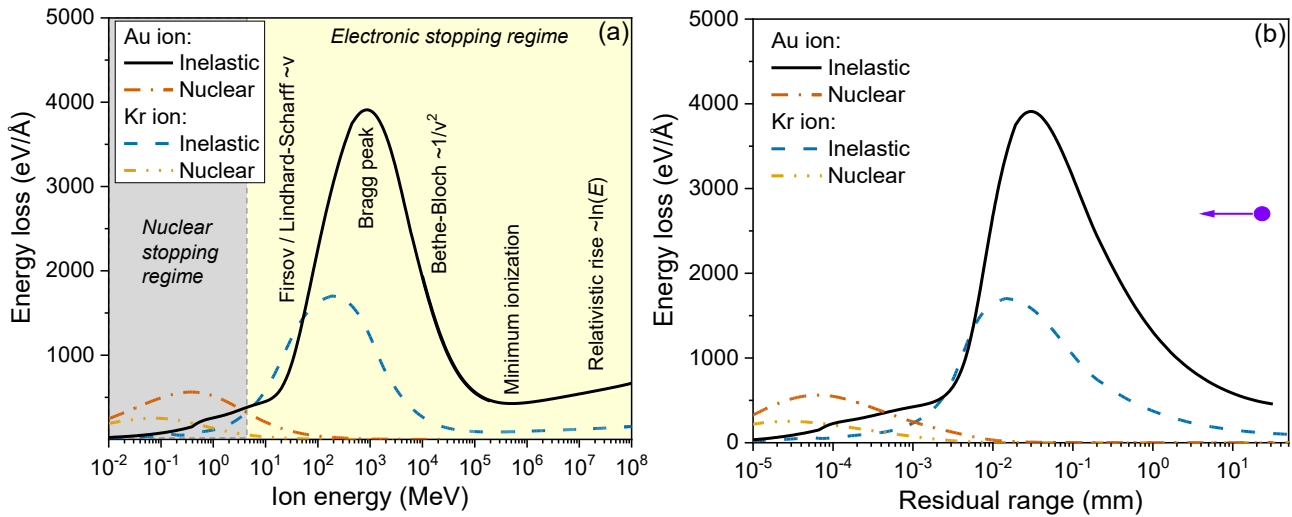

*Figure 1. Nuclear and inelastic energy losses of Au and Kr ions in $Al_2O_3$, extracted from SRIM [4] for nonrelativistic energies and from TREKIS [5,6] for relativistic ones, (a) as a function of incoming ion energy, or (b) residual range along the ion trajectory (direction of ion penetration is schematically shown with an arrow).*

When a high-energy ion (say, 10 GeV gold ion) enters a target, it starts losing its energy primarily to the excitation of the electronic system. Below the minimum ionization dip, a decrease of the energy leads to a faster increase of the electronic energy loss (also called linear energy transfer, LET), until its Bragg peak is reached. The ion velocity at the Bragg peak is roughly equal to that of atomic electrons in the target. The electronic energy losses of the heaviest ions (e.g., Au, Bi, U) achieve values of 40-50 keV per nanometer (4000-5000 eV/Å) of their trajectory at the Bragg peak.

At even lower energies, deeper along the path (see Figure 1), inelastic energy loss decreases and becomes subdominant with respect to the nuclear energy loss at the ion energies below ~1 keV/nucleon. At the very end of the ion track, a nuclear cascade takes place, and the ion comes to rest.

The practical importance of the ion radiation effects is twofold. On the one hand, there are many dangers associated with exposure to fast ions. Cosmic rays pose danger to space missions [7,8]. Energetic ions directly break DNA bonds [9] and produce reactive species and free radicals that interact with genetic material indirectly [10]. Ion irradiation also leads to soft errors and single event upsets in computer equipment [11]. Prolonged exposure of matter may create macroscopic damage in materials, which is crucial to take care of in nuclear materials applications and nuclear waste handling [12]. Radiation safety is thus one of the major concerns in any operation with heavy ions.

On the other hand, usage of the swift heavy ion (SHI) irradiation in a controlled way opened up possibilities of materials modification with unprecedented precision, allowing for nanometric design and producing huge benefits to technology and medicine. Since ion impacts create nanometric-diameter etchable tracks, it enables the creation of nano-pores and nano-membranes [13,14]. Chemical etching of SHI tracks in polymers allows to controllably open and radially enlarge nano-pores, which enables applications of polymers in particle detectors and filtration membranes [15,16]. Nanoscale modifications of matter create regions with altered electronic properties, which is a way of creating quantum dots [17,18] or nano-electronics [19]. Medical applications include proton and ion therapy for cancer treatment [20,21]. Swift ions deposit the main part of their energy in a localized region, at the Bragg peak as shown in Figure 1, which is located deep inside of the target. This allows to damage inoperable tumors, e.g., inside a patient's brain.

Despite the long history of SHI irradiation research, there is still a lack of a detailed understanding of the fundamental phenomena governing track formation [2]. That fact precludes the control of the material response and tailoring of the ion parameters to practical needs.

In a laboratory, ion accelerators are used to create charged ions with controlled high kinetic energies, e.g., at GSI (Darmstadt), JINR (Dubna), IMP (Lanzhou), GANIL (Caen), LHC at CERN (Switzerland), RIKEN (Japan). Precisely knowing the ion charge and energy allows to characterize damage caused by ions and to systematically study the effects of SHI irradiation in various matter [22].

Semi-empirical rules in choosing appropriate ion parameters for particular goals resulted in great progress in terms of practical applications of SHI irradiation. The semi-empirical models developed since the 1990s allowed to estimate track diameters in various materials under various ions irradiation with satisfactory precision [23,24]. Nowadays, inelastic thermal spike (i-TS) [23,24], which is a version of the two-temperature model (TTM [25,26]), is most commonly used. It assumes that the electrons and the atoms may be described as two coupled systems with different temperatures within a solid around the ion trajectory. For each of them, an equation of heat diffusion is used with an empirically fitted electron-ion coupling parameter describing energy transfer from excited electrons to atoms in a track. Having the electron-phonon coupling strength as a free parameter, the model can reproduce experimentally measured track radius in various materials [2]. This simple estimate helps to design experiments and applications.

However, the major problems in the description of SHI tracks are extreme levels of excitation and ultra-short temporal and spatial scales. Under such conditions, the macroscopic approaches fail. For example, the thermo-diffusion model used in i-TS is inapplicable in the tracks due to nonequilibrium conditions and ballistic propagation of the excitation front at the early stage [27,28]. Unfortunately, the communities working with SHIs have overly relied on semi-empirical models, which hindered the progress of a deeper understanding of mechanisms governing track formation.

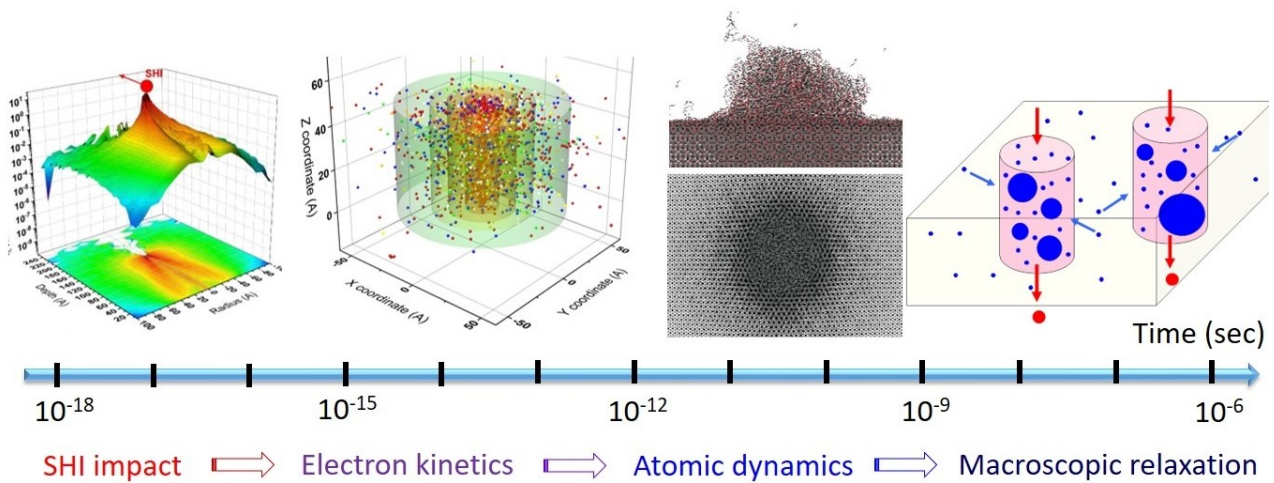

*Figure 2. Schematic illustration of characteristic timescales of an SHI track creation problem.*

The problem of the SHI track can be solved by taking into account the multiscale origin of the track kinetics. An SHI impact induces a sequence of processes starting from the energy deposition into the electronic system, to the final formation of the observable material modifications. A typical timescale of material response to an ion impact spans over ten orders of magnitude, see an illustration in Figure 2. An SHI passes an interatomic distance of a target within attosecond timescales ($\sim 10^{-18}$ s), exciting a number of electrons around its trajectory. During such times, the atoms and electrons of the target are essentially immobile and fixed in space. Electrons in the target excited by an SHI, called delta-electrons, fly outwards from the SHI trajectory, creating secondary cascades of ionizations and transferring energy to the atomic lattice. Excited electronic states in the conduction and valence bands typically occur and decay *via* a number of channels at the femto- to pico-second timescales, $10^{-15}$ s to $10^{-12}$ s [5]. Deep shell holes, created by an SHI impact, decay *via* Auger or radiative mechanisms also during femtoseconds [29]. Atoms react to the transferred energy from the excited electrons and valence holes during $10^{-13}$ s to $10^{-12}$ s [28], which may result in phase transitions and damage formation [30]. Partial or full recovery of the transiently disordered area around an SHI trajectory typically takes $\sim$100 ps ($10^{-10}$ s) [30]. At nanoseconds ($10^{-9}$ s), macroscopic relaxation occurs with the onset of the kinetics of defects and deformation, which typically may relax by the time of microseconds ($10^{-6}$ s) or even longer, macroscopic times [31]. In biological samples, the processes triggered by irradiation may persist for days or even years (e.g. cell death or mutations) [32].

Each stage forms initial conditions for the subsequent one, and thus requires a precise description and fine understanding of the involved processes and their cross-influence when building a model describing the effects of SHIs in the matter.

There are methods for treating each of the abovementioned stages individually [33]. For example, *ab-initio* methods are applicable to SHI-created conditions, such as time-dependent density functional theory (TD-DFT [34,35]). They allow treating ion penetration and formation of the primary excited electrons but are limited to systems of a few tens of atoms with modern-day computer capabilities.

Femtosecond electron kinetics may be traced with *ab-initio* techniques like nonequilibrium Green functions [36,37], or simplified approaches such as Boltzmann's kinetic equation [38], the Fokker-Planck equation [39], or a Monte Carlo method [40,41]. Energy exchange between the electronic and the atomic system may be modeled with such methods as the "surface hopping" method [42,43] and related approaches combined with Boltzmann collision integrals [44], or coupled electron-ion dynamics (CEID) [45]. Simplified thermodynamic-based methods are often used to describe this exchange, such as the above-mentioned TTM.

Atomic dynamics is usually modeled with help of the classical molecular dynamics (MD) simulations [46], or their combination with the TTM [47]. To trace defect kinetics, approximated methods are employed, such as the kinetic Monte Carlo (KMC) methods [31,48]. At high fluences and energy depositions, resulting in macroscopic damage such as material ablation and plasma formation, hydrodynamic approaches and similar tools are used, e.g. Particle in Cell (PIC) codes [49].

Although these methods may be appropriate to treat each stage of track formation and material response to irradiation separately, no rigorous model exists to date capable of treating the entire timespan of the problem with good precision. New approaches are required to describe the problem. One of the most practical, easy-to-implement, and sufficiently precise methods is the so-called multi-scale models, which acquire more and more popularity in the last decade [10,50,51]. A multi-scale or a hybrid model combines a few different approaches, each of them applicable at its own spatial/temporal scale, with a proper overlap and exchange of information between them [33]. This allows covering many orders of magnitude in space and/or time, preserving a required level of precision throughout the simulation.

In the paper, we review the relevant processes taking place in the ion track formation phenomenon across multiple timescales. Since track formation may be divided into a sequence of steps, each of them will require its own approach. We describe each stage in the context of the appropriate models, discussing their respective advantages and shortcomings. After describing each of them, we will consider the current state-of-the-art in modeling: namely, combined and multiscale models [33]. We will pay particular attention to the interconnection of the models applied at different stages of the track evolution, physical justifications, and limitations. Limits of validity and shortcomings of the existing methods will be discussed, outlining future directions of research. We will point out the strengths and weaknesses in the current understanding of the track formation processes, and emphasize the points where new theory and model development is required, and which experimental data are missing.

This review is structured as follows. We start with a description of a single ion impact process taking place at subfemtosecond timescales, such as ion charge equilibration and excitation of target electrons in Section II. Section III reviews the processes of the nonequilibrium electron kinetics. Section IV focuses on the various channels of energy transfer to the atomic system. The response of the atomic system is reviewed in Section V. Relaxation of the atomic system with the formation of the observable tracks and related effects are described in Section VI. Section VII briefly describes residual macroscopic effects in the track surrounding such as defect kinetics in the track halo. We unify all the stages of track formation in the central chapter of this review, Section VIII, describing multiscale models with an accent on the interconnection between different models. We then proceed to the description of the effects of high ion fluence, resulting in track overlap in Section IX. Section X focuses on the surface effects and differences they introduce in comparison to processes taking place in the bulk. Section XI illustrates the post-mortem analysis of tracks, namely a model of wet chemical etching. The concluding summary and outlook are then presented in Section XII.

## II.        Subfemtosecond timescales: ion impact

An SHI experiences various interaction channels in a solid: electronic or inelastic scattering, which excites electrons of the target, nuclear or elastic scattering transferring kinetic energy to target atoms, and radiative energy losses via emitting photons (Bremsstrahlung and Cherenkov radiation). Depending on the ion energy, different channels dominate at different parts of the SHI trajectory.

Investigating the interaction of ions with matter in the electronic stopping regime, accelerators deliver ions with energies up to some GeV, and initial charge states up to a nearly fully stripped ion. For example, for a gold ion (Figure 1), the energy of 100 GeV corresponds to the ion velocity of ~2.3x10$^8$ m/s or 76% of the speed of light in a vacuum. It results in the SHI passing a typical interatomic distance in a target within the time of ~1 attosecond [52], when even the electrons of the target, except perhaps for deep-shell electrons of heavy elements, have no time to move. The scattering of an SHI with target electrons may then be described as pairwise interaction between individual particles. This is known as the Bethe-Bloch regime of ion stopping [53], which forms the right shoulder of the Bragg curve in Figure 1: in the case of gold ion, 1-100 GeV. The energy loss in this regime is approximately proportional to $1/v^2$, with $v$ being the ion velocity [53]. At energies about 0.5-1 TeV ($10^6$ MeV) minimum ionization occurs, after which relativistic rise of energy loss takes place, which scales as ~ln($E$), where $E$ is the ion kinetic energy. This happens due to the increased polarization of the substance around the ion trajectory, caused by the relativistic stretching of the interaction region perpendicular to the ion trajectory [54]. This effect saturates reaching an energy loss plateau (known as the 'Fermi plateau', not shown) originated from the effect of the finite size of the projectile nucleus [55] eliminating a divergence tendency of the energy loss with the increase of the particle velocity. This holds up until the ultra-relativistic effects kick in, at energies significantly higher than those shown in Figure 1. There, another energy loss channel becomes dominant: Bremsstrahlung photon emission and radiative energy loss of the ion – this effect for heavy ions will only take place at energies above those available at modern accelerators. Thus, ultra-relativistic effects are not considered in detail in this review.

With a decrease of the SHI velocity along its trajectory, effects of motion and mutual interactions of target electrons change the nature of the SHI-electrons interaction: from pairwise scattering on immobile electrons, it turns into the passing of the ion through an electronic liquid. This manifests itself as friction and forms the left shoulder of the Bragg curve of energy loss in Figure 1, ~5-1000 MeV for Au ion. This is known as the Firsov / Lindhard-Scharff regime of ion stopping, within which the energy loss of an ion is linearly proportional to its velocity $v$ [56,57].

At even lower energies (below ~5 MeV for a gold ion, Figure 1), nuclear stopping becomes dominant, meaning the energy transfer to the nuclei of the target atoms instead of electrons takes place. It results in knocking of target atoms out from their equilibrium positions, creating atomic cascades, and leaving atomic defects in the target [3,58,59].

During the entire trajectory of a swift heavy ion in a target, it may also capture or lose electrons, changing its own charge state [60]. That is important because the cross-section of scattering and the energy loss of an ion is proportional to the square of its charge. We will consider all of these effects below in more detail.

## II.A.    Ion charge state

Losing and capturing electrons leads to charge exchange between a heavy projectile and a media. A simple consideration helps to understand the meaning of this process: an electron of the swift heavy ion, whose orbital velocity is slower than the velocity of the ion, cannot keep up with it and will be lost in the media. In turn, electrons of the media are attracted to the penetrating ion and may attach to it.

The charge state equilibration of swift ions along their path and the associated charge oscillations can be simulated, for example, using the Monte Carlo method, or by solving rate equations for electron populations of the SHI energy levels such as:

$$\frac{dF_Z(x)}{dx} = \sum_{k \neq Z} F_k(x)\sigma_{k,Z} - F_Z(x) \sum_{k \neq Z} \sigma_{Z,k} \,, \qquad \sum_k F_k(x) = 1 \,. \qquad (1)$$

where $F_Z$ is a fraction of ions with charge state $Z$ to be found at the depth $x$ (which is defined in the units of surface density $x=n_{at}L$ with $n_{at}$ being the target atomic density and $L$ the SHI penetration depth); $\sigma_{k,Z}$ are the cross-section of the charge state change from $k$ to $Z$, which corresponds to the electron loss for $Z>k$, or electron capture for $Z<k$ [61]. The fractions are assumed to be normalized to unity, so they can be interpreted as distributions or probabilities to find an ion in a charge state $Z$.

Both methods, Monte Carlo and rate equations can be found in the literature. A Monte Carlo simulation sampling process of electron loss and capture was implemented in such codes as ETACHA [60] and its recent extension to heavier elements [62]. The rate equation solutions are used in codes GLOBAL [63], CHARGE [63], and BREIT [64], and were also written in terms of a matrix formalism in Ref. [65]. A detailed review of the experiments and theoretical and numerical methods of calculation of an SHI charge evolution has recently been published in Ref. [61].

Such methods work well at high ion energies, at the right shoulder of the Bragg curve. At even higher ultra-relativistic energies, ions are fully stripped. However, at lower energies, at the left shoulder of the Bragg curve, they may not work so well due to the solid-state effects of a target. Further corrections need to be introduced to account for them, such as target-density effect and shell corrections [61].

After some traveled distance, the processes of electron capture and loss eventually even out, reaching a steady state – an equilibrium charge with some fluctuations around it [61]:

$$\bar{Z} = \sum_k k F_k(x \to \infty). \tag{2}$$

This definition assumes that the SHI does not lose its energy along its path. In reality, the equilibrium charge will evolve along the trajectory due to the deceleration of the ion.

If an SHI enters a material with a charge higher than the equilibrium charge state, the equilibration occurs when an SHI captures enough electrons in its deep shells so that an additional electron will be captured only to higher lying states, whose orbital velocity is smaller than the speed of the SHI. This means that such an additional electron will be lost very fast. If an SHI enters the target with a charge below the equilibrium charge, it will lose extra electrons to reach equilibrium. An electron loss is typically a much faster process than capture, and the equilibration depth of ions is, correspondingly, shorter.

A typical example of charge state equilibration of sulfur ion in carbon foil is shown in Figure 3 [66]. An ion of sulfur with the starting energy of 2 MeV/a.m.u. and various initial charge states entering a carbon foil quickly loses its electrons (increases its charge) if starting with a charge below the equilibrium one ($\bar{Z} = 12.68$), but it takes longer (a larger depth) to capture electrons and equilibrate if starting with a charge above the equilibrium one.

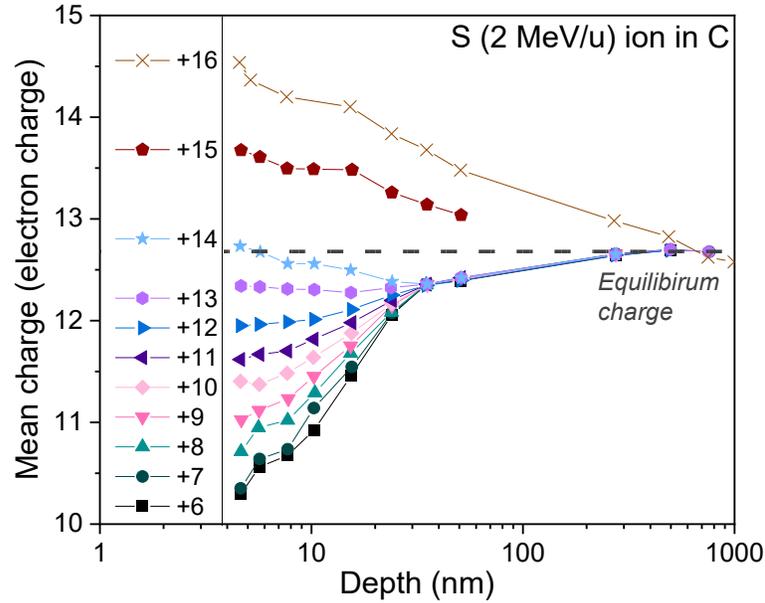

*Figure 3. Mean charge of S ion with starting energy of 2 MeV/a.m.u. and different incoming charge states from +6 to +16 in a carbon foil target. Note the log scale of the depth. Data taken from Ref. [66].*

The charge equilibration depth of an ion starting with a below-equilibrium charge state in a solid does not typically exceed a few hundred nm [60,67], which is much shorter than the total ion penetration depth (~10-100 μm). That means, an ion has an equilibrium charge on the major part of the trajectory within the bulk. In experiments, deceleration foils are often used to control the SHI energy, which simultaneously will equilibrate the ion charge before its arrival at the surface of the target.

The values of the reached equilibrium charge depend mainly on the ion velocity or energy, whereas dependence on the parameters of the target is rather weak. There have been proposed a number of different methods to evaluate the equilibrium charge of an ion in a media [68]. Perhaps the most accurate empirical estimate was proposed in Ref. [69], which was implemented in the CasP code [70], to reproduce the equilibrium charge state in agreement with experimental values over a wide range of targets and SHI velocities:

$$\bar{Z} = \frac{Z_{ion}(8.29x + x^4)}{0.06x^{-1} + 4 + 7.4x + x^4} \, , x = c_1 \left(\frac{\tilde{v}}{1.54c_2}\right)^{1+1.83/Z_{ion}} ,$$

$$c_1 = 1 - 0.26 \exp\left(-\frac{Z_t}{11} - \frac{(Z_{ion} - Z_t)^2}{9}\right), \tag{3}$$

$$c_2 = 1 + 0.03\tilde{v}\ln(Z_t) \, , \tilde{v} = Z_{ion}^{-0.543}\frac{v_{ion}}{v_B}.$$

Where $Z_{ion}$ is the atomic number of the SHI, $Z_t$ is the atomic number of the target element, $v_{ion}$ is the ion velocity, and $v_B$ is the Bohr velocity. This expression allows for evaluation of the mean equilibrium charge of an ion in an arbitrary target, without solving the system of equations (1)-(2). Similar models were proposed e.g. in Refs. [71] and [72].

It is important to keep in mind the distinction between the *equilibrium* charge and an *effective* charge of an ion often mentioned in the literature [61]. Whereas the former is a real average ion charge that may be measured in experiments, the latter is a fictitious charge that does not correspond to the real one but serves other purposes. Namely, the effective charge is empirically adjusted to reproduce correct ion energy losses within a chosen model – the linear response theory (first-order Born approximation). In practice, effective charge allows an easy rescaling of the energy loss calculations between different elements. Reliably calculated energy loss for one ion with a charge $Z_1$ (e.g., protons) enables estimating the stopping of another ion with a charge $Z_2$ at equal velocity *v* as:

$$S_e(Z_2, v) = S_e(Z_1, v)(Z_2/Z_1)^2. \tag{4}$$

The scaling rule follows from the dependence of the energy loss on the ion charge, as will be seen below in Section II.C.

One of the most common effective charge models was developed by Bohr [1], and later adjusted by Barkas [73]:

$$Z_{eff}(v) = Z_{ion}\left[1 - \exp\left(-\frac{v_{ion}}{v_0}Z_{ion}^{-\frac{2}{3}}\right)\right], \tag{5}$$

where $v_0$ = *Ac*, with *c* being the speed of light in a vacuum. In Bohr's original model, *A* = $\alpha$ = 1/137, whereas the Barkas model assumes *A* = 1/125, significantly improving the agreement of the energy loss calculated within the linear response theory with experiments. This model is very convenient due to its simplicity since it depends only on the ion atomic number and velocity. More complex expressions for the effective charge, such as the Brandt-Kitagawa model depending on the properties of the target [74], are also used in the standard codes simulating ion ranges in solids, such as SRIM [4], GEANT4 [75], and FLUKA [76].

So, we can conclude that the practical recommendation is, for the evaluation of a realistic ion charge after passing a certain thickness of matter, the average charge should be used. For calculations of the inelastic energy loss of an ion within the linear response (first-order perturbation theory), the effective charge can be used, as will be discussed below in Section II.C.

## II.B.    Nuclear energy loss

Energy transfer to target atoms/ions without excitation of electrons forms the nuclear energy loss channel of an ion (*nuclear stopping*) which is significant for ion energies below ~0.1 MeV per nucleon. In such a scattering event, the total kinetic energy of the system is conserved, it only redistributes between the projectile and the atoms of a target.

Being a subject of radiation physics, the regime of nuclear stopping is well studied [58]. At low energies corresponding to the left shoulder of the Bragg curve, the processes of ion penetration through a solid may be modeled with high precision using *ab-initio* techniques, such as time-dependent density functional theory (TD-DFT [34,35]), density-functional-theory molecular dynamics (DFT-MD [77]), or its simplified version such as tight-binding molecular dynamics (TBMD [78]). Other approximate methods to model SHI nuclear stopping and materials response to it include classical molecular dynamics (MD) simulation, Monte Carlo (MC) method with binary collision approximation (BCA) and kinetic MC (KMC) for modeling target atomic cascades [48].

The nuclear energy losses are small for swift heavy ions decelerating in the electronic stopping regime, see Figure 1. Nonetheless, for some applications it may be important for at least three reasons: at the end of a trajectory of an SHI the nuclear stopping always dominates (see Figure 4), as long as the ion is inside of material; at the surface, nuclear scattering induces experimentally observable sputtering effect (a part of it, at least [79]); nuclear scattering may drastically alter the SHI trajectory in close collisions which deflect the ion momentum considerably, however rare such an event might be.

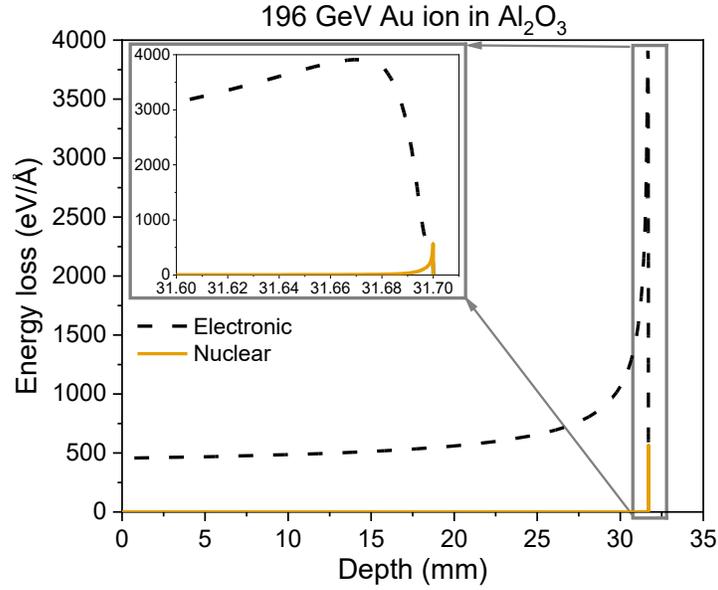

*Figure 4. Inelastic and nuclear energy loss along the trajectory of 196 GeV gold ion in Al₂O₃ according to SRIM [4].*

Atomic motion and interaction between an ion and a target atom can be very well approximated by classical mechanics, assuming Newtonian equations of motion with an appropriate interaction potential [80]:

$$M_i \frac{d^2 \boldsymbol{R}_i}{dt^2} = -\frac{\partial V(\{R_{ij}\})}{\partial \boldsymbol{R}_i}. \qquad (6)$$

Here $M_i$ is the mass of a simulated particle (an ion or a target atom), $\boldsymbol{R}_i$ is its coordinate vector and $R_{ij}$ is the distance between a pair of atoms $i$ and $j$, $V(\{R_{ij}\})$ is interaction potential (collective potential energy surface or a pairwise potential, depending on all atoms $j$ involved). Equations of motion (6) for the ion-matter scattering may be solved using molecular dynamics (MD) simulations [80]. The set of Eqs.(6) is solved for all the atoms in the simulation box by numerical discretization of time into time steps, and propagating coordinates of all atoms accounting for the dynamical change of the potential of interaction among them.

In the case of *ab-initio* MD simulations, the potential is calculated from the first principles using, for example, DFT or TB Hamiltonian (more on this point will be discussed in Sections III and IV). However, the problem may be simplified, considering that significant energy exchange between a fast projectile and ions of the target will only take place at very short distances (close collisions). Thus, it is often sufficient to account only for ion-ion repulsion, and the potential of interaction may be considered independent of the electronic states of the involved ions. That means, an empirical potential may be set as a fitted function to reproduce the exact interaction potential in the electronic ground state. A screened Coulomb potential is often used for modeling ion-ion scattering. The most common expression for screening is the universal Ziegler-Biersack-Littmark (ZBL) potential [4]:

$$V(r) = \frac{Z_{ion}Z_t e^2}{r} \, \Phi\left(\frac{r}{a}\right),$$

$$a = \frac{0.8854 \, a_0}{Z_{ion}^{0.23} + Z_t^{0.23}},$$

$$\Phi\left(\frac{r}{a}\right) = 0.1818 \exp\left(-3.2\frac{r}{a}\right) + 0.5099 \exp\left(-0.9423\frac{r}{a}\right)$$

$$+ 0.2802 \exp\left(-0.4028\frac{r}{a}\right) + 0.02817 \exp\left(-0.2016\frac{r}{a}\right).$$

(7)

where $Z_{ion}$ is again the atomic number of an impinging ion, $Z_t$ is the atomic number of a target ion, $r$ is the interaction distance, $a_0$ is the Bohr radius, and $e$ is the electron charge. The ZBL potential was constructed by fitting the screening function parameters $\Phi(r/a)$ to theoretically obtained potentials for a large variety of pairs of different elements, and thus is indeed quite universal [4]. There are, however, a lot of other potentials also applied to simulate ion-ion scattering in different materials, see e.g. Ref. [58]. Since only the repulsive part of the potential is included, ZBL potential cannot describe the material properties, but only the ion-ion scattering.

In practical implementations of ion-ion scattering, one must take special care of the fact of a very fast SHI in a steep potential: the time-step of the MD simulation needs to be chosen sufficiently small not to lead to numerical instabilities. A practical solution may be a usage of an adaptive time step, e.g. $\Delta t_{new} = min(k_t/v; E_t/(Fv); 1.1\Delta t_{old})$, where $v$ is the recoil velocity with a chosen proportionality constant $k_t$; $F$ is the total force the recoil atom experiences, and another proportionality constant $E_t$ [81].

Another important aspect is, to model an entire SHI trajectory in a material, it is not possible to treat the millimeter-size target with MD with present-day computers. Instead, one may employ a dynamical simulation box, following the ion trajectory with the target atoms entering and leaving the box as needed [82]. It allows tracing the entire SHI trajectory, however, not the detailed response of the target.

A further simplification often used for the scattering of an SHI with energy above ~1 MeV/nucleon is an assumption of a target to be a uniform homogeneous arrangement of atoms. Under this approximation, it is unnecessary to trace the motion of all target atoms; instead, it suffices to sample randomly interaction points of an SHI. Between the scattering events (points of interaction), an SHI travels along a straight line, changing its direction of motion only at a scattering event. At these points of interaction, a change of the SHI momentum is sampled probabilistically, according to its scattering cross-section. This method is known as the binary collision approximation (BCA [48,58]), and it belongs to the Monte Carlo (MC) methods [83–85]. The cross-section of scattering may be obtained in the classical limit by integration of the potential (e.g., Eq.(7)) over the impact parameter (all possible distances between an SHI and a target atom).

Within the MC method, the cross-section of scattering, $\sigma$, fully defines the scattering process [40,41,85]. The mean free path – the distance a particle travels between successive collisions, $\lambda$, – is defined as follows:

$$\lambda^{-1} = \sigma n_{at} = n_{at} \int_{W_-}^{W_+} \int_{Q_-}^{Q_+} \frac{d^2\sigma}{dWdQ} dWdQ, \qquad (8)$$

here $n_{at}$ is the density of target atoms; $W_{\pm}$ are the minimal and maximal transferred energy ($W = \hbar\omega$); the recoil energy $Q$ is defined by the transferred momentum in a collision as $Q(1 + Q/2m_tc^2) = \hbar^2 q^2/(2m_t)$ (in the nonrelativistic limit $Q = \hbar^2 q^2/(2m_t)$, with $\hbar q$ being the transferred momentum and $m_t$ is the scattering center (atom) mass, see below, Eq.(15).

In the nonrelativistic limit, the differential scattering cross section becomes [86]:

$$\frac{d\sigma}{dt} = \frac{-\pi a^2}{2} \frac{f(\sqrt{t})}{t^{3/2}}, t = \epsilon \frac{E_{ion}}{E_{max}},$$

$$f(x) = \frac{d}{dx}\big(xS_n(x)\big), \qquad (9)$$

$$\epsilon = \frac{32.53 \, m_t Z_{ion}}{Z_{ion} Z_t (M_{ion} + m_t)\big(Z_{ion}^{0.23} + Z_t^{0.23}\big)}$$

Where the screening coefficient $a$ is defined by Eq.(7); $\epsilon$ is the reduced energy variable; $M_{ion}$ is the SHI mass, $m_t$ is the scattering center (atom) mass; the maximal transferred energy, $W_+$, in the non-relativistic limit is:

$$W_+ = E_{max} = \frac{4M_{ion}m_t E_{ion}}{(M_{ion} + m_t)^2}, \qquad (10)$$

The function $S_n(x)$ is the nuclear energy loss function of an SHI (nuclear stopping power) [58]:

$$S_n(E_{ion}) = \frac{8.462 \times 10^{-15} Z_{ion} Z_t M_{ion}}{(M_{ion} + m_t)(Z_{ion}^{0.23} + Z_t^{0.23})} S_n(\epsilon),$$

$$S_n(\epsilon) = \frac{0.5\ln(1 + 1.1383\epsilon)}{(\epsilon + 0.01321\epsilon^{0.21226} + 0.19593\epsilon^{0.5})}.$$

(11)

At the relativistic energies, the cross-section of elastic (nuclear) scattering, and the corresponding energy loss, can be obtained, e.g., from the Mott or Wentzel-Moliere screened scattering cross sections [87,88] (renormalized by the ion charges [89]):

$$\frac{d\sigma}{d\Omega} = \left(\frac{Z_{ion} Z_t e^2}{p_{ion} v_{ion}}\right)^2 \frac{1}{(2\eta + 1 - \cos(\theta))^2},$$

(12)

here $\Omega$ is the solid angle, $p_{ion}$ is the incident ion momentum, $\theta$ is the scattering angle, and $\eta$ is a screening parameter [40].

Concluding this section, within the electronic stopping regime of an SHI, nuclear stopping is usually negligible and may be ignored in reliable models. In cases when it is important, such as at the left shoulder of the Bragg curve where the nuclear stopping may be comparable with the electronic one, or modeling an entire SHI trajectory where the very end of it is dominated by the nuclear cascades, or to take into account rare elastic scattering events in the electronic stopping regime, Monte Carlo methods within the binary collision approximation may be used. For an even more detailed analysis, molecular dynamics simulations may be employed with ZBL ion-ion interaction potential.

Further, we will be focusing mainly on the electronic stopping regime; as for the nuclear stopping regime and induced effects, the reader may find detailed information e.g. in Ref. [58].

## II.C.    Electronic energy loss

Electronic energy loss (or *electronic stopping*) of an ion refers to inelastic energy transfer to media, exciting electrons of the target. In metals, an arbitrary amount of energy may be transferred to an electron, as there are electronic energy levels just above the Fermi level. In band gap materials (semiconductors, insulators), only energy larger than the band gap of the material may be transferred to an electron, exciting it from the valence band or deep atomic shells to the conduction band. This process is usually referred to as electron ionization (impact ionization) [90]. In such an inelastic scattering event, the kinetic energy is not conserved in the system — a part of this energy is lost to overcome the ionization potential (in the case of ionization of an electron from a deep atomic shell) or the band gap (in a case of valence electron ionization). Such an event also leaves an electronic hole in the shell an electron is emitted from. Interaction and energy transfer may also take place simultaneously for many electrons of a dense media forming collective excitations, plasmons.

An ion penetration through a solid may be modeled with time-dependent *ab-initio* methods (such as TD-DFT) very precisely [91–94]. TD-DFT methods are tracing the movement of the classical ion and all the target atoms in a simulation box in the evolving potential set by the quantum electrons [33]. DFT methods are based on the substitution of electrons by non-interacting quasiparticles evolving in an effective potential including exchange-correlation functional, which is unknown and requires an approximation. Even the simplest exchange-correlation functionals deliver good quantitative results, however, at a high computational cost [95].

For example, ion-irradiation-induced direct damage of dry DNA duplex by energetic protons and α-particles was presented in [96]. Based on the TD-DFT approach, formation of holes after irradiation was observed in a large-scale DNA model system. The modeling demonstrated that the maximal number of holes were generated at ion energies below the Bragg peak, whereas experiment showed the maximal damage above it. This discrepancy amplifies the importance to trace all the stages of damage formation, from an SHI impact to final material response at longer timescales.

One of the key parameters describing an ion penetration in a media is its electronic energy loss, $S_e$, or the stopping power, which is accessible in experiments. It can be extracted from TD-DFT calculations directly as the average of a retarding force acting on the moving SHI, $\vec{F}_{ion}$, or as its average energy loss [94]:

$$S_e(E_{ion}) = -\langle \vec{F}_{ion} \vec{v}_{ion} \rangle \frac{1}{v_{ion}} = -\langle \frac{dE_{ion}(t)}{dt} \rangle \frac{1}{v_{ion}}, \qquad (13)$$

where the force can be obtained as the Ehrenfest force formed by all the contributing particles [94]. Electronic energy loss calculated in Ref. [93] with this method revealed interesting insights into the initial excitation kinetics: e.g., a passing ion drags target electrons behind it creating a comet-like tail. The standard TD-DFT method with the Ehrenfest approximation describes the left shoulder of the Bragg curve at low ion energies, however, the right shoulder can only be described if non-adiabatic effects are included, e.g. by means of the correlated electron-ion dynamics (CEID) method [97]. Non-adiabatic effects mean that energy transfer between electrons and ions occurs due to electron transitions between energy levels accompanied by momentum transfer into the atomic system of a target (e.g., elastic electron scattering on atoms or electron-phonon coupling).

To the best of our knowledge, no *ab-initio* method has yet been applied to the relativistic regime of an SHI.

An additional difficulty with a practical application is that DFT-based simulations often use pseudopotentials for core electrons (i.e. they replace deep shell electrons and nucleus with an effective potential [98]) to reduce the number of electrons in the simulations and to be able to include a sufficient amount of atoms in the simulation box. As a consequence, a constrain on the impact parameter appears – it should be chosen so that the ion trajectory does not pass close to atomic nuclei, thus neglecting a contribution to the energy loss from core shells, which may be significant (as will be discussed below, see Figure 6).

Since the dynamics of all the electrons of the projectile *and* all the target atoms is traced, it makes the *ab-initio* computations very time-consuming. In practice, it is limited to the modeling of a few tens of atoms up to the times of a few femtoseconds with modern-day computers. Due to high demands on computational resources and, in some cases, model limitations, *ab-initio* methods are usually employed as supportive tools for better understanding of a particular physical process or to provide a parametrization for less computationally demanding approaches. Delivered from TD-DFT the most accurate information on the ion penetration effects allows to validate other models and identify their limitations [94]. To extend DFT-based methods to larger systems, further simplifications are used, such as the linear-scaling density functional theory (LSDFT). It employs a simplified algorithm of DFT Hamiltonian diagonalization *via* splitting the large simulation box into overlapping smaller ones [99]. Within this approach, at the cost of a moderate accuracy loss, one can simulate some thousands of atoms. Such methods are yet to be applied to the irradiation problem.

The Monte Carlo method [75,76] is the most commonly used simplified approach for SHI penetration. The same Eq.(8) can be used with the cross-section of the inelastic scattering replacing the nuclear scattering one. The electronic stopping power of a projectile with energy $E$ can also be calculated (analogously to the nuclear stopping power) *via* the differential cross section of scattering as follows [6]:

$$S_e(E) = n_{at} \int_{W_-}^{W_+} \int_{Q_-}^{Q_+} W \frac{d^2\sigma}{dWdQ} dWdQ. \tag{14}$$

The integration limits defining the maximal and minimal allowed momentum transfer in the relativistic case are defined as follows, assuming scattering on a free particle [41]:

$$Q_{\pm} = \sqrt{\left( \sqrt{E(E + 2Mc^2)} \pm \sqrt{(E - W)(E - W + 2Mc^2)} \right)^2 + (m_t c^2)^2} - m_t c^2, \qquad (15)$$

here $M$ denotes the mass of the incident particle (here equals the SHI mass, but the same can be applied to other particles, e.g., electrons). In the non-relativistic case ($E \ll min(M, m_t)c^2$), the limits of the transferred momentum, Eqs.(15), simplify to the form [100]:

$$Q_{\pm} = \frac{M}{m_t} \left( \sqrt{E} \pm \sqrt{E - W} \right)^2. \qquad (16)$$

The lower limit of the transferred energy ($W_-$) and the upper limit ($W_+$) for scattering on a target particle (an electron, in this case) are defined by the following formulae:

$$\begin{cases} W_- = I_p \\ W_+ = \dfrac{2m_t c^2 E(E + 2Mc^2)}{2m_t c^2 E + (Mc^2 + m_t c^2)^2}, \end{cases} \qquad (17)$$

where $I_p$ is the ionization potential of the atomic shell an electron is being ionized from. The upper limit is written here for a free particle. For scattering on a bound particle with a given ionization potential, the expressions are more complicated, see Ref. [6], but the free-particle approximation works very well for $W_+ \gg I_p$. In the case of ionization from the valence band $I_p = E_{gap}$. Here, $E_{gap}$ is the material band gap, which is equal to zero in the case of metals. In the non-relativistic limit, the upper limit simplifies to $W_+ = E_{max}$ from Eq.(10).

The cross-section of scattering of a projectile on a complex correlated system (a solid) is difficult to calculate. The question of a nonperturbative closed solution or at least a simple and efficient model of an SHI scattering in solids without empirical adjustable parameters is still open.

Most often, the perturbation theory is used to evaluate the inelastic scattering cross-section, with only the leading term included – the first-order Born approximation. It further assumes plane waves of the incident particle [101]. It is also known as the linear response theory due to its connection to the complex dielectric function (CDF) of the material, which will be revealed below.

In a general relativistic case, a double differential cross section is expressed in terms of a longitudinal and a transverse contribution [54]:

$$\frac{d^2\sigma}{dWdQ} = \left( \frac{d^2\sigma}{dWdQ} \right)_l + \left( \frac{d^2\sigma}{dWdQ} \right)_{tr}. \qquad (18)$$

The longitudinal term is responsible for the Coulomb scattering, whereas the transverse one is a result of virtual photons exchange [41,54]. Eq.(18) can be expressed in the following terms [54]:

$$\frac{d^2\sigma}{dWdQ} = \frac{2\pi Z_{eff}^2(E)e^4}{m_t c^2 \beta^2}\left[\frac{|F_n(q)|^2}{Q^2(1+Q/(2m_t c^2))^2} + \frac{|\beta_\perp G_n(q)|^2}{(Q(1+Q/(2m_t c^2))-W^2/(2m_t c^2))^2}\right]\left(1+\frac{Q}{m_t c^2}\right), \quad (19)$$

here $\beta = v/c = \sqrt{1-(1+E/Mc^2)^{-2}}$ is the incident particle velocity normalized to the speed of light in vacuum $c$, $\vec{\beta}_\perp = \vec{\beta} - (\vec{\beta}\cdot\vec{q})\vec{q}/q^2$ is the component of $\vec{\beta}$ perpendicular to the transferred momentum $\vec{q}$, $Z_{eff}(E)$ is the effective charge of the incident particle (as discussed above in Section II.A), and $a_0$ is the Bohr radius.

The longitudinal contribution contains $F_n(q)$ which is the inelastic form factor [54] (index $n$ denotes the energy state corresponding to the transferred energy W=$\hbar\omega$), squared modulus of which is proportional to the dynamic structure factor (DSF) usually denoted as $S(\omega,q)$ [101]. With some approximations, the transverse contribution $G_n(q)$ can also be expressed in terms of $F_n(q)$ [41]; this term vanishes in the nonrelativistic limit, and the longitudinal term defines the interaction.

The DSF can be recast in terms of the so-called generalized oscillator strengths [41], and is connected to the complex dielectric function of the solid ($\varepsilon(\omega,q)$) via the fluctuation-dissipation theorem [102]:

$$|F_n(q)|^2 \equiv \frac{S(\omega,q)}{\hbar} = \frac{q^2}{n_{at}4\pi^2 e^2}\frac{1}{1-\exp(-\hbar\omega/(k_B T))}Im\left(\frac{-1}{\varepsilon(\omega,q)}\right), \quad (20)$$

which is valid under the assumption of local thermal equilibrium of the homogeneous target; $T$ is the temperature of the solid, and $k_B$ is the Boltzmann constant. Note that the CDF also implicitly depends on the target temperature.

Thus, within the linear response theory, the problem reduces to defining an appropriate CDF, or the loss function $Im\left(\frac{-1}{\varepsilon(\omega,q)}\right)$. Even though it may be calculated with *ab-initio* methods such as DFT, it is in practice still rather time-consuming to do it sufficiently precise [103]. There are a few models proposed for the CDF evaluation from the experimental (or *ab-initio* calculated) optical coefficients (refraction and transmission coefficients [104,105]), allowing to reconstruct $Im\left(\frac{-1}{\varepsilon(\omega,q=0)}\right)$ [106–109]. Having the optical coefficients, the CDF can then be analytically continued to the entire plane of $q$>0, via a few different methods [106–109].

Approximating the loss function in Eq.(20) with the Dirac delta-function centered around a mean ionization potential of a target, $\langle I_p \rangle$, and integrating Eqs.(14,19) produces the well-known Bethe-Bloch formula [110,111] for inelastic ion energy losses of an SHI at the right shoulder of the Bragg peak [54]:

$$S_e(E) = \frac{4\pi Z_{eff}^2 Z_t n_{at} e^4}{m_e c^2 \beta^2} \left[ ln\left( \frac{2m_e c^2 \beta^2}{\langle I_p \rangle (1 - \beta^2)} \right) - \beta^2 \right]$$

Note that Eqs.(14,19) allow for evaluation of the stopping power without an assumption of a mean ionization potential, using precise CDF peaks instead [6].

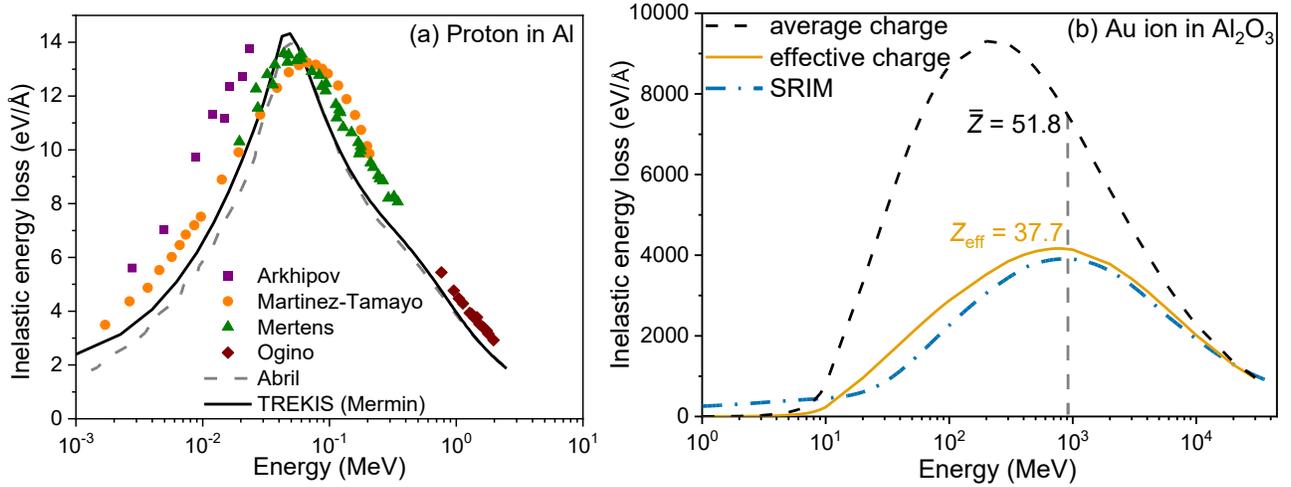

*Figure 5. (a) Proton energy losses in aluminum, calculated with TREKIS using Mermin-like CDF [5], the same calculations by Abril et al. [108], experimental data from Refs. Al [112–115]. (b) Inelastic energy loss of Au ion in Al$_2$O$_3$ calculated with equilibrium average charge, Eq.(3), vs. effective charge, Eq.(5), calculated with TREKIS [5], compared with the reference data from SRIM [4]. Charges are marked at selected energy of 900 MeV (Bragg peak) shown with the vertical line.*

A very convenient way to reconstruct the loss function from the optical coefficients, and extend it to non-zero $q$ region, was proposed by Ritchie and Howie [106]. It approximates the optical coefficients with Drude-Lorentz oscillators. The method was later extended by replacing the Drude oscillators with Mermin-like dielectric function [108], providing a better agreement of the resulting mean free paths and stopping powers with experimental data. Further improvements of the approach include Coulomb-field corrections to the first Born approximation — potential energy gained by the incident electron in the field of an atom (or a molecule) can be accounted for [116,117]. A detailed description of various models of CDF can be found in the review [32]. More details and examples on reconstructing CDF from optical coefficients will be discussed in Section VIII.

Properly selected CDF (with coefficients from Ref. [108]) with the most advanced model currently available – the Mermin model – reproduces energy losses of protons, see example in Figure 5. For heavy ions, it is not the case, and the effective charge, discussed in Section II.A, must be used to obtain a reasonable ion energy loss, see Figure 5b. An actual average charge (Eq.(3)), within the linear response theory, produces the stopping power drastically overestimating the reference one. It is thus clear that the effective charge (Eq.(5)) is used as an *ad hoc* parameter. The underlying physical reason for this will become clearer from the following considerations.

The total mean free path of an ion is dominated by the scattering on the valence band of the material, whereas scattering on the deeper atomic shells is much rarer forming much larger mean free paths (Figure 6). However, scattering on deeper shells contributes noticeably to the stopping power – e.g., nearly 30% of the energy loss of Kr ion in $Al_2O_3$ (see Figure 6b). This is because, in an event of scattering on deeper shell electrons, an ion loses at least the energy equal to the ionization potential of the shell. Thus, although scattering events on deep shells electrons are rarer than interaction with the valence band, a lot more energy is transferred in each of them, resulting in a comparable energy loss.

A projectile may interact with many particles at once, and its reach of the potential (impact parameter) extends beyond the nearest neighbor distance. It becomes obvious if we look at the mean free path of heavy ions, Figure 6. An SHI interacts with many electrons of the same atom, and its potential created by a high charge reaches a few-neighbor distance around its trajectory. This makes the mean free path much shorter than the mean interatomic distance in $Al_2O_3$ (~2 Å), and even shorter than the average inter-electronic distance (~0.2 Å), clearly indicating ion interaction with more than the nearest electron at once.

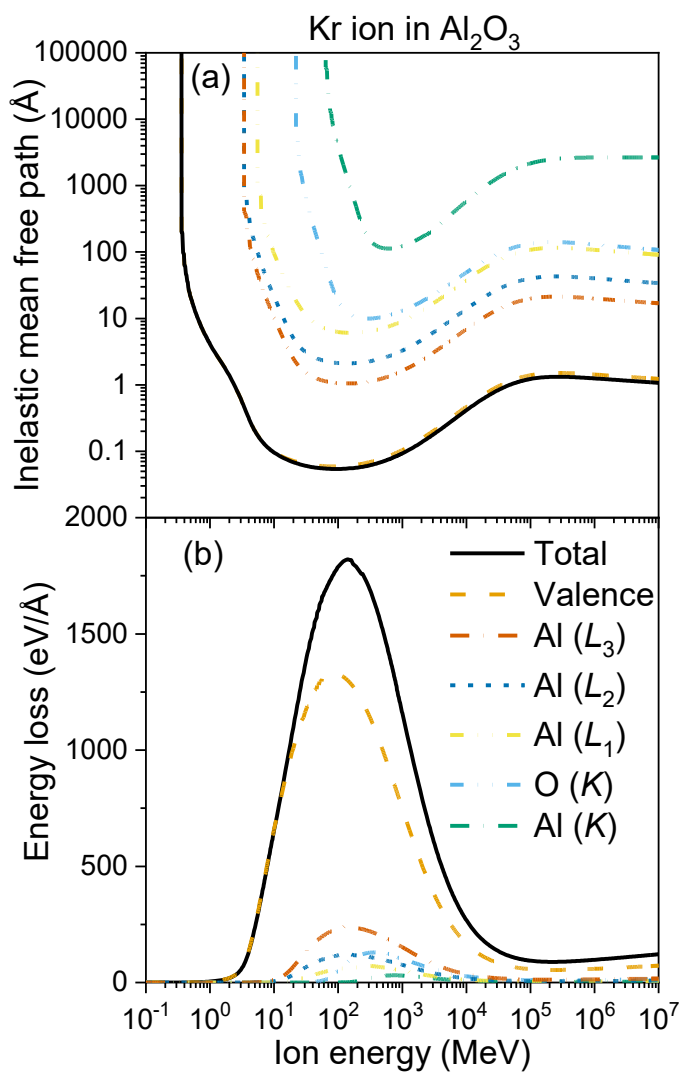

*Figure 6. Total and partial inelastic (a) mean free path and (b) energy loss of Kr ion in Al₂O₃ of scattering on various shells and valence band of the material, calculated with TREKIS [5,6].*

It leaves multiple-ionized target atoms after an SHI passage. Figure 7 shows the experimental spectra of photons emitted from target ions with different numbers of holes left in Si atoms in fused SiO₂ after a swift Ca ion impact. They are measured via spectroscopy of the photon emission due to radiative decays in such atoms (more details on the processes of hole decays will be discussed in Section III.B). They provide information about the distribution of charges of target atoms in a track core – the region excited by an SHI directly – thereby allowing to test models of ion interaction with a solid target [52].

The multiple ionized states are created by an SHI, but not by further kinetics of electrons, and thus can clearly be distinguished.

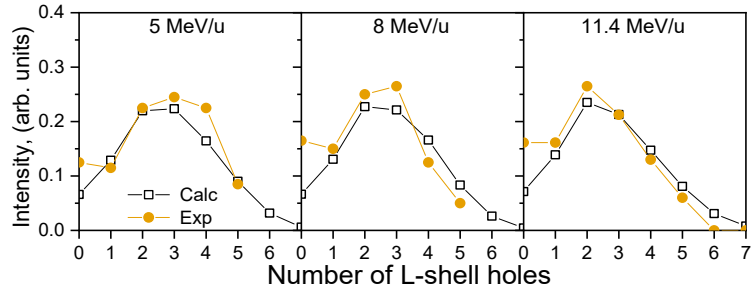

*Figure 7. Experimental and calculated spectra of radiative decays of holes in K-shell of Si atoms with a given number of holes in L-shell in silica irradiated with Ca ions with different energies (5 MeV/a.m.u., 8 MeV/a.m.u. and 11.4 MeV/a.m.u.). Adapted from Ref. [52].*

The ionization potential of an ion is larger than that of a neutral atom [118]. It depends on the particular configuration of electron populations on all the shells of an ion, but the general tendency is that the ionization potential increases with the increasing charge (ionization degree) of an ion [119]. That means, in a multiple-ionized ion of a target, ionization potentials increase, making it more difficult to ionize each successive electron [119].

Due to this difference, using atomic ionization potentials, thus, produces more ionizations (and higher energy losses) than it would if the ionization potential changes were accounted for. This kind of non-linear response of the target seems to be a reason for the need to adjust the energy losses within the linear response theory – which is done *via* using an effective charge.

In MC simulations, often the method of *condensed history* (or condensed collisions) simulations is used [75,76]. In this approach, only the close collisions (with the energy transfer above a certain threshold) are modeled in detail, whereas distant collisions (with a small energy transfer) are averaged along the SHI trajectory and treated as a deceleration with the stopping power $S_e(E)$ calculated with Eq.(14). Such a calculation method significantly speeds up a computation, enabling very efficient modeling of an SHI transport, but a loss of information on the created electrons does not allow for modeling of the target response.

In contrast, an *event-by-event* (or analog) MC simulation traces each scattering event in detail, which is sufficient for tracing material response [120]. Transferred energy in a scattering act is sampled according to the following expression:

$$\gamma\sigma = \int\limits_{W_-}^{W} \int\limits_{Q_-}^{Q_+} \frac{d^2\sigma}{dWdQ} \, dWdQ \tag{21}$$

here $\gamma \in [0,1)$ is a random number; Eq.(21) must be solved for the transferred energy $W$. In some cases, the differential equation allows for analytical integration and a closed solution for $W$ [6,41]. In general, a numerical solution is required [5], or further approximations to the electron spectrum should be employed [41]. An electron is emitted with the kinetic energy $E_e = W - I_p$.

An example of a spectrum of electrons created by an impact of Au ion with the energy 2187 MeV in LiF (before electron cascades start) is shown in Figure 8. This typical spectrum follows the Rutherford law at large energies (significantly above the plasmon peak energy $\hbar\omega_p$), scaling as $\sim 1/E_e^2$. At lower energies, a noticeable deviation from this law takes place. It follows from the fact that at low energies, collective electron effects (screening, plasmon formation) define the scattering event, whereas at transferred energies $W \gg \hbar\omega_p$, the complex dielectric function tends to unity, producing unscreened potential [121]. Scattering on an unscreened Coulomb potential, thus, follows the well-known Rutherford shape.

The recoil angle of a target particle with the mass $m_t$ (an electron, $m_t = m_e$, or an atom) after receiving energy $W$ from an incident particle with the kinetic energy $E$ and mass $M$ is:

$$\cos(\theta_r) = \frac{\sqrt{W}(E + m_t c^2 + M c^2)}{\sqrt{E}\sqrt{E + 2Mc^2}\sqrt{W + 2m_t c^2}}, \tag{22}$$

or in the nonrelativistic case:

$$\cos(\theta_r) = \sqrt{W}\frac{(M + m_t)}{\sqrt{4Mm_t E}}.$$

Thus, the angular dependence of the electron spectrum is $\sim 1/\cos^4(\theta)$, which has a maximum at the perpendicular direction with respect to the SHI trajectory. That means, the majority of electrons have low energies (large impact parameters) and are emitted perpendicularly to the SHI velocity, whereas a small minority of electrons with high energies (from head-on collisions) are emitted along the SHI trajectory.

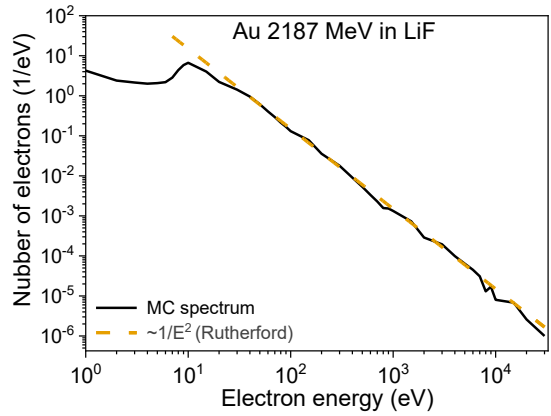

*Figure 8. The spectrum of electrons generated by Au 2187 MeV ion in LiF, calculated with Monte Carlo code TREKIS* [5]. *A characteristic shape of a Rutherford scattering ($\sim 1/E_e^2$) is shown for comparison.*

It is important to keep in mind that transport Monte Carlo simulations rely on Markovian approximation [33]. Each scattering event is assumed to be instantaneous and independent of other scattering events. This is valid in a case when the duration of an individual event is much shorter than the free flight time, which is satisfied for close collisions of electrons with energies above some ~50 eV. Distant collisions, especially of slow electrons, transfer a small amount of energy, which according to the quantum speed limit theorem requires sufficiently long times [122]. For example, a phonon emission requires a time proportional to the inverse phonon frequency. It is still an open question if and how finite duration of scattering and energy exchange may be incorporated into MC and other Markovian methods (a possible solution for the case of Boltzmann scattering integrals was proposed in Ref. [123]).

Summarizing, low-energy ion transport can be precisely described with *ab-initio* methods such as TD-DFT but at a high computational cost. Fast calculations based, e.g., on the Monte Carlo method, require reliable cross sections. An absolute majority of available cross sections rely on the first-order Born approximation (linear response theory) and thus use semi-empirical adjustments at low ion energies, such as an effective charge of an SHI. At high energies, linear response theory based on the longitudinal complex dielectric function works very well and can provide reliable results for SHI energy loss and spectra of excited electrons. At relativistic energies, the effects of the transverse dielectric function contribution should be taken into account, and the finite size of the SHI nucleus plays a very important role for ultra-relativistic projectiles. Created spectra of SHI-excited electrons have a characteristic shape of $\sim 1/E_e^2$, except for the low energies where collective (plasmon) effects dominate and make the spectra deviate from this dependence. A majority of electrons with relatively low energies, traveling perpendicular to the SHI trajectory, are created, with a few but highly energetic delta-electrons, initially directed along the SHI trajectory. Simultaneously, valence- and core-shell holes with multiple-charged ions are left in the target. This sets the initial conditions for further electrons and holes kinetics.

## II.D.    Other processes

Apart from phenomena originating from the nuclear and electronic energy loss of an SHI, other effects manifest themselves under certain conditions.

The effect of ion channeling can occur when the incident direction of an ion beam is aligned with a particular axis of a crystal [124]. In this case, an ion travels in-between the atomic planes without meeting a significant electron density. It reduces the SHI energy losses and allows traveling within this "channel" much larger distances than its typical range. This geometric effect is found in various experimental applications and may be important to study for practical purposes [82,125].

Materials with noticeable anisotropy will also manifest geometrical effects in particle transport. A particle propagation along different axes will be different, which requires tracing cross sections dependent on the crystallographic orientation [126,127].

In non-ideal materials with impurities, direct electron excitation into impurity levels may occur. This channel of excitation is usually of minor importance since it is rare and accounts for very little energy loss of a projectile. Analogously, the formation of excitons (bound electron-hole states within the band gap of material) by a direct impact is usually negligible [128].

An SHI is often treated as a point-like charge traveling through a solid. Accounting for the finite size of the ion (a nucleus with electrons around) usually leads to only minor modifications of its interaction with target atoms, most pronounced around the Bragg peak [129].

The nuclear fragmentation of an SHI may take place during its travel in a solid. Such an effect is extremely rare, but alters all further kinetics drastically, instantaneously changing the mass, charge, and direction of motion of the projectile [130,131]. It may be important to take this effect into account if modeling entire trajectories of multiple ions, which makes such an effect statistically noticeable, depending on the nucleus of the SHI and target atoms [132].

During an SHI scattering, complex processes may take place, such as the creation of so-called convoy electrons. These electrons are not fully captured by the SHI but follow it along, carrying some energy out of the target [133]. They are essentially analogous to electrons in the Rydberg states of an ion [134]. Attracted electrons may also re-scatter off the ion and be re-emitted with different velocities. This process is known as the "Fermi shuttle" [135]. These effects can often be neglected as they have only a little impact on the SHI penetration and target response. However, they may create additional features in the spectra of emitted electrons, and thus may be observable in dedicated experiments [135].

Bremsstrahlung photon emission is an additional channel of energy loss of a particle, becoming noticeable at ultra-relativistic energies [136,137]. As will be discussed below, for electrons the Bremsstrahlung effect becomes dominant at energies of some GeVs. As the intensity of this radiative energy loss scales inversely proportional to the projectile mass, even for protons it is negligible up to the energies of some TeVs. At lower energies, Bremsstrahlung plays a minor role and usually can be neglected, unless it is of particular experimental interest [138].

To conclude, there is a multitude of immediate effects induced by an SHI impinging on solid targets. Most of those discussed in this subsection may often be neglected. It is, nonetheless, important to keep them in mind, since they may manifest under certain conditions.

## III.    Femtosecond timescales: electronic system

Electronic kinetics starts in the wake of an SHI immediately after its passage. It tends to equilibrate the excited electronic system and consists of a collection of complex and intertwined processes, taking place typically at a femtosecond timescale (~1-10 fs). This scale is much shorter than that of atomic dynamics. Although experiencing new forces caused by the excited electrons, the atomic ensemble remains essentially frozen during this stage of the electronic kinetics.

This "age of electrons" involves kinetics of fast primary δ-electrons, electronic cascades, ensembles of generated secondary electrons, deep shell holes, valence holes, and photons [28] as well as initial changes in the interatomic potential [139]. Their kinetics governs the spatial spreading of the excitation from the ion trajectory and prepares parameters of the electronic system providing energy transfer into the atomic system at longer times.

High-energy-electron transport excites secondary electrons, holes, and photons, and transfers kinetic energy to atoms. Created photons, although typically having only a small amount of the deposited energy, bring it far outwards from a track core. Deep-shell holes decay via the Auger-channel, producing secondary electrons, or radiative channel, producing secondary photons. Transport of the valence holes is similar to electron transport: it involves energy transfer to atoms (will be discussed in the next Section), and may also excite secondary electrons in narrow-band materials [140,141].

All of these processes transiently change the electronic distribution function towards the so-called 'bump-on-hot-tail' distribution within femtoseconds after the ion passage [142]. It consists of a majority of slow electrons with nearly thermalized energy distribution, whereas a minority of high-energy electrons forms a long tail that is far from equilibrium.

As electrons form the attractive part of the interatomic potential, which keeps atoms of a solid together under normal conditions, heating low-energy electrons to high electronic temperatures changes the potential, triggering an atomic response to it: acceleration of atoms and structure instability ("nonthermal effects") [139]. The changes in the electron distribution function and atomic positions strongly affect the electronic structure (band structure) of a material [143,144]. As slow electrons are mainly concentrated within short distances from an SHI trajectory, these effects may be important in the track core.

We will discuss all of these effects below in detail to see how they eventually affect the kinetics of the electron ensemble and atomic dynamics, leading to observable material modifications.

### III.A.    Transport of excited electrons (up to 10 fs)

*Ab-initio* techniques such as TD-DFT, or nonequilibrium Green function (NEGF [145,146]) formalism allow tracing transport of electrons as quantum-mechanical objects with high precision. However, they are only used these days for a description of SHI impact on very thin films (graphene [133]). They are, again, limited in time to only a few femtoseconds with present-day computational power.

Computationally efficient approximations and alternative approaches are being actively developed. One of such methods is the orbital-free density functional theory (OFDFT) [147]. It implements the original idea of the DFT that the total energy of the electronic system is a functional of a single variable: the electron density [148]. Since the exact expression of this functional is unknown, a number of approximations have been proposed [149]. Depending on an approximation, OFDFT allows simulating systems with a number of atoms ranging from $10^4$ to $10^6$ [150]. Such an advance, however, resulted in a significant loss of accuracy, and currently, OFDFT produces reasonable results only for metals [147].

The most successful in terms of application to an SHI track simulation is the electron force-field method (eFF) [151]. It uses classical molecular dynamics to simulate electrons smeared in space (as floating Gaussians) and atoms. For the description of electron-electron and electron-nuclei interaction, phenomenological potentials parametrized *via* DFT are used. Within the eFF approach, one can simulate many thousands of atoms and electrons. This method works well only for light atoms up to carbon. For heavier atoms, it requires some model improvements and a usage of pseudopotentials for core electrons, which limits the description of the early-stage electron kinetics in the SHI track. The heaviest element simulated by now was silicon: a crack creation in bulk silicon [152], and SHI track formation in bulk and a slab of Si were studied [153]. However, the capability to reproduce experimental results with this method is yet unclear.

Another promising *ab-initio* method is based on Bohmian mechanics, which is significantly faster than DFT-MD and similar approaches, with some approximations for the interaction potential [154]. This method has not yet been applied to the SHI problem, but will probably find a wide range of applications in the future.

Apart from the high computational cost, DFT and DFT-based approaches (except for the TD-DFT and NEGF) have another crucial disadvantage when applied to the electronic transport in SHI tracks: they are limited to systems in thermodynamic equilibrium [35], which is not the case at the femtosecond scale of the process. In order to solve the problem of electron transport from the SHI trajectory, one has to apply the kinetic-equation-based numerical approaches.

After an SHI impact, as discussed in Section II.C, electrons with the spectrum $\sim 1/E^2$ are created. The majority of these electrons are "slow". They do not fly far away from the SHI trajectory, concentrating around it at distances shorter than 10 nm at sub-10 fs timescales and contributing to the formation of the interatomic potential there. Thus, we do not discuss their effect on energy spreading in this section (see Section IV). In contrast, the electrons at the long high-energy tail of the distribution (δ-electrons) will transport energy away from the track core, performing secondary scattering. In dielectrics and semiconductors, the scattering of these electrons mainly occurs *via* two channels – elastic one on target atoms and inelastic scattering redistributing this energy *via* electronic cascades exciting secondary electrons and creating deep-shell and valence holes – in full analogy to the SHI scattering from the previous section. The 'elastic' channel, which is also often called 'quasi-elastic' [155], conserves the total kinetic energy in its exchange between the scattering particles (an incident electron and an atom/phonon).

The spreading and scattering of δ-electrons can be described well in the framework of the one-particle approximation and the linear response theory (Eqs.(12,14-20)). Since it is based on the first-order Born approximation, it is limited to electron energies above some tens of eV. However, in practice, even lower energies may be described satisfactorily, if interpreted carefully, keeping in mind that the quantum nature of electrons as wave-packets instead of classical point-like particles becomes important [32].

The transport MC method is not well-suited to describe low-energy strongly interacting particles (although it is possible in principle, it becomes very inefficient [156]). Ensemble simulations are a better way to study such systems. In a semi-classical limit, electronic behavior may be traced with the Liouville equation [157,158], which in the case of a single-particle distribution simplifies to the Fokker-Planck [39,159] or Boltzmann [160,161] equations. These methods allow tracing the evolution of the electronic ensembles, and their energy exchange with atoms. Unfortunately, they are also computationally time-consuming and have been rarely applied in the SHI community [162].

Figure 9, Figure 10, and Figure 11 present the mean free paths (MFP) for elastic and inelastic scatterings of electrons in alumina, calculated with the CDF-based cross sections in MC code TREKIS-4 [163]. Figure 9 illustrates that the energy is mainly transported out from the track core by electrons with energies larger than 50-100 eV – they have the MFP increasingly larger than ~1 nm.

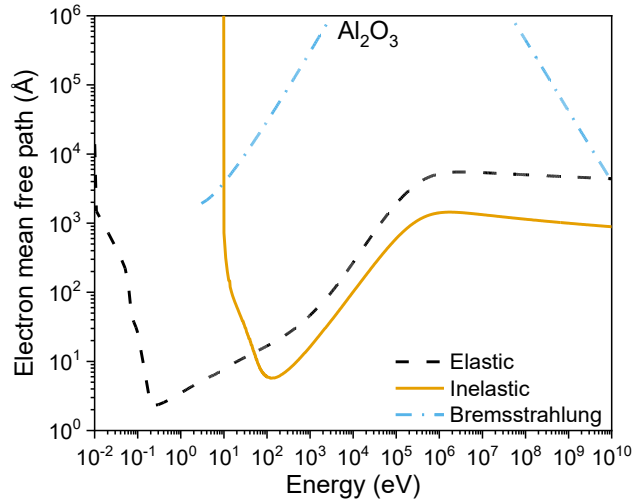

*Figure 9. Electron mean free paths of different processes in Al₂O₃, calculated with TREKIS [5,6,163]. Bremsstrahlung cross section is calculated according to Ref. [41].*

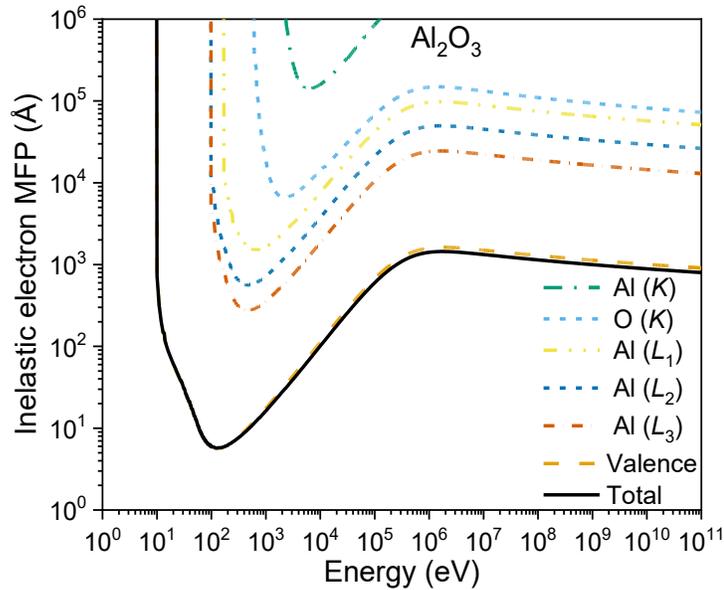

*Figure 10. Inelastic electron mean free path in Al₂O₃, calculated with TREKIS [5,6,163]. Total MFP, scattering on the valence band electrons, and on all deep shells, are shown.*

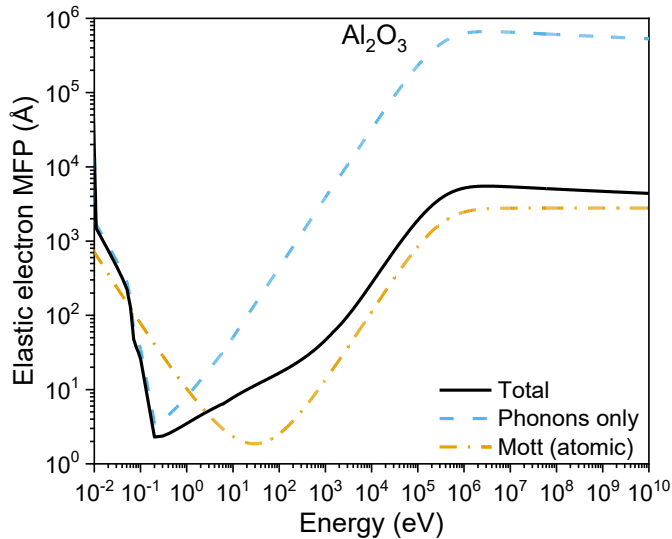

*Figure 11. Elastic electron mean free path in Al₂O₃, total and estimated scattering on phonons, compared with atomic Mott's scattering cross section (combined for the compound using Bragg additivity rule [41]), calculated with TREKIS [5,6,163].*

These electrons are also the most effective in producing new generations of electrons, valence and core holes due to impact ionizations, as seen in Figure 10. The secondary electrons and holes start their own transport and scattering, producing more electrons *via* impact ionizations. This process is known as electron cascade, which changes the state of the electronic ensemble and opens new channels of its relaxation kinetics. Electron inelastic scattering is dominated by the scattering on the atomic shells with the lowest ionization potential smaller than the electron energy (Figure 10). The probability of the scattering on each atomic shell/band is proportional to its relative cross section of scattering (or inversely proportional to its MFP). Thus, most of the excitations of secondary electrons in the cascade will be from the valence band of alumina, some excitation of L-shells of Al atoms will be present, and on rare occasions, electrons will excite K-shells of Al₂O₃ in SHI tracks.

Figure 11 illustrates that the scattering of the fast electrons ($E_e$ > 100 eV) can be well described by Mott's scattering cross section with the modified Molier screening parameter (Eq.(11)) [40]. Underestimation of Mott's atomic scattering in Figure 11 seems to stem from the Bragg additivity rule [41] used to construct the total elastic scattering with the atomic cross section for the binary compound (alumina), which may produce results deviating from the exact solution [164]. The instantaneous regime of scattering on individual lattice atoms frozen in their current positions is a good approximation for these short-wavelength electrons, whose Compton wavelength is smaller than the interatomic distance [101,165,166].

Similar to SHIs, elastic scattering on atoms is dominant at low electron energies (below ~50 eV) – correspondingly rescaled by the mass ratio of an electron and an SHI. Scattering on phonons describes the coupling of slower electrons ($E_e < 10$ eV) with lattice atoms [167]. However, the shorter-than-10-fs timescale of this stage of electronic kinetics does not allow the formation of even the highest frequency optical phonons. At such short times, this scattering channel of slow (but hot!) electrons is only starting to form. We return in detail to the problem of different modes of elastic electronic scattering in the next Section.

Photon emission due to the Bremsstrahlung process plays a significant role only at relativistic energies (above some ~MeV for an electron). Its cross-section starts to grow (MFP decreases) and becomes dominant at energies above some GeVs. At nonrelativistic energies, Bremsstrahlung emission is associated with low-intensity low-frequency photons and does not contribute significantly to electron energy loss, although spectra of Bremsstrahlung photons may be experimentally observable [138].

In each scattering event, an electron will deflect from its previous trajectory by an angle defined by the transferred energy (sampled according to Eq.(21)) [41]. In a nonrelativistic case, it is $\cos(\theta) \sim 1 - W/E$, which is almost isotropic for slow particles, and has a strong preferential direction forward for fast incident particles. Thus, fast delta-electrons first scatter with minor deflections from their initial direction (ballistic transport, with travelled distance proportional to time, $d \sim t$), but their behavior becomes more diffusive (random change of the direction in each scattering event resulting in Brownian motion with $d \sim \sqrt{t}$) as they lose their energy [168,169].

In a solid, an electron having its initial momentum defined by the recoil angle Eq.(22) will not necessarily be emitted with the same angle. On its way out, it will interact with the media, which may alter its direction of motion. An emitted electron acquires an angular distribution, which is defined by the complex dielectric function and the double-differential scattering cross section associated with it [167]. On average, the momentum of an emitted electron is still approximately proportional to the square root of energy, but with a broadened distribution. This dependence of the allowed transferred energies and momenta (or recoil energies Q, W) is known as the Bethe surface, and $W=Q$ is called the Bethe ridge [41,54].

It is important to point out that in standard MC simulations, often energy cut-offs are used to stop the tracing of electron transport. That means, an electron is artificially stopped once its energy drops below a predefined value, usually chosen as 10 eV [170]. This method of simulation allows to estimate the radial dependence of the deposited dose around the SHI trajectory, redistributed due to electron transport [171,172]. Even though it is a standard practice to use energy cut-offs, thusly calculated *cumulative* doses do not resemble energy density distribution at any time instant after SHI energy deposition [11]. For quantitative modeling of SHI track creation, it is important to trace the realistic evolution of the spatio-temporal energy density distribution around an SHI trajectory, as will be discussed in detail in Section VIII.

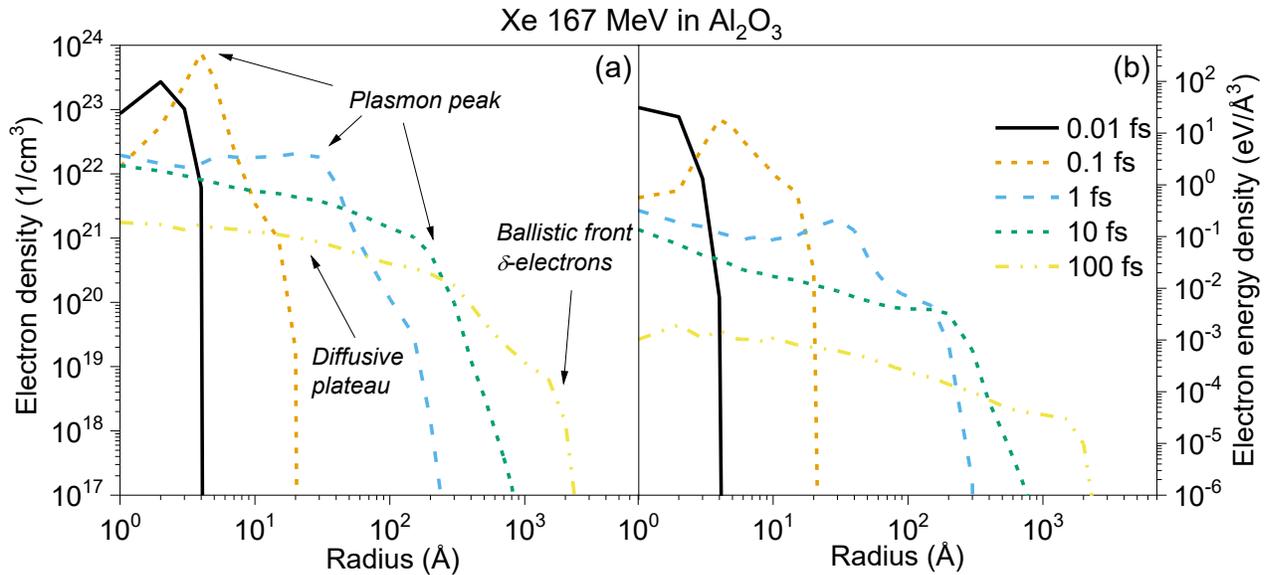

*Figure 12. Radial distributions of the density of excited electrons (a) and their energy density (b) around the trajectory of Xe (167 MeV) in Al₂O₃ ion at different times after the impact, calculated with TREKIS [5].*

A typical example of the evolution of the density and energy density of excited electrons after an impact of Xe ion with 167 MeV energy in $Al_2O_3$ is shown in Figure 12 [5]. Immediately after an SHI impact, all created electrons are located within its impact parameter, a few angstroms around the ion trajectory. The electron transport starts with the ballistic regime. The delta-electrons, produced by an SHI, fly away fast, forming a front of electron density and energy density propagation. This finite speed of propagation of excitation cannot be described with the diffusion equation [173].

Behind it, the second front of propagation appears which is related to the plasmon peak. Many electrons with nearly the same energy are created due to the plasmon resonance excitation [169], which start to travel outwards from the SHI trajectory. Their transport shows dissipative-wave-like behavior [5,169]. In its wake, they leave slow electrons, whose diffusive behavior forms a plateau-like distribution, as expected from diffusive transport in the cylindrical geometry [174]. A bimodal distribution in the electronic ensemble forms before 10 fs after the ion impact. It consists of: (a) fast electrons forming the ballistic front of the spreading excitation, and (b) a large number of "slow" electrons spreading diffusively in the proximity of ion trajectory.

Energy transport exhibits the same features even more pronouncedly, Figure 12b. By the time of ~10 fs, the kinetic energies of most of the electrons are insufficient to perform secondary ionizations ($E_{kin} < E_{gap}$, counted from the bottom of the conduction band). The electron cascades after ~10 fs can still occur only far from the ion trajectory (they are nearly over by the time of ~100 fs). The electron transport does not stop after 10 fs but continues mainly as a slow diffusion process.

Inelastic scattering of electrons creates secondary generations of electrons, leaving holes in the valence band and deep atomic shells. Figure 13 presents a cumulative radial distribution of holes in LiF after Pb ion (2300 MeV) impact. In this plot, all created holes are shown, without accounting for their decays. Valence holes in this plot are considered fixed in space for illustration; in reality, they are mobile and will transport energy in space too. This sets the stage for the deep and valence holes kinetics, which will be discussed in the next section.

Note that in all the considerations in this Section, we did not account for the interaction of the excited electrons among themselves, or with the created holes. Excited electrons and holes would interact *via* two different modes: long-range Coulomb attraction/repulsion, and close-range scattering.

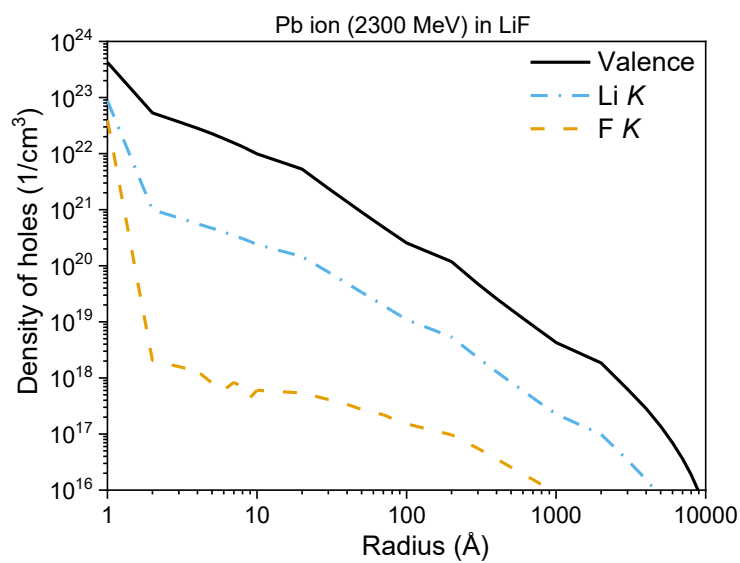

*Figure 13. Cumulative radial distribution of holes in various shells in LiF after Pb-ion impact (E=2300 MeV), calculated with TREKIS* [174].

The first one, an effect of the fields created by excited electrons and holes, may lead to a slowing of the electron transport outwards of the track core (and corresponding speeding up of valence holes transport). It is currently unclear how important this effect would be; however, the following considerations suggest that it is not of major importance. The fastest electrons are practically unaffected by those small fields, and will quickly escape from the effective interaction region. The slowest electrons, which are affected, may exhibit ambipolar diffusion with the valence holes [175,176] – considering that they are already in a diffusive mode of transport, it may not alter their behavior significantly. Thus, it is expected that the effects of created fields should not have a major impact on the initial kinetics of electrons. The exception is finite-size samples, such as thin films, where the emission of electrons may strongly be affected by the induced fields and attraction to holes (since holes cannot be emitted out of the material to follow them) [133].

The second effect, the scattering among the excited electrons and with created holes, seems to be also of minor importance since the density of the excited particles is much lower than the density of scattering centers (atoms and electrons) present in a solid target. Thus, both of the effects are typically neglected in simulations of SHI track creation. Nevertheless, these effects of interactions of excited electrons among themselves and with created holes need to be studied in the future to validate these considerations.

In addition, there are other material-specific channels of electrons interaction. For example, in organic and biological materials, it is important to account for molecular excitations, and dissociative electron attachment [177]. Those channels, absent in inorganic solids, play an important role in organic matter. They may lead to molecular breaks and observable damage [178]. For practical purposes, it may also be important to consider electrons scattering on existing defects, such as grain boundaries in polycrystalline materials or dopants in semiconductors [83].

In conclusion, the dominant channels of scattering for excited electrons are elastic and inelastic scattering on target atoms/electrons. Scattering turns the transport of primary and secondary electrons from ballistic to diffusive. It typically progresses within a few tens of femtoseconds, and elastic scattering plays the most important role in randomizing the angular velocity distribution [169]. Elastic scattering of electrons also exchanges kinetic energy with atoms, heating the atomic system. Inelastic scattering of electrons creates secondary electrons, forming electron cascades. It is the major channel of energy redistribution among electrons. The secondary electrons vastly outnumber the primary delta-electrons produced directly by an SHI, thereby defining the cascade size and shape, and evolution of the electron density. Electron cascades typically last for some ~100 fs (for those induced by SHIs with energies around the Bragg peak), but in the track core, they are typically nearly over within ~10 fs. The deep-shell and valence holes are also created in such cascades, initiating complex hole kinetics. Bremsstrahlung emission of photons by electrons only plays a noticeable role at relativistic energies, whereas at lower energies this channel of energy loss may be neglected.

### III.B.    Kinetics of electron holes

In an impact ionization, a secondary excited electron is emitted with a kinetic energy $E_{kin,2} = W - I_p$, where $I_p$ is the ionization potential of the shell it is being emitted from. Ionization potentials for various shells in elements across the Periodic Table are shown in Figure 14. Note that the ionization potentials of atoms in a chemical environment (a solid) are shifted with respect to the atomic ones, due to the change of the valence shells into a valence band which affects the screening of core electrons. Calculations of electronic energy levels for solids are possible with all-electrons DFT or Hartree-Fock methods [179–181], but they are time-consuming. In practice, almost all MC simulations use atomic ionization potentials to approximate those in solids.

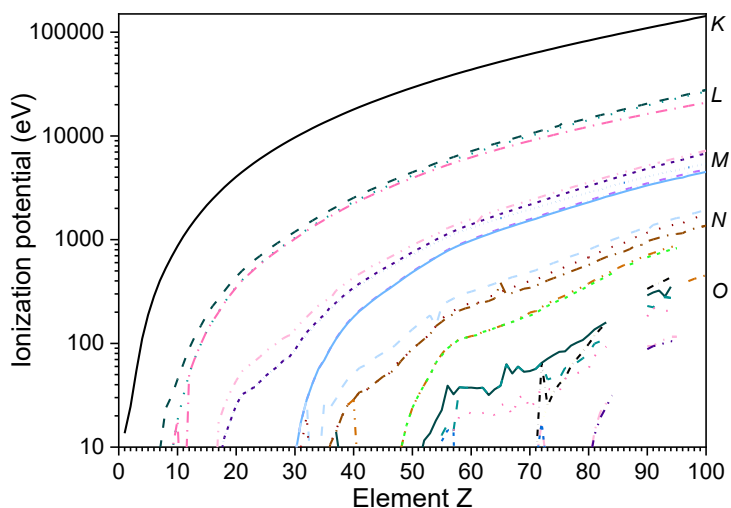

*Figure 14. Ionization potentials of different shells (marked on the right) of isolated neutral atoms in the ground states across the Periodic Table, extracted from Ref.* [182].

All deep-shell holes created by an SHI or electron impact in atoms quickly decay, typically at femtosecond timescales, see Figure 15. For all shells of all elements, except for the K-shell, Auger or Coster-Kronig (which is an intra-shell Auger decay where an electron participating in the process comes from a subshell of the same shell that the decaying hole resides in [183]), decays are much more probable than radiative decay. For K-shell holes, light elements are more probable to decay *via* the Auger channel, whereas for elements above Z~30, radiative decay overtakes[c].

In an Auger (and Coster-Kronig) decay, a new electron is emitted, and two new holes are left in the upper shells of the atom. The kinetic energy of an Auger electron is defined by the energy differences of the electronic energy levels, participating in the decay: $E_{kin,A} = I_{p,h} - I_{p,1} - I_{p,2}$; where $I_{p,h}$ is the ionization potential of the decaying hole, and $I_{p,1/2}$ are those of the participating shells.

Similarly, radiative decay (or fluorescence) emits a photon with energy: $\hbar\omega = I_{p,h} - I_{p,1}$, and only one new hole is created, whereas an old hole is filled – essentially, it means a hole jumps into an upper shell.

---

[c] A very useful compilation of atomic data, such as ionization potentials, decay times and probabilities, etc. can be found in EPICS2017 database (its EADL part) [190]. The EPDL part of the same database contains data on cross sections of interaction of photons with atoms, and EEDL has the data on electrons interactions.

All new holes will also undergo their own decays until they float into the valence (or the conduction) band of the material.

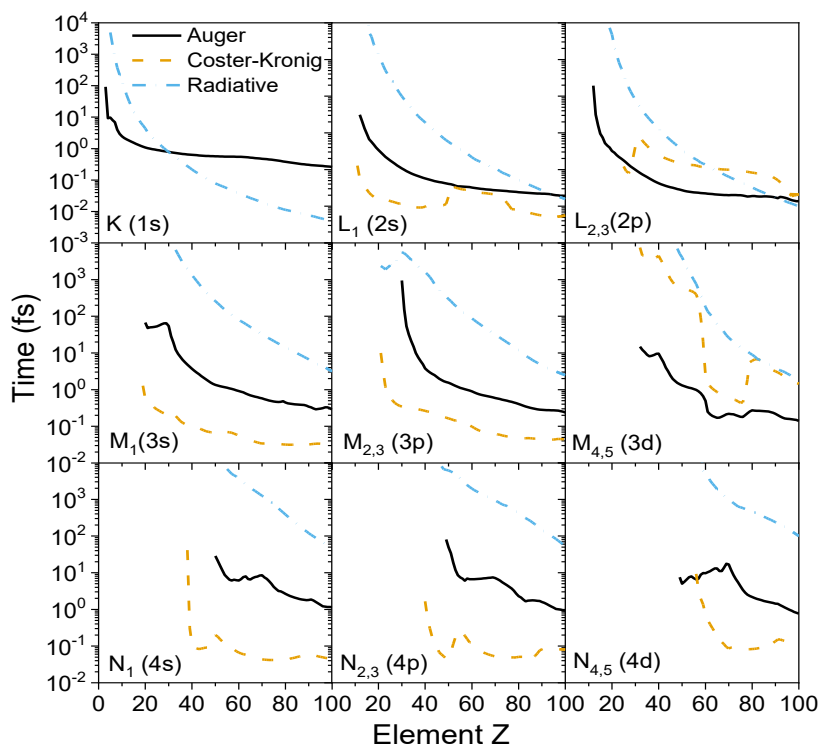

*Figure 15. Auger, Coster-Kronig, and radiative characteristic decay times in various shells of elements across the Periodic Table, extracted from Ref.* [29]*.*

In solids, also inter-atomic Auger decays are possible [184]. In such a process, one or both electrons participating in an Auger decay may come from neighboring atoms, instead of the atom containing the decaying hole (sometimes called Knotek-Feibelman process) [185]. In such a case, energetically forbidden transition in an isolated atom may become allowed in molecular/solid environment – the processes known as Interatomic Coulombic Decay (ICD [186]), Excitation-Transfer-Ionization (ETI [187]), and related processes. Accounting for these kinds of processes is necessary, since target atoms in proximity to the SHI trajectory may be stripped, as discussed above in Section II.C [188]. Unfortunately, there are almost no data available on the characteristic decay times in solids, so in practice all decays in MC models are characterized by the atomic Auger (or Koster-Kronig) times [41,75,76,189,190]. In the MC simulation, the decay time is sampled with the exponential distribution: $t = -t_0 \ln(\xi_h)$, where $\xi_h \in [0,1)$ is a random number, $t_0$ is the characteristic decay time [163].

All deep-shell holes eventually end up in the valence/conduction band, typically within a few femtoseconds (except for very light elements, in which the decays may take a few tens of fs). In contrast to deep-shell holes, valence holes in solids are mobile. They behave as positive charges with an effective mass defined by the band structure of the material [83].

If the width of the valence band is larger than the band gap of the material (see an example in Figure 16), valence holes can scatter inelastically. This process sometimes is also called 'Auger decay of valence holes', since it resembles the actual Auger decay in the sense that the valence hole jumps up into a higher-lying state of the valence band, and an electron is promoted into the conduction band. Holes will perform such impact ionizations as long as their kinetic energy (counted down from the top of the valence band) is larger than the HOMO-LUMO (highest occupied molecular orbital – lowest unoccupied molecular orbital) band gap of the material. Strictly speaking, the limit is somewhat larger than the band gap, and typically lies within the interval between $E_{gap}$ and $3/2 E_{gap}$, depending on the incident particle energy [6], and particular dispersion law in the material band structure [191].

The MC simulation of holes transport requires taking into account the effective mass of holes, $m_{h,eff}$, which correspondingly rescales the scattering cross sections [169]. Within an 'effective one-band' approximation proposed in Ref. [192], the effective mass of a valence-band hole or a conduction-band electron can be worked back from the density of states of the material, $D(E)$:

$$E_h = \frac{\hbar^2 q_h^2}{2 m_{h,eff}}, q_h(E_h) = \sqrt[3]{\frac{6\pi^2}{s} \int_0^{E_h} D(\epsilon)\, d\epsilon} \qquad (23)$$

here, the hole's momentum $q_h$ is defined within an effective single band depending on the energy in the valence band $E_h$, $s$=2 is the spin degeneracy. This approximation assumes a uniform and homogeneous material – which is consistent with the assumptions used in the MC method.

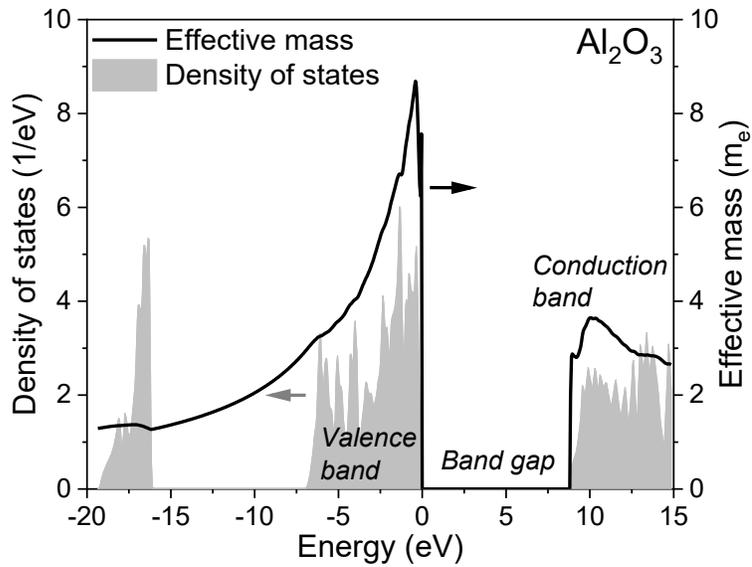

*Figure 16. The density of states in Al₂O₃ calculated with tight binding code XTANT [193], and an effective mass of valence holes or conduction electrons calculated within the effective one-band approximation, Eqs.(23), in units of the free electron mass.*

An example of the effective mass of the holes in $Al_2O_3$ is shown in Figure 16. Since the effective mass within the effective one-band approximation depends on the energy, the mass conservation does not hold in a scattering event. This should not confuse the reader, since the energy (and momentum, if required [191]) is still conserved.

The cross sections of scattering of a valence hole are the same as for electrons, rescaled by the mass ratio (also in corresponding momentum integration limits). Since valence holes are typically heavier than free electrons, elastic scattering delivers more energy to atoms than electrons of the same energy. Thus, valence band kinetics is a very important channel in energy redistribution, which must be taken into account in SHI tracks simulations [140,141]. This energy is distributed mainly around the SHI trajectory, forming a larger energy density than that produced by electrons alone, see Figure 17.

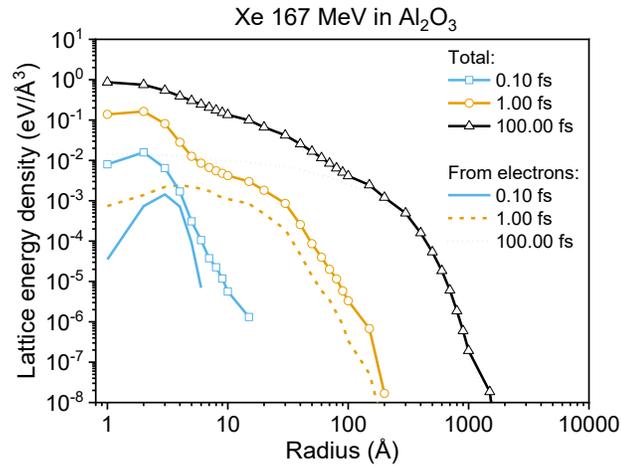

*Figure 17. The radial density of the lattice energy in a track of Xe 167 MeV ion in Al₂O₃, transferred from electrons only, and total transferred due to elastic scattering of electrons and valence holes* [169].

A pair of carriers, a valence hole and a conduction-band electron, have the potential to recombine and release the energy, equal to the band gap (potential energy of such a pair of carriers) plus their kinetic energies. A large amount of energy, deposited by an SHI, is transiently stored as the potential energy of produced electron-hole pairs [140,141]. For convenience, the potential energy may be attributed to valence holes.

Figure 18 illustrates the temporal evolution of the energy fractions in different subsystems of Al₂O₃ after Xe (167 MeV) ion impact [140]. The figure demonstrates well-separated stages of the track kinetics: (a) the electronic kinetics dominates by 10 fs after the ion impact ("age of electrons"); (b) the electron-hole ensemble accumulates a large amount of the deposited energy by the time of 10 fs. Nearly 60% of all the energy deposited by the SHI is accumulated as the potential energy of holes (the "age of holes"). This energy can transfer into the atomic system *via* channels different from that of electron-atom or hole-atom scattering. We will return to this point later in a discussion of atomic heating in Section IV.C. (c) energy transfer to lattice atoms *via* electrons and holes scattering takes place after 10 fs – this stage will be discussed in the next Section. The difference between the total atomic heating and heating solely by electrons demonstrates the importance of the valence holes' elastic scattering. By the time of ~100-200 fs, neither electrons nor holes have significant kinetic energy left.

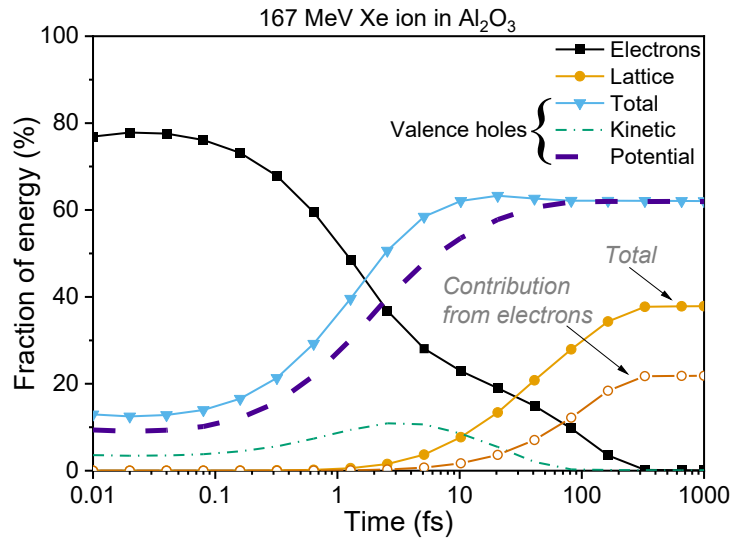

*Figure 18. Fractions of energy accumulated in each subsystem of Al₂O₃ after the impact of Xe 167 MeV ion, calculated with TREKIS* [140]. *Energy transferred to lattice atoms from electrons (hollow circles), and the total one (from electrons and valence holes; full circles) are shown for comparison.*

Mobility of valence holes has another important consequence. If holes are not bound to their parent ions, in contrast to plasma, the Coulomb repulsion in solids will repeal the holes instead of lattice ions. As soon as holes pop up into the valence band, they become the subject of Coulomb interaction, without dragging ions behind. This is another argument in support of the negligibility of the Coulomb explosion effects in solids. Only the deep-shell holes are bound to their parent ions, thus the lifetime of Coulomb explosion is limited by the lifetime of those holes (and plasmon frequency screening the charges). As discussed above, they decay within a few femtoseconds or even faster.

In all the considerations, we neglected the effects of long-range fields acting between the holes, or between holes and excited electrons. It was done on the same grounds as for the electrons, discussed in Section III.A.

The valence holes may recombine with electrons *via* various channels. They may emit photons, excite another electron (three-body recombination process), transfer energy to an atom of a target, or in a small-band-gap material to collective modes, phonons. In some materials, far from the track core, valence holes and electrons may form a quasi-stable couple, an exciton. An exciton may then also exhibit complex kinetics, which includes its transport, self-trapping, and recombination [128,194]. All these effects may be important for the formation of point defects in the track halo, which will be discussed in Section VI.B.

To summarize, deep-shell holes created by an SHI or electrons impact ionization quickly decay – predominantly via Auger-decay channel emitting an electron. With a lower probability, they may decay radiatively, emitting a photon. In each decay, a hole jumps into a higher energy level, eventually ending up in the valence band of the material. Valence holes are mobile, and they behave similarly to excited electrons but with their own effective mass, which may be effectively constructed from the DOS of the material. Inelastic scattering of valence holes is possible in materials, in which the valence band is wider than the band gap, thereby exciting secondary electrons. Elastic scattering transfers a significant amount of energy to atoms, most notably in the nearest proximity of an SHI trajectory. A part of the energy is transiently stored as the potential energy of electron-hole pairs, which may be released at longer times *via* various recombination channels (including band gap collapse, see Section IV.C).

### III.C.    Photon transport

Photons, created either *via* Bremsstrahlung emission by electrons or *via* radiative decay of deep-shell holes, then propagate in the target until escaping from its surface or being absorbed by electrons in the material. A photon may experience a few different types of interaction. Long wavelength (low-energy) photons mostly scatter off the lattice atoms via Rayleigh scattering (elastic scattering channel which does not change photon energy, only its direction of motion) [41]. For photon energies above the band gap of the material, photoabsorption by the valence electrons will take place. In a photoabsorption event, a photon disappears and an electron is emitted with its energy equal to the difference between the photon energy and the ionization potential of the electron level an electron is emitted from: $E_{kin} = \hbar\omega - I_p$. With an increase of the photon energy, deep-shell atomic electrons start to absorb photons, if the photon energy overcomes an ionization potential of this shell. At energies above some ~10 keV, photons start to scatter on atoms via inelastic Compton scattering channel [41]: a photon loses a part of its energy and continues propagation with a longer wavelength, while an electron is emitted.

At relativistic energies above $\hbar\omega = 2m_e c^2$, the creation of electron-positron pairs becomes possible [41].

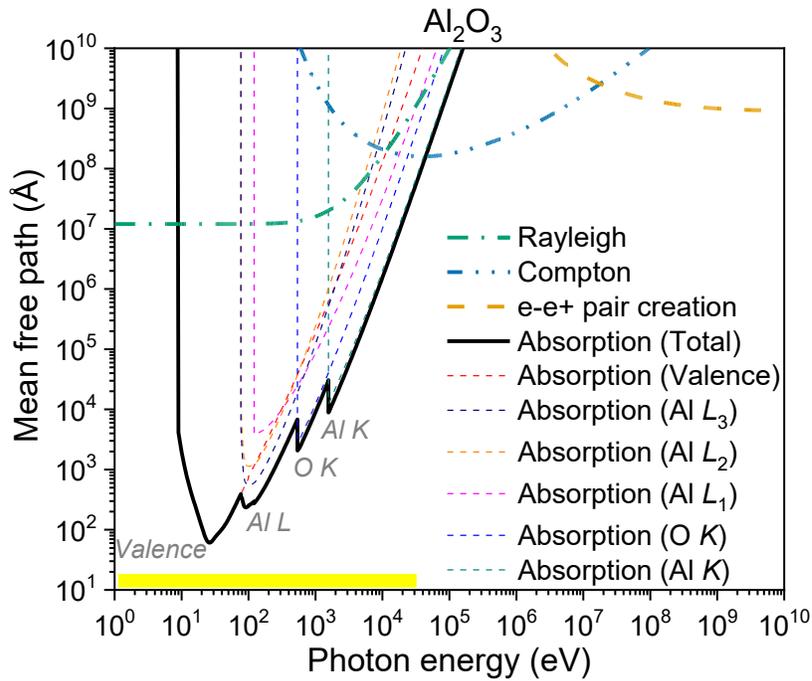

*Figure 19. Photon mean free paths of various processes in Al₂O₃. Rayleigh, Compton, and electron-positron pair production cross sections are calculated, according to Ref. [41]. Photoabsorption cross sections by different atomic shells are taken from the EPICS database [190]. Valence band photoabsorption cross section is reconstructed from the optical coefficients from Ref. [104]. The yellow bar schematically shows the region of energies typical for those of the SHI problem.*

Mean free paths of these processes in $Al_2O_3$ are shown in Figure 19, giving an estimate of their relative probabilities. From this figure, we can see that for typical energies of photons produced in a track of an SHI around the Bragg peak, the photon absorption is dominant by many orders of magnitude. It is a typical situation for all materials. In contrast to the inelastic scattering of electrons (cf. Figure 10), photons are preferentially absorbed by the deepest shell possible, whose ionization potential is below the photon energy [41]. This creates the characteristic peaks in the total photoabsorption cross section and mean free path.

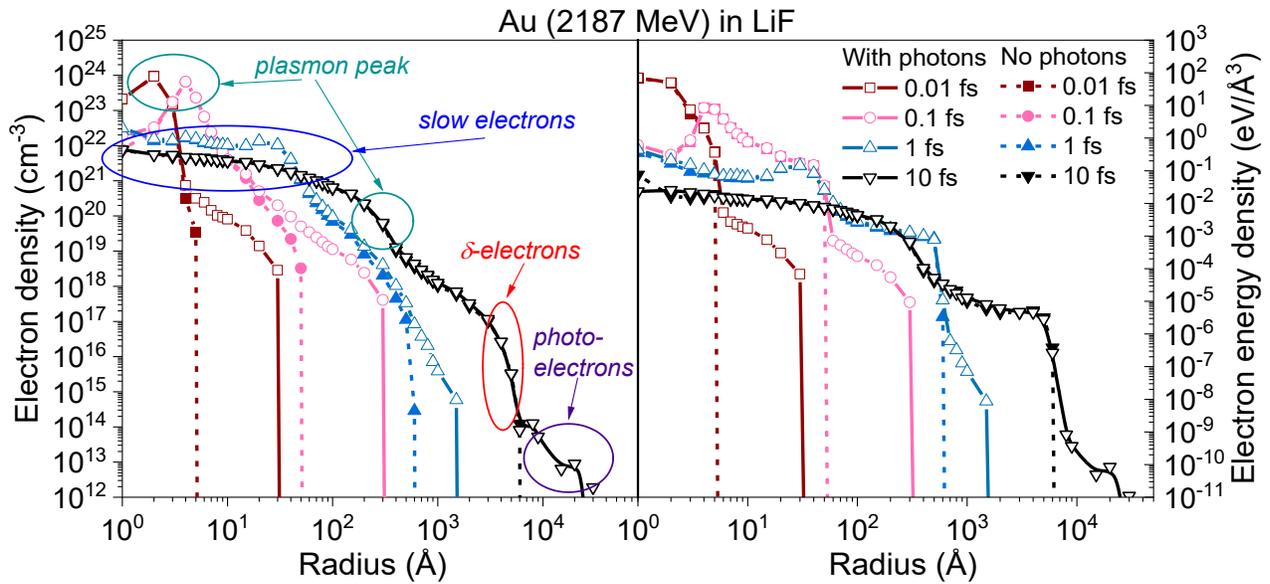

*Figure 20. The radial distributions of electrons calculated without radiative decays of deep holes vs. with these decays and induced photon transport at different times after passage of Au 2187 MeV ion in LiF [169].*

Since the electron Bremsstrahlung and radiative decay of deep shell holes are both rather improbable at considered energies, photon transport plays only a minor role in the track creation problem. Photons carry out a small portion of energy far from an SHI trajectory, see Figure 20 [169]. Photons travel much faster and farther away from the track center than electrons, creating an additional excitation front. Their absorption creates a long radial tail of secondary electrons, but their density and energy density in this tail are orders of magnitude lower than those produced by δ-electrons.

In conclusion, in an SHI problem, photon transport usually plays a minor role, bringing an insignificant fraction of energy far away from the track core. Photoabsorption is the dominant channel of photon interaction with the media in the relevant energy range. However, as it may create some defects in the structure at large distances, photon-related effects may be important for experiments or applications dealing with such defects.

## IV. Electron to lattice energy transfer: 10 fs to ps

After about ten fs after an ion passage, the bimodal 'bump-on-hot-tail' energy distribution of electrons arises around the ion trajectory [142] (also, see Figure 20). It consists of slow electrons in the track core (~10 nm) with a nearly thermalized energy distribution, whereas a minority of high-energy electrons forms a long tail that is far from equilibrium. The ensemble of fast electrons locates far from the SHI trajectory and has low particle and energy densities. Its contribution to an increase of the energy of lattice atoms is negligible after these times. On a 10-fs timescale, "slow" electrons with energies of ~10 eV accumulate most of the energy transferred by the ion to the target and provide energy transfer to atoms (see Figure 18).

The extreme initial localization of the ensemble of slow electrons (~10 nm) results in high gradients of its energy density leading to a fast drain of the electronic energy from the ion trajectory. This energy flow cools down the electronic system in the ion track within a hundred-femtosecond timescale. Significant heating of materials in SHI impacts required to form observable tracks suggests extremely fast energy transfer from the electrons to target atoms within this ultrashort cooling time of the electronic system.

Thus, a reliable model of an SHI track formation must self-consistently describe the two most important effects and their interplay: energy transport (in the electron, hole, and atom systems), and energy transfer to the atoms. It is particularly important to identify the mechanism responsible for the extremely fast heating of atoms.

In this Section, we review existing models of track heating and discuss their advantages and disadvantages. Special attention is paid to the "nonthermal" mechanism of the energy transfer to the atomic system, which realizes at extreme excitations of the electronic system.

### IV.A. Two temperature (thermal spike) model

The two-temperature model (TTM [195], also called the inelastic thermal spike, i-TS), or its extended versions, is the most often-used model describing the transport of the excitation and energy transfer to the atomic system in an SHI track. It is based on two macroscopic coupled parabolic equations of heat diffusion in the electronic and atomic ensembles, which have the simplest form *in metals* [25,26,33]:

$$c_e(T_e)\frac{dT_e}{dt} = \nabla(\kappa_e(T_e, T_a)\nabla T_e) - G(T_e, T_a)(T_e - T_a) + Q_s,$$

$$c_a(T_a)\frac{dT_a}{dt} = \nabla(\kappa_a(T_e, T_a)\nabla T_a) + G(T_e, T_a)(T_e - T_a)$$

(24)

Here $T_{e,a}$ are the electronic and atomic temperatures, $c_{e,a}(T_{e,a})$ are the temperature-dependent electronic and atomic heat capacities, $\kappa_{e,a}(T_e, T_a)$ are the electronic and atomic thermal conductivities, and $Q_s$ is a source term of energy delivery into the electronic system. The electron-ion coupling parameter, $G$, is the key one guiding the energy flow between electrons and atoms (per volume $V_0$):

$$G(T_e, T_a) = \frac{1}{V_0(T_e - T_a)}\frac{dE}{dt}$$

(25)

Microscopically, the model assumes that the scattering of electrons defines electronic energy transport and its transfer to the atoms.

Unfortunately, this, formally self-consistent and simple, classical TTM has flaws making its application to ultrafast SHI tracks questionable.

Firstly, in an appropriate application of a TTM-based model, the electronic system must be divided into high-energy electrons, which will form a source term $Q_s$ for energy deposited into low-energy electrons [168,196]. The fast electrons cannot be described with the thermo-diffusion equation, as discussed in Section III.A. The impossibility of the TTM to define $Q_s$ forces one to use an *ad-hoc* function of a "deposited dose" as an external (fitting) parameter of the model. A realistic source term $Q_s$ can be defined, e.g., by a separate MC simulation tracing the fast electrons and their coupling with the slow electrons [47,168].

Another issue is, at the initial stage of their spreading, slow electrons form a front moving with a finite speed. The movement of an excitation front should also be taken into account when describing heat spreading through the atomic system at a picosecond timescale. This is not accounted for in the Eqs.(24) which use the Fourier law for the heat flux that assumes its infinite propagation velocity. To circumvent the problem, the parabolic heat diffusion Eqs.(24) should be replaced with the hyperbolic dissipative wave equations (Catteneo equation [173]) providing finite velocities of heat transport at femtosecond timescales for the electronic system and picoseconds for the atomic system.

In band gap materials (semiconductors and insulators), simple Eqs.(24) cannot describe the electronic system, since the electron density in the conduction band is highly inhomogeneous and is defined by the excitation level. In SHI tracks, valence holes accumulate at least half of the deposited energy of the electronic ensemble by 10 fs after the ion impact. In such a case, Eqs.(24) need to be supplemented with additional equations tracing the evolution of the electron-hole density, sometimes called the nTTM [175,197].

Application of macroscopic coefficients of the heat capacities, thermal conductivities, and the threshold parameters of phase transitions, *e.g.* the melting temperature, to highly excited material states at the nanometric spatial and picosecond temporal scales is dubious. In the i-TS applied to SHI tracks, it is often assumed that detected regions of damaged structure coincide with those where these thresholds were achieved. No relaxation of the damaged structure is assumed during the track cooling.

Thermalization of the atomic ensemble at times as short as 100 fs, as assumed in Eqs.(24), is also questionable. To avoid this problem and to trace phase transitions and track formation, the atomic heat diffusion equation can be replaced with MD (solving Eqs.(6)), simulating the atomic system in sufficient detail [198,199]. Such models are known as coupled TTM-MD, which are nowadays a standard simulation tool for modeling ultrafast energy deposition (either by femtosecond lasers or by swift ions) [26,200]. In this approach, the energy exchange from electrons must be delivered to individual atoms in some way.

The simplest way to do that is velocity scaling, in which at each time step of MD simulation, the energy provided from electrons is distributed to atoms *via* rescaling their velocities by a proportionality factor $\xi$. It essentially assumes that the electrons form a friction force that may accelerate or decelerate atoms depending on a difference between the electronic and ionic temperatures [199]:

$$M_i \frac{d^2 \mathbf{R}_i}{dt^2} = -\frac{\partial V(\{R_{ij}\})}{\partial \mathbf{R}_i} + \xi M_i \mathbf{v}_i \,,$$

$$\xi = \frac{G}{N_{at}} \frac{(T_e - T_a) V_0}{\sum_j M_j v_j^2} \tag{26}$$

here $\mathbf{v}_i$ is the velocity of $i^{th}$ atom with mass $M_i$, $V$ is the volume of the MD simulation box, $N_{at}$ is the number of atoms in the simulation box (note the additional term cf. Eq.(6)). This Eq.(26) replaces the second equation (24) describing the atomic system [26,199]. More advanced approaches instead of simple velocity scaling are also used, e.g. based on inhomogeneous Langevin thermostat [197].

TTM-MD models resolve the problem of ionic temperature and can describe the relaxation of a damaged structure during track cooling. But they have no method of calculating a realistic initial spatial profile of the electronic temperature $Q_s$, and adjust the electron-phonon coupling parameter $G$ to provide necessarily fast atomic heating in SHI tracks fitted to reproduce the damaged area size detected in the experiments [201].

Due to application of the fitting parameters, the two-temperature thermal spike model fundamentally lacks predictive power. This problem demonstrates that it is not possible to build a self-consistent model describing track formation without a microscopic quantitative knowledge of the mechanisms governing its kinetics.

On the other hand, TTM helps point out fundamental effects of electrons-atoms energy exchange in SHI tracks. Coupling parameters extracted from the TTM by fitting the calculated SHI track sizes to experimental data are orders of magnitude larger than those measured in laser-irradiation experiments and consistently calculated with various models, see Figure 21.

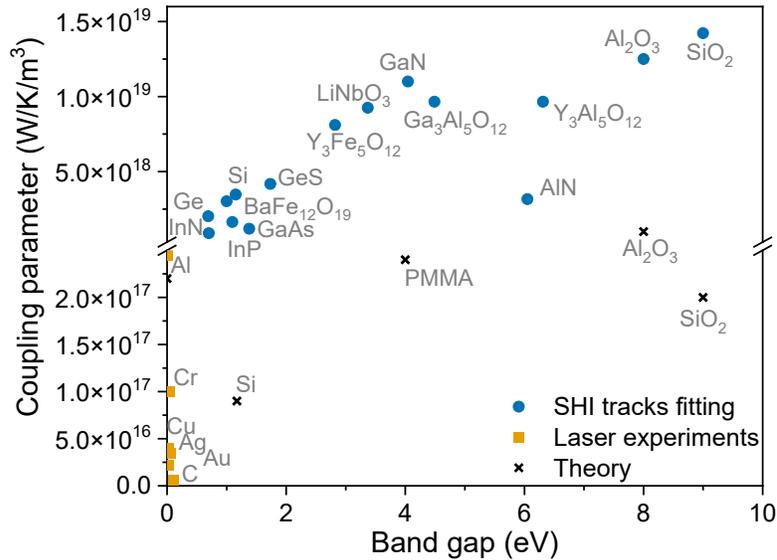

**Figure 21.** *Fitted electron-ion coupling parameter in various materials. Those, extracted from the SHI irradiations data from Ref. [201], are shown with circles. Squares are the parameters measured in laser experiments (taken from Ref. [44] and references therein). Crosses are calculated coupling parameters at the laser damage threshold doses [44,202].*

### IV.B.    Scattering of electrons on atoms: non-adiabatic energy exchange

Most of the modern microscopic models of SHI track excitation use algorithms explicitly or implicitly assuming the scattering of electrons and holes on the atomic and electronic ensembles. Electron scattering provides energy and momentum transfer to the atoms causing finally equilibration between the electronic and ionic systems [26]. Also, the scattering-based models can self-consistently describe spatial spreading of the excitation from the ion trajectory.

The scattering-based momentum/kinetic energy exchange is referred to as the electron-ion coupling or the "electron-phonon coupling", although in general, any atomic displacement caused by electronic scattering (not only phononic modes) triggers electron transitions and momentum and kinetic energy exchange between the two systems.

In the one-particle approximation, the electron-ion coupling parameter $G$ is defined *via* [160]:

$$G(T_e, T_a) = \frac{1}{V_0(T_e - T_a)} \sum_{i,j} \varepsilon_j I_{e-a}^{ij} \tag{27}$$

Where summation runs over all electronic states $i$ and $j$, and the scattering integral $I_{e-a}^{ij}$ is the time derivative of the electron distribution function,

$$\frac{df(\varepsilon_i, t)}{dt} = \sum_j I_{e-a}^{ij}, \tag{28}$$

The quantum mechanical methods such as TD-DFT [35] or CEID [45,203] allow evaluating this nonadiabatic electron-ion coupling, exchanging energy between electrons and atoms. For example, the CEID method traces the evolution of the electron density matrix, including the electron transitions (off-diagonal elements of the density matrix) induced by the changes in atomic positions, modeled with MD simulations [204].

A simple approach to calculating electronic transition probability was proposed in *ab-initio* femtochemistry by Tully, known as the "surface hopping" [42,43]. The method allows evaluating probabilities of electron transitions between adiabatic energy surfaces of a molecule *via* overlap of their wave functions of states $i$ and $j$, $|\psi_i(t_0)\rangle$ and $|\psi_j(t)\rangle$ at different timesteps of a Born-Oppenheimer (BO, adiabatic) MD simulation [42,43]. Changes in the electronic wave functions induced by atomic displacements allow to obtain the so-called coupling vector: $\vec{d} = \langle i|\nabla|j\rangle$. For solids, it can be used to obtain matrix elements (or probabilities $W_{ij} = \left|\langle\psi_j(t)|\psi_i(t_0)\rangle\right|^2$) entering the scattering integral $I_{e-a}^{ij}$ [44]:

$$I_{e-a}^{ij} = w_{ij} \begin{cases} f(\varepsilon_j)[2 - f(\varepsilon_i)] - f(\varepsilon_i)[2 - f(\varepsilon_j)]e^{-\varepsilon_{ij}/T_a}, \text{for i} > j \\ f(\varepsilon_j)[2 - f(\varepsilon_i)]e^{-\varepsilon_{ij}/T_a} - f(\varepsilon_i)[2 - f(\varepsilon_j)], \text{otherwise} \end{cases} \tag{29}$$

$$w_{ij} = \frac{dW_{ij}}{dt} \approx 2\frac{\left|\langle\psi_j(t)|\psi_i(t_0)\rangle\right|^2}{\delta t}$$

Here again $f(\varepsilon_i)$ is the electron distribution function, normalized to 2 due to spin degeneracy; $\varepsilon_{ij} = \varepsilon_i - \varepsilon_j$; the time derivative is approximated with the finite difference method for the molecular dynamics time step $\delta t$, and the wave functions are calculated correspondingly on two consecutive steps: $t_0$ and $t = t_0 + \delta t$. The exponential terms result from the Maxwellian distribution of the atomic ensemble, and, in general, may be replaced with an integral of the transient nonequilibrium atomic distribution.

This method allowed to calculate of electron-ion coupling parameters in multiple materials [44], but it requires tracing the state of the electronic structure alongside MD simulation with such methods as DFT or TB. Unfortunately, peculiarities of the track kinetics pose problems with practical realizations of this possibility, requiring too large simulation box sizes to address with *ab-initio* techniques.

When the atomic system may be approximated as a perfect periodic crystal with harmonic interatomic potential (small atomic oscillations), and at timescales longer than the characteristic period of atomic oscillations (inverse phonon frequency), the electron-atom coupling reduces to the electron-phonon one [101,205]. It is often then computed within the Eliashberg formalism, which uses phonon spectral function to evaluate electron-phonon coupling matrix elements [206–208]. However, it has recently been noticed that this formalism, originally developed for superconductors, may produce overestimated results at high electronic temperatures [44,209,210]. Let us also note that none of the underlying assumptions of the phonon approximation is satisfied in an SHI track [28,166,211]: the processes taking place are much faster than the characteristic phonon time, the atomic structure is changing fast from the perfect crystal, the atomic motion is not harmonic, so the phononic approximation does not hold.

More advanced approaches, beyond the Eliashberg formalism, have also been developed, such as the coupled-modes approach (treating coupling between collective modes of electrons and ions: plasmons and phonons), or models that include self-consistently the quantum mechanical and statistical nature of the electronic ensemble [212–214]. In particular, models accounting for nonequilibrium effects warrant a special note. Nonequilibrium phononic distributions were accounted for in Ref. [210], which improved agreement with the experimentally measured electron-ion energy exchange rate. Nonequilibrium electronic distribution is predicted to have a significant effect on the coupling rate [160,192].

If exact electronic wave functions are unavailable, further approximations can be made. E.g., the wave functions of fast electrons can be well-approximated as plane waves [101,165,166]. Also, the first-order

Born approximation is valid for a scattering of such electrons with energies larger than ~10-100 eV on target atoms. It allows describing the electron-ion scattering within the dynamical structure factor (DSF) formalism discussed in Sections II.C and III.A.

Being the Fourier transform of the spatio-temporal atomic pair correlation function, the DSF contains information about the dynamical modes of the atomic ensemble, e.g., about the phonon spectrum [165]:

$$S(\mathbf{q}, \omega) = \frac{1}{2\pi N} \int \exp(i\mathbf{q}\mathbf{r} - i\omega t)\, G(\mathbf{r}, t)\, d\mathbf{r}dt$$

$$G(\mathbf{r}, t) = \sum_i^N \sum_j^N \langle \delta[\mathbf{r} + \mathbf{R}_i(t) - \mathbf{R}_j(0)] \rangle \qquad (30)$$

Here, $G(\mathbf{r}, t)$ is the atomic pair correlation function, the brackets sign $\langle ... \rangle$ denotes averaging over the atomic ensemble, $N$ is the number of atoms in the simulation box, $\mathbf{R}_i(t)$ is the coordinate of the $i$-th atom at time $t$. If there is more than one kind of atoms in the system, in the expression (30) the DSF must take into account partial atomic correlation functions [51].

The DSF determines cross-sections which take into account the collective response of the atomic ensemble to excitations induced by a scattered particle [101]:

$$\frac{\partial^2 \sigma}{\partial \Omega \partial (\hbar \omega)} = |V(\mathbf{q})|^2 \frac{k_f M^2}{4\pi^2 k_i \hbar^5} S(\mathbf{q}, \omega) \qquad (31)$$

here $\vec{q} = \vec{k}_j - \vec{k}_i$, $k_i$ and $k_f$ are the initial and final wave vectors of the incident particle (before and after a scattering act), $M$ is this particle mass, and $V(\mathbf{q})$ is the Fourier transform of interaction potential between the particle and a single atom in a target.

Application of the DSF formalism to elastic scattering of electrons on atoms allows to distinguish between scattering modes, recall Figure 11. Scattering of the fast electrons (short wavelengths) does not depend on the atomic structure and collective dynamics – it reduces to the Mott's cross-section of scattering on an individual atom. Electrons with wavelengths comparable with the interatomic distance scatter on individual atoms frozen in their structure [101,165,166]. At low energies (the electron wavelengths much larger than the interatomic distance) involved atoms are interacting among themselves and may redistribute energy *during the scattering act* [101], which alters the way and the amount of energy that the atomic ensemble receives from an incident electron. It then reduces to the scattering on collective atomic dynamical modes (phonons), which may depend on the crystallographic anisotropy [127].

The DSF allows for a straightforward evaluation with help of classical molecular dynamics simulations [215,216]. Within the formalism, full atomic dynamics may be accounted for *via* MD simulations, thereby allowing to calculate electron-ion coupling in an arbitrary system, including amorphous or melted states.

Averaging the energy exchange calculated with the DSF-based cross sections (Eqs.(30,31)) over an ensemble of quantum electrons provides the scattering integral [211,215]:

$$I_{e-a}^{DSF} = -\frac{4}{(2\pi)^5 \hbar^2} \int d\vec{k}_i d\vec{k}_j |V(\vec{q})|^2 \times$$
$$\left[ f(\vec{k}_j)\left(1 - f(\vec{k}_i)\right) S(-\vec{q}, -\omega) - f(\vec{k}_i)\left(1 - f(\vec{k}_j)\right) S(\vec{q}, \omega) \right] \tag{32}$$

the electron distribution function $f(\vec{k}_i)$ is now written in the momentum space of a free electron $\vec{k}_i$.

An example of an electron-ion coupling parameter in aluminum calculated with the above-mentioned models is shown in Figure 22. They produce close results at low electronic temperatures, but the rise of the DSF-based coupling parameter at electronic temperatures above ~10000 K is overestimated. The agreement with the data from laser experiments is also reasonable, considering a wide discrepancy among different experiments. More discussion on this point can be found in Ref. [44], together with coupling parameters for many metals across the Periodic Table.

With a typical coupling parameter (example in Figure 22), heating of the atomic system by electrons takes a few picoseconds [44]. The coupling parameter has a general tendency to decrease with an increase of the atomic mass [44], and materials made of heavy elements exhibit even slower coupling that may take tens of picoseconds [223]. There are experimental indications that electron-ion coupling may be even slower than the most advanced theories predict [223–225]. Such timescales are too long in comparison with the cooling time of electrons in the vicinity of the SHI trajectory (see Figure 21) since, as was discussed in Section III.A, transport of electrons outwards from the track center decreases their energy within ~100 fs.

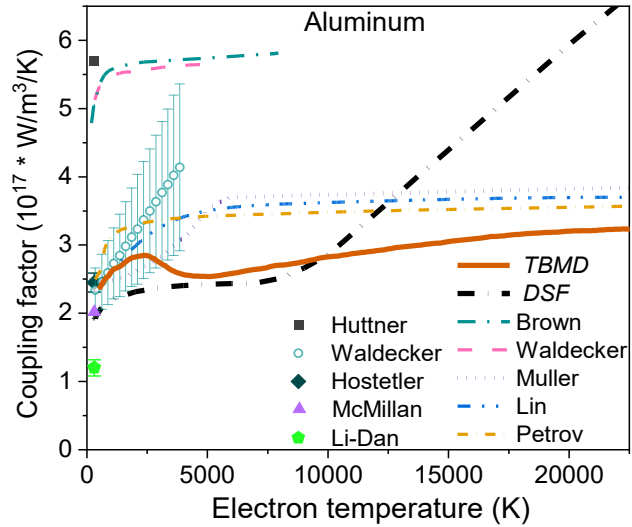

*Figure 22. Electron-ion coupling parameter as a function of electron temperature in aluminum calculated within the XTANT-3 TBMD approach (Eq.(29)) [44]; DSF-based approach by Gorbunov et al. (Eq.(32)) [215]; compared with other estimates by Lin et al. [208], McMillan [217], Allen et al. [218], Brown et al. [209], Petrov et al. [219], Muller et al. [192], Waldecker et al. [210]. Experimental data for comparison are from Huttner et al. [220], Waldecker et al. [210], Hostetler et al. [221], and Li-Dan et al. [222].*

Summarizing, the methods of calculation of the electron-phonon coupling parameter predict electron-phonon temperature equilibration times significantly longer than the time of electronic cooling in an SHI track. Thus, the models that use this parameter need *ad-hoc* adjustments, fitting its value to reproduce track radii detected in the experiments [23,226,227]. The fitted values exceed by more than an order of magnitude those calculated by various methods and measured in the laser experiments (see Figure 21, [139]). The problem can be resolved by taking into account that electron-phonon coupling is not the sole channel of energy exchange between excited electrons and atoms in a solid, which will be discussed in the next Section [139].

## IV.C.    Adiabatic energy exchange between electrons and ions

Unlike the ballistic electrons, the low-energy electrons are involved in the formation of the interatomic potential in a solid. They form the attractive part of the potential, which keeps atoms of a solid together under normal conditions [228]. A change in the electronic system state due to strong excitation forms a new spatial distribution of the electron density altering interaction of neighboring atoms. The atomic ensemble then experiences different forces than those in the equilibrium state. That results in the acceleration of atoms, attempting to find their new equilibrium positions [139].

It is known from laser experiments that in some materials, electronic excitation may lead to ultrafast atomic lattice destabilization even at room atomic temperature – "nonthermal melting". It occurs *via* the breaking of interatomic bonds induced by the electronic excitation instead of the increase of the atomic temperature. It takes place in covalently bonded semiconductors [229–231], ionic crystals [232,233], oxides [202,234], and polymers [235,236]. Nonthermal phase transitions may be regarded as a universal effect taking place in non-metallic crystalline targets upon energy deposition faster than the electron-phonon coupling. More complex nonthermal effects may occur in metals [237,238]. Whether nonthermal effects play a role in amorphous materials is still an open question – it has only been shown that amorphous carbon does not seem to have any significant contribution from nonthermal effects near the damage threshold, and the damage is formed thermally at picosecond timescales [239].

In contrast to the *kinetic* energy (momentum) exchange between the electronic and atomic system described in the previous Section, the nonthermal process manifests a conversion of the *potential* energy of the excited material into the kinetic energy of the atomic system ("nonthermal heating" [139]). It forms a distinct channel of the transfer of the energy deposited to electrons by a passing ion into the atomic system of a target.

A possibility of atomic heating *via* mechanisms other than electron-phonon coupling was also discussed in the literature in terms of the Coulomb explosion [240], assuming transient charge separation in the SHI track. As was discussed in Section III.B, unlike in gases/plasma, the positive charge in solids is not bound to parent ions. The uncompensated charge in an SHI wake is neutralized within a few femtoseconds, as was measured experimentally [67,241]. In contrast, "nonthermal effects" is a more general case of electronic influence on the interatomic potential, which does not require an unbalanced charge.

Nonthermal effects may be well described with ab-initio methods such as DFT-MD or TBMD within BO adiabatic approximation. It assumes that electrons are much lighter and thus faster than ions, so they instantly adjust to (a) any atomic displacement following their guiding dynamics [242], and (b) the changes in the band energy levels induced by the atomic motion [196,243]. The latter means that no electron transitions between these levels take place.

Within BO, there is no coupling (scattering) of electrons to the atomic motion, including atomic vibrations (electron-phonon coupling), thus this approximation cannot capture thermal effects [42,244]. As a result, the electronic and atomic temperatures cannot equilibrate within the BO approximation [244]. The methods of calculation of the non-adiabatic (non-BO) coupling were discussed in the previous Section.

Figure 23 demonstrates the temporal evolution of the kinetic atomic temperature (*i.e.* the equivalent temperature defined by the kinetic energy content, see more on this point in Section V.A) in $SiO_2$ and $Al_2O_3$, modeled with the tight-binding molecular dynamics code XTANT-3 [139]. The figure shows that after the deposition of the dose of 5 eV/atom, the kinetic atomic temperature starts to increase (the same effect in other materials was shown in [139]). The rate of its increase rises with the increase of the dose. At doses close to or above the nonthermal damage threshold dose (reported e.g. in Ref. [202]), the atomic temperature increases within ~100 fs. This increase in the kinetic energy of atoms does *not* occur due to electron-ion (electron-phonon) coupling, as this channel of energy exchange is absent within the BO simulations. Atoms are heated purely due to nonthermal effects – the acceleration caused by changes in the atomic potential energy surface due to the electronic excitation [139,245].

Section III.A showed that electrons cool down and travel outwards from the track core within ~ 100 fs at the doses above a few eV per atom, but the valence holes, which are typically heavier than conduction band electrons, last longer in the track core [140]. These times are sufficient if not for the completion of a nonthermal phase transition, at least for significant nonthermal energy transfer from the electronic to the atomic system.

The dependence of the electron-phonon coupling parameter on the atomic temperature is nearly linear [44]. A strong kick to atoms due to nonthermal heating increases their kinetic energy thus additionally increasing the electron-ion coupling. In turn, this enhanced coupling increases the acceleration of atoms. A faster rate of atomic displacements results in faster changes of the electron-ion Hamiltonian, leading to a self-amplifying process.

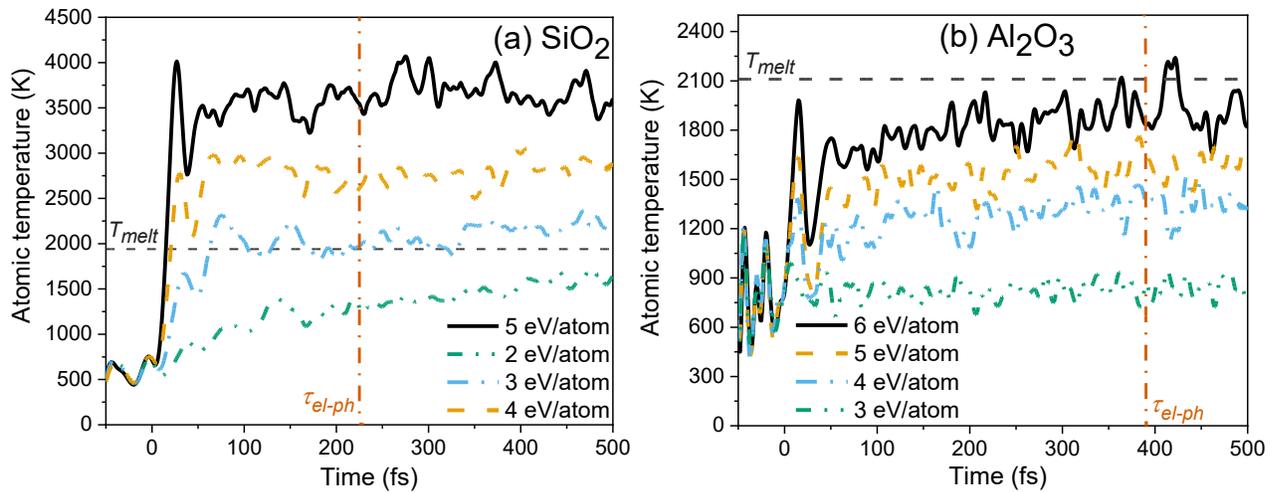

*Figure 23. Atomic kinetic temperature increase after ultrafast energy deposition into the electron system in SiO₂ (a) and Al₂O₃ (b). Thermodynamic melting temperatures are shown as grey dashed lines for comparison. The characteristic times of the electron-phonon coupling extracted from SHI track data with help of i-TS (from Figure 21) are shown with orange dash-dotted vertical lines.*

Understanding the governing mechanisms of the atomic heating in SHI tracks, it can be concluded that the early estimates of the "electron-phonon coupling" extracted from the fast ion tracks parameters, such as in Refs. [23,201] (Figure 21) must be interpreted as reflecting the rate of nonthermal increase of the atomic kinetic energy, with a contribution of elastic scattering, and not as a real electron-phonon coupling parameter (Figure 22) [139].

The results of the laser-irradiation experiments confirm the mechanism of an extremely fast conversion of the potential energy into the kinetic energy of atoms with possible disordering on the timescales relevant to the SHI track problem [246–248].

An increase in the electronic temperature may lead to large changes in the band structure of a material [249,250]. In Ref. [249] it was demonstrated *via* the Keldysh diagram technique [37] that at electronic temperatures of a few eV, the bandgap of insulators may decrease by several eV. It must also be noted that DFT calculations with local density approximation (LDA) predict erroneous band gap increase with the electronic temperature [249]. At ambient conditions, LDA functional significantly underestimates band gaps of insulators and semiconductors, however, with the increase in the electronic temperature, the electron gas becomes more uniform and the LDA functional predictions become more accurate. The band gap shrinkage, or, more general, modifications of the band structure, is even more sensitive to atomic motion.

The potential energy of an atom in the second quantization tight binding formalism can be approximated as a contribution of the ionic repulsion and attraction to electrons [251,252]:

$$V(\{R_{ij}(t)\}, t) = E_{rep}(\{R_{ij}(t)\}) + \sum_i f\left(\varepsilon_i(\{R_{ij}(t)\}, t)\right)\varepsilon_i, \tag{33}$$

where the potential $V$ depends on distances between all the atoms in the simulation box $\{R_{ij}(t)\}$, $E_{rep}$ is effective ion-ion repulsion term (containing all contributions apart from the electronic band energies), and $f_i$ is fractional electron occupation numbers (distribution function) on the transient molecular orbitals $\varepsilon_i = \langle i|\widehat{H}(\{R_{ij}(t)\})|i\rangle$ (or electronic band structure, the eigenstates of the electronic Hamiltonian $\widehat{H}$ that depends on all atomic positions in the simulation box, and $|i\rangle$ is an eigenvector of this Hamiltonian) [228].

From Eq.(33) it can readily be seen that the state of the electronic system directly affects the interatomic potential. Excitation of electrons changes the distribution function $f$ due to an increase in the electronic temperature. In turn, that changes the interaction among atoms, which may destabilize the atomic structure [229,253]. This is the cause of the atomic acceleration shown in Figure 23 in response to the electronic excitation.

Any atomic motion alters the Hamiltonian that depends on all the atomic positions (as well as electronic populations [254]). The eigenstates of the Hamiltonian – the material band structure – thus also evolve in time, during the material response to the deposited energy. For example, Figure 24 demonstrates the temporal behavior of the electronic energy levels in $Al_2O_3$ after deposition of the dose of 5 eV/atom (the energy was deposited during 10 fs at the time of 0 fs *via* increase of the electronic temperature) [235]. This figure shows that the material quickly turns from insulating into a metallic one *via* the collapse of the band gap. The same effect takes place in other covalently bonded materials [193,255].

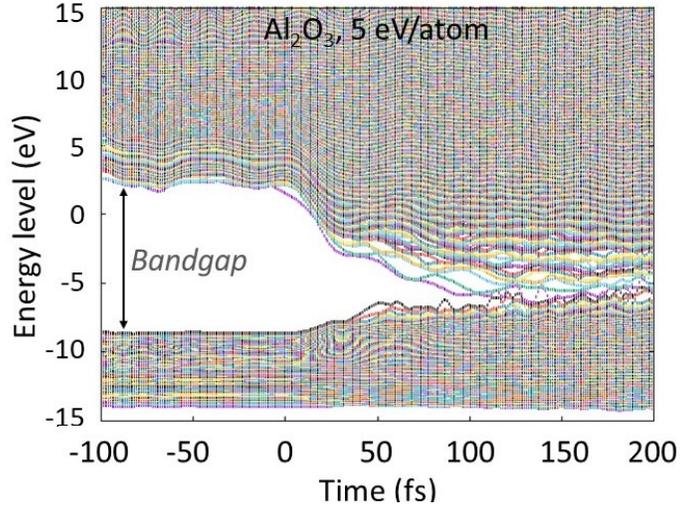

*Figure 24. Evolution of electron energy levels (molecular orbitals) in Al₂O₃ after ultrafast deposition of 5 eV/atom dose* [245].

A deposited dose in the nearest proximity of an SHI trajectory may reach even higher values [256]. One thus may expect similar or even stronger transient changes of the band structure in SHI tracks.

Changes in the band structure will induce changes in other electronic properties such as electron transport coefficients and cross sections of scattering. However, as the band structure change is predicated on the electronic temperature or, more generally, the excited electron distribution function, it is progressing as the electron cascades are progressing [193]. The changes in the band structure become more important as the electron cascades finish [257]. Thus, the majority of a cascade is completed before significant changes of the electronic band structure take place.

When the band gap collapses, the conservation of the total energy demands that the changes of the potential energy between the initial (before the band gap collapse) and final (after the collapse) states must ultimately correspond to the changes in the kinetic energy of atoms (Figure 23). Thus, Eq. (33) and the energy conservation point that the increase of the kinetic energy of atoms in such a process can be estimated as follows:

$$\Delta E_{kin} = \Delta V\big(\{R_{ij}(t)\}, t\big) \approx \sum_i f^{in}(\varepsilon_i^{in} - \varepsilon_i^{fin}), \tag{34}$$

here the $\varepsilon_i^{in}$ and $\varepsilon_i^{fin}$ are the initial and final (after the atomic motion and accompanying band gap collapse) energy levels (cf. Figure 24). In this process, there is no contribution of electron hopping between the levels. This is purely the BO effect because the electron distribution function $f^{in}$ does not change [242]. It clearly distinguishes this nonthermal atomic heating from the electron-phonon (non-BO) effect.

Eq. (34) suggests that the transfer of the excess energy of electron-hole pairs to the atomic kinetic energy at times about 100 fs after the ion passage can model the nonthermal atomic heating by tracing the spatial distribution of the formed electron-hole pairs. This will be discussed in detail in Section VIII.A.

In summary, for an SHI track creation problem, the majority of energy is delivered from excited electrons to atoms of the material *via* three distinct channels: (a) scattering of ballistic electrons/holes [5], (b) coupling between excited low-energy electrons/holes with atoms (electron-phonon coupling), (c) nonthermal forces induced by modified interatomic potential accelerating atoms. They all are completed within some ~100 fs. The ballistic scattering channel "(a)" provides an insufficient amount of energy to atoms [120]. The electron-ion coupling parameter "(b)" may be reliably calculated with various methods, such as non-adiabatic *ab-initio* MD, or DSF formalism based on classical MD, requiring no fitting parameters. However, the electron-phonon coupling channel is too slow to provide necessary energy transfer in SHI tracks (in contrast to longer-lived laser spots). The nonthermal atomic acceleration "(c)" provides necessary additional energy transfer at such a short timescale, forming a crucial mechanism of atomic heating in swift heavy ion tracks.

## V. Picosecond timescales: atomic kinetics from ~100 fs to 1-10 ps

The nonequilibrium electron kinetics at femtosecond timescales forms the initial conditions for further atomic dynamics of a target. The energy delivered to atoms in (quasi-) elastic scattering by fast electrons and valence holes, together with the nonthermal forces induced by electronic excitation, will trigger an atomic response. An example of the energy transfer to the lattice *via* these channels in $Al_2O_3$ and Bi 700 MeV or Xe 167 MeV ions impact is shown in Figure 25. The contributions of the electrons and holes scattering are most pronounced in the track center, whereas the contribution of the potential energy (associated with the nonthermal acceleration of atoms via band gap collapse, see Section IV.C) is more spread radially.

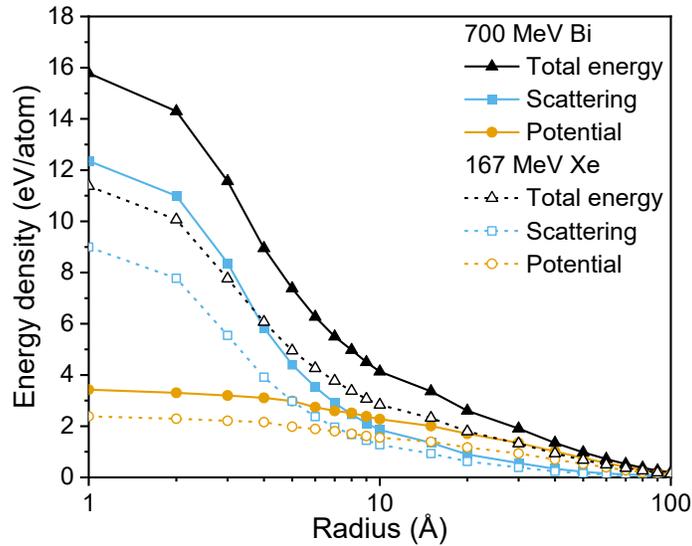

*Figure 25. Radial density of the lattice energy in tracks of Bi 700 MeV and Xe 167 MeV ions in Al₂O₃ at 100 fs including potential energies of valence holes.*

Atoms, provided with kinetic and increased potential energy, may transiently form a damaged region within a few-nanometer radius around an SHI trajectory at the picosecond timescale. A fast increase of the atomic kinetic energy in a localized region around an SHI trajectory also increases pressure, emitting shock waves out of the track core [258,259]. It may transiently lower the density of the material in a track region, and even create voids, which later may recover or stay permanently [47]. In near-surface regions, it results in a flow of atoms, which may be emitted or attracted back to the surface and form observable surface defects (hillocks or craters, will be discussed separately in Section X).

Relaxation of the energy transferred from electrons/holes to atoms can be modeled with the classical molecular dynamics (MD) method. Since by the time of ~100 fs electronic excitation in an SHI track already dissipates outwards from the track core, the interatomic potential returns to its unexcited form [163], which can be used in the standard MD simulation scheme.

As was demonstrated in Ref. [163], to reproduce the final material modification in a track, this initial stage of ~100 fs when electronic kinetics develops in parallel with the atomic one, does not require a simultaneous tracing of both systems: the atomic dynamics may be started to be simulated with MD after the 100 fs, using the energy transferred from electrons as initial conditions. This significantly simplifies the modeling, since Monte Carlo simulations of the electronic processes may then be decoupled from the MD simulation of the atomic response.

In this section, we will describe in more detail these processes lasting from ~100 fs up to 1-10 ps.

### V.A.  Atomic response to heating/excitation (up to 1 ps)

Atomic heating (the kinetic energy) profile created within ~100 fs by scattering of electrons and holes and nonthermal interaction between the electronic and atomic ensembles in the nearest proximity of the SHI trajectory (cf. Figure 25) start to disorder atomic structure at the timescales of a hundred fs. An example is shown in Figure 26.

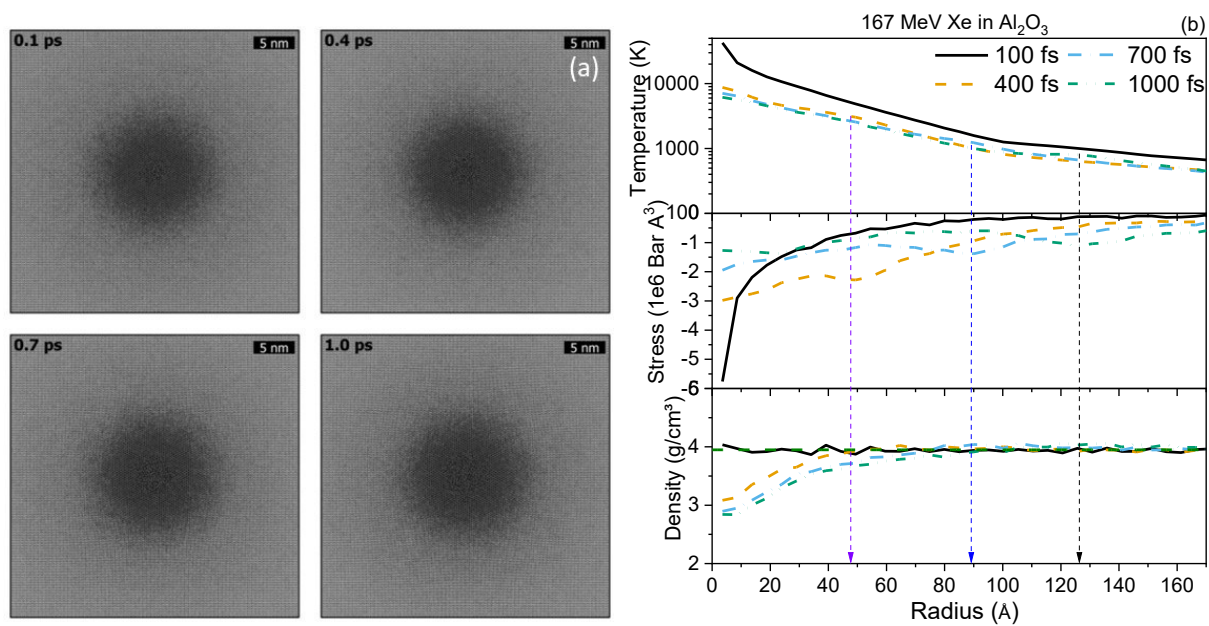

*Figure 26. (Left panel) Atomic snapshots of disordering Al₂O₃ within the first 1 ps after 167 MeV Xe ion passage, calculated with help of classical molecular dynamics [256]. (Right panel) Corresponding atomic temperature (top), stress (middle), and density (bottom) radial profiles at various time instants. Vertical arrows point at the position of the pressure wave peak at the corresponding times. A green horizontal dashed line indicates the ambient density.*

The transiently disordered region is a few nm in radius, having a cylindrical symmetry in the bulk. It has a completely disordered low-density central region which increases in diameter from about 3 to 5 nm surrounded by a severely damaged cylindrical layer of about 15 nm in diameter within the shown 1 ps. The main features of this damage are already seen in the snapshots within 100 fs and change only slightly during the 1 ps timescales.

The increase of the kinetic energy of the atomic system, propagating from the track center outwards, may lead to local nonequilibrium within the atomic ensemble [260]. Receiving an ultrafast strong kick, an atomic distribution function may transiently depart from the equilibrium Maxwellian one describing the atomic system with temperature. Within the MD simulations, it is possible to identify equilibrium conditions by a comparison between the kinetic ($T_{kin}$) and configurational ($T_{conf}$) temperatures [261]. The kinetic temperature represents the average kinetic energy in the system, whereas the configurational temperature is connected with the average potential energy:

$$T_{kin} = \frac{1}{k_B N_{at}} \sum_{i=1}^{N_{at}} \frac{M_i v_i^2}{2} \ , \quad T_{conf} = \frac{1}{k_B N_{at}} \sum_{i=1}^{N_{at}} \frac{|\nabla_i V|^2}{\nabla_i^2 V} \tag{35}$$

here $V$ is an atomic potential, the gradient of which with respect to the coordinates of $i$th atom returns the force acting on that atom. In the case of equilibrium, according to the virial theorem, the time averages of $T_{kin}$ and $T_{conf}$ coincide and are equal to the thermodynamic temperature [262]. Out of equilibrium, they differ, and thus may serve as a criterion for identifying the equilibration in atomistic simulations [263,264]. In SHI tracks, the propagation of temperature and pressure leads to transient nonequilibrium in the atomic system, however, it does not seem to be a strong effect [260].

Atomic response to a nearly instantaneous (~100 fs) heating, compared to times of the final track formation, also results in the creation of pressure waves, emitted radially outwards from the track core, see Figure 26 [120,265]. The pressure waves reflect in some deviation of the temperature profiles from a smooth monotonous decreasing function. The waves help to dissipate the initial excitation, thereby reducing the atomic temperature [120]. This relaxation is accompanied by a decrease of the atomic density in the center of the track within a few nm radius already before 1 ps, and a slight increase around it in the track periphery region [226,256]. In some materials, it may even lead to the creation of voids – transiently or permanently [47,259]. This density profile may be detectable in an experiment if solidifies in this state [226]. The cooling and resolidification processes will be discussed in Section VI.A.

The effects of pressure are especially important for biological matter since even impacts in surrounding water can create such strong pressure waves that damage nearby strands of DNA [265]. It also may transport damage fragments such as free radicals away from the track center farther and faster [258].

Simulations of temperature and pressure waves propagation require large simulation boxes, which are currently impossible to model with *ab-initio* MD. On the other hand, simpler models such as thermodynamic equations (TTM or i-TS) do not account for the finite speed of signal and changes in the equation of state of a material at high pressures and evolving densities [27]. In contrast, such effects are perfectly well captured by classical MD simulation tools, if the used potentials are applicable under such conditions (high temperatures and pressures). For modeling extreme conditions and ultrafast phase and structure transformations, it is important to ensure that the interatomic potential reproduces correctly the response of a material to compression and stretching in a wide range of strains [266]. For polymers and biological materials, it is important to ensure that the force fields used allow for a proper description of chemical reactions – the so-called reactive potentials. E.g., ReaxFF and REBO/AIREBO-based force fields are commonly utilized for the reactive description of organic materials [267].

Nowadays, most of the empirical potentials are fitted to data obtained with DFT by methods such as the force-matching method [268], or the stress-matching method [269]. More recently, self-learning and machine-learning methods are gaining momentum as the most accurate methods of development of MD potentials in physics [270–273] and chemistry [274,275]. A review of various MD potentials appropriate for modeling SHI-irradiation effects can be found, e.g., in the book [33].

Semi-empirical MD potentials are usually fitted to ground-state *ab-initio* simulations (unexcited materials at zero electronic temperatures). It is very difficult to develop a classical MD potential dependent on the electronic temperature (to be used within the TTM-MD model described above, see Eqs.(24-26)), and so far it has been accomplished only for silicon [270,276] and some metals [277,278]. However, for the SHI track formation problem, electron-temperature-dependent potentials may not be necessary. As we have seen above (cf. Figure 18), the initially strongly excited electronic system cools down after some 100 fs, and ground state potentials become applicable to describe the atomic response to this ultrafast energy transfer.

To conclude, in SHI tracks, a response to an increase of the atomic potential energy converts into kinetic energy, which may trigger atomic disorder at extremely short timescales – by the time of a few hundred femtoseconds. It is accompanied by the emission of shock waves radially outwards from the track core, and a transient reduction of material density. These effects are well described with classical molecular dynamics simulations using semi-empirical potentials. Such potentials are usually fitted to ground-state properties of materials, which is sufficient for the SHI track creation problem, as long as the potential reproduces properties of atomically hot matter at high pressures.

## V.B.    Different zones of tracks

The previous section demonstrated that above a certain threshold in the stopping power of an SHI, a considerable disorder can transiently occur within a few-nanometer radius around the SHI trajectory. Later, after atomic cooling and relaxation, this region may form permanent damage called a *track core*. It may have a transitional region from the damaged track core to an undamaged material with its own distinct structure and density sometimes called a *track periphery* [226,279]. This complex structure of a track was first discovered in $SiO_2$ [226], and later shown to be present in other materials. However, whether the transient damage will recover or remain depends on atomic relaxation at longer timescales, which will be discussed in Section VI.

Not all materials form a track core with an atomic disorder or severe damage [2]. Some materials do not form *post-mortem* observable tracks upon SHI irradiation – typical examples include metals, some crystalline narrow band gap semiconductors (e.g., silicon), and nonamorphizable insulators (e.g., diamond, LiF). There may be various reasons for materials not to form a track core: (a) due to the small elastic scattering rate of fast electrons and valence holes it may not deliver sufficient energy to the atomic system around an SHI trajectory to disorder it (this seems to be the case e.g. in silicon [280]); (b) nonthermal forces may be absent or not lead to atomic heating (which is a typical case for metals [281]); (c) transient disorder may occur but further atomic relaxation recovers the damage at longer timescales (which is the case for simple insulators [30]).

Apart from atomic disorder, there are different channels of possible damage. Even in materials that form a track core, thermal (and nonthermal) heating in the vicinity of a projectile trajectory is insufficient to cause damage at large radii, but pressure waves may create strains and defects. It may also affect preexisting defect ensembles in the target [59,227,282]. At relativistic ion energies, excited delta-electrons may possess sufficient energy to knock out a target atom from its equilibrium position by a direct impact (elastic scattering). Atoms may also be knocked-out by the SHI via nuclear stopping scattering, creating atomic cascades.

In wide-band gap materials under an SHI irradiation, a prominent channel of point defects formation occurs *via* recombination of electrons and holes, directly or *via* the excitonic mechanism [128,283]. An exciton is a bound state of a conduction band electron and a valence band hole mutually attracted by the Coulomb field and behaving in a correlated manner. An exciton energy level lies inside the band gap of material [128]. Such quasiparticles may be moving freely (mobile excitons), or be spatially localized (self-trapped excitons) [284,285].

Excitons may decay *via* various channels, such as photon emission, emission of phonons, or creation of point defects. The latter mechanism is only possible in wide-band gap materials, where exciton energy is larger than the cohesive energy of atoms. In such a case, the energy released by an exciton decay is sufficient to knock an atom out of its equilibrium position and thereby create a Frenkel pair of point defects: a lattice vacancy and an interstitial defect [128,194]. Electron energy levels of these defects and their complexes can absorb visible light, hence they are called "color centers". This makes a normally transparent material acquire a color [128].

These defects are typically localized within a radius of a few tens of nm around an SHI trajectory, forming the so-called *track halo* [286,287]. The initial conditions for the formation of a radial profile of the track halo containing color centers are defined by the profile of formed excitons, which, in turn, depends on the valence holes and conduction electrons transport [288]. The typical time before self-trapping of excitons is from a few hundreds of fs (e.g. in $SiO_2$ [289]), to about a picosecond (e.g. in LiF), to 50-100 ps (MgO and $Al_2O_3$ respectively [289]), and is strongly dependent on material and radiation properties. In particular, the higher the excitation density, the faster the trapping process, so in some materials (e.g. NaCl) self-trapping characteristic time cannot even be defined meaningfully [289].

Various point defect types (interstitials and vacancies, charged and neutral, mobile and immobile, single-atom and aggregated) exhibit complex kinetics and interactions. Neutral defects that are immobile may trap a charge (an electron or a hole) and become charged and mobile [290]. Defects may recombine back into a pristine material or form agglomerates. Mobility and stability of various defects are also sensitive to the temperature of material [288,291].

The kinetics of excitons and various point defects may be described with chemical balance equations, accounting for spatial diffusion and reactions for each type of defect and active chemical species, including electrons, valence holes, and excitons. These equations may be solved numerically [290], or sampled with help of the kinetic Monte Carlo method [292]. We will come back to this point in the next section since the typical characteristic timescales of such kinetics are from some 10 ps to nanoseconds or longer [292].

The creation of local defects and related effects play an especially important role in biological matter, since it may lead to the formation of damage in an extended halo around an SHI trajectory [293]. Since a biological sample may be damaged by water radiolysis, the extended region of defects in the surrounding matter may significantly enhance the damage [294].

Apart from a point defect, collective localized defects may also form dislocation loops, such as prismatic or screw dislocations. A dislocation loop is a missing or an extra layer of atoms between the regular atomic planes of the material. It can form when interstitial atoms or vacancies cluster together [295], or during recrystallization of transiently disordered track core [48,256,296]. Dislocation loop creation can modify atomic properties even in materials usually considered radiation resistant, without creating distinct track cores such as in metals [297].

In conclusion, the surrounding region of an energetic ion trajectory consists of various types of damage. In the immediate proximity of a few nanometers, a complete atomic disorder takes place above a certain threshold of the energy deposition, forming the track core. At larger distances, a defected structure of the material is formed or modified, known as the track halo. The halo, depending on the particular material, may contain point defects such as color-centers, dislocations, accumulated stresses, and various radicals and molecular fragments in biological targets and polymers. Transient kinetics of the track core can be well modeled with classical MD. The track halo may be modeled with such methods as kinetic Monte Carlo or chemical balance equations. Both cases, modeling a track core or a halo, require detailed information on the electronic kinetics to set reliable initial conditions.

## VI. Subnanosecond timescales: atomic relaxation

### VI.A. Atomic cooling and recovery

After the initial increase of the atomic temperature and pressure and the formation of the temperature and pressure waves, a slower cooling of the atomic system takes place via diffusive energy transport outwards from the track [120]. An example of this cooling in $Al_2O_3$ after 167 MeV Xe ion impact is shown in Figure 27 (see Figure 26 for the zoom onto the first picosecond). It has been calculated with the classical MD, which is capable of modeling of dynamics of millions of atoms spanning up to nanosecond timescales with modern supercomputers.

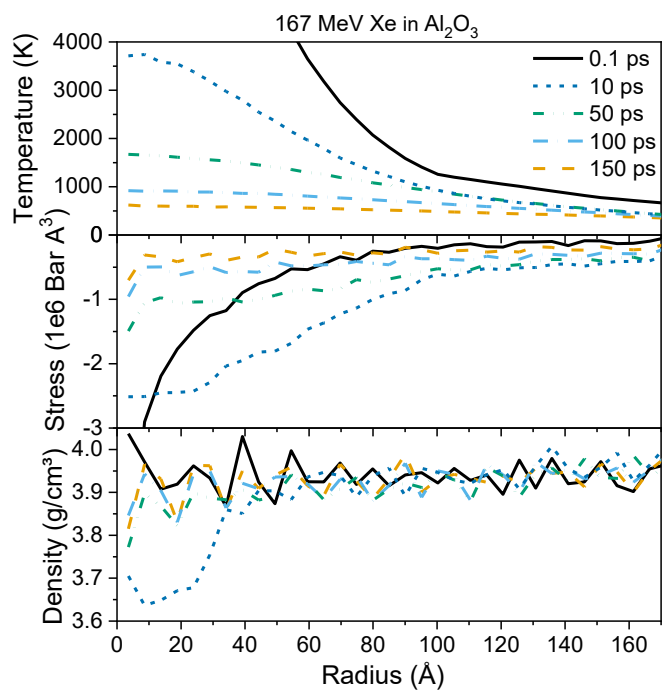

*Figure 27. Relaxation of the atomic temperature (top), stress (middle), and density (bottom) radial profiles in $Al_2O_3$ after 167 MeV Xe ion impact.*

Figure 27 demonstrates that the atomic temperature in $Al_2O_3$ relaxes within some hundred picoseconds to values close to the room temperature at a track periphery. At the same time, pressure/stress relaxes too, leaving only small residual stress in the system, most pronounced in the very track core. In the case of $Al_2O_3$, the density nearly recovered its original value at this time. This is not always the case, and in some materials, a low-density track core may be observed, with an overdense periphery [226]. In extreme cases such as e.g. in amorphous germanium, even voids can be formed in the track core after an SHI impact [47].

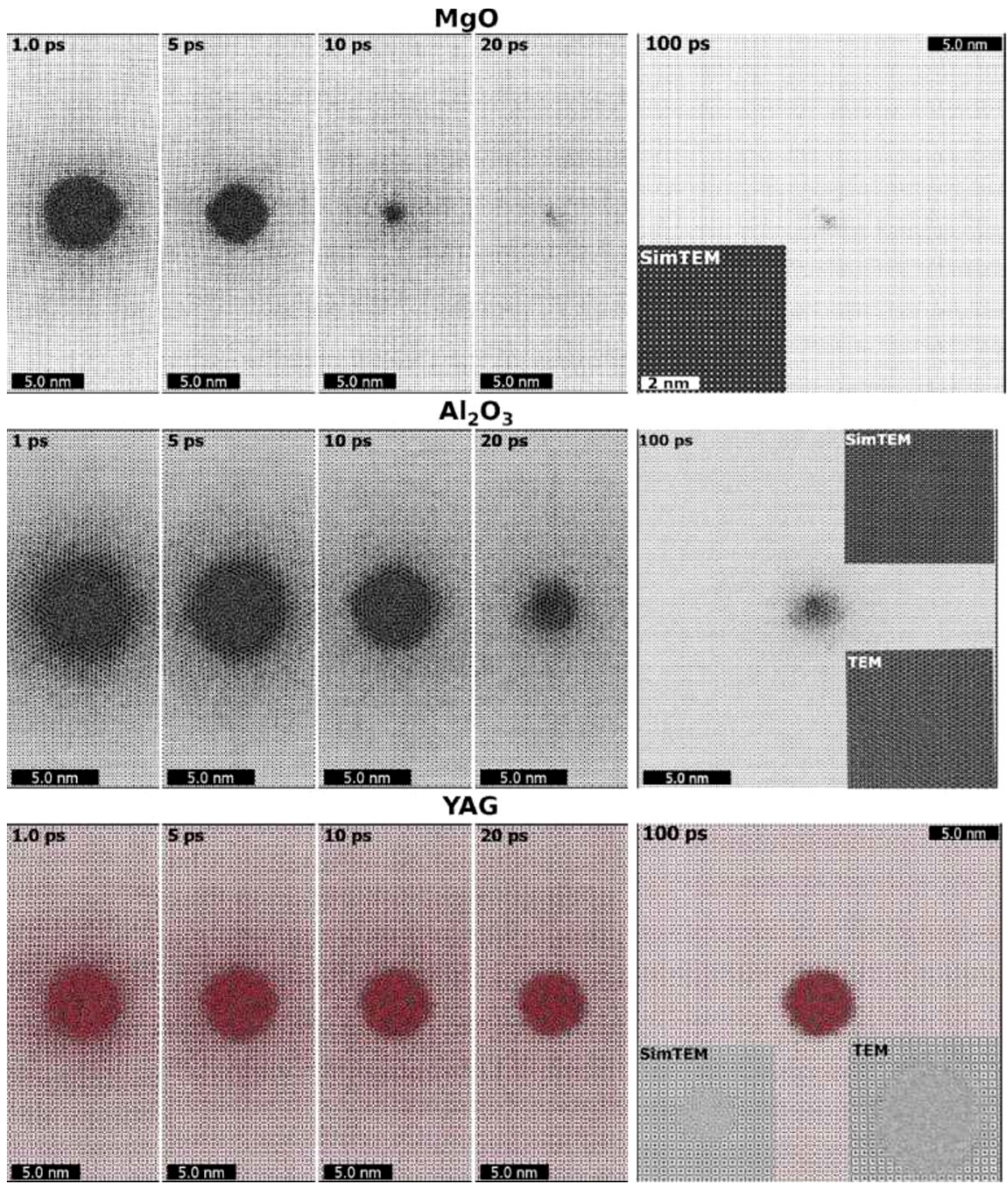

Figure 28. Snapshots of MD supercell of MgO, Al₂O₃, and YAG at different times after the passage of 167 MeV Xe ion (left panels). Comparison of the final produced tracks with those measured in experiments with TEM [256].

During this relaxation, Al₂O₃ undergoes partial recrystallization of the disordered region: the initial disorder shrinks over the time of ~50 ps, leaving a smaller stable track afterward, see Figure 28 [30,256].

This recrystallization behavior strongly depends on the particular material: e.g., MgO, also shown in Figure 28, exhibits no final track despite having nearly the same transient disorder after an ion impact. In contrast, yttrium aluminum garnet (YAG) shows the final track of nearly the same radius as the transiently disordered region, with almost no recrystallization, Figure 28. This illustrates the importance of recrystallization for the formation of observable tracks: depending on the material properties it may drastically change the observable track diameter with respect to the transiently achieved disorder/melting [30]. The ability of a material to recrystallize in an SHI track depends on its complexity: simpler structures are easier to recover within the short time of track cooling than more complex ones [30].

The relaxation kinetics is more complicated in polymers. Long linear chains crystalline polymers react differently to the deposited dose, e.g., forming elliptical and wedge-shaped tracks, whilst tracks in amorphous polymers look like circular dots [298]. A typical polymer heat conductivity is much lower than that of inorganic dielectrics (compare 137 W m$^{-1}$ K$^{-1}$ for silicon or 550 W m$^{-1}$ K$^{-1}$ for diamond vs. 0.03-0.8 W m$^{-1}$ K$^{-1}$ for typical polymers), so a heated region of a track with the radius of 4-5 nm exists longer than 100 ps [15]. Threshold temperatures of damage formation in polymers are lower than those in inorganic crystalline materials [299]. Lightweight hydrogens are likely to detach from their parental atoms breaking and changing chemical bonds under high irradiation doses [235,236,300,301]. Specific processes such as cross-linking, different kinds of polymerization, diverse monomeric composition, various degrees of crystallinity, and so on provide a wide variety of polymeric species with different properties. These features impose limitations on their atomic dynamics simulation, as will be discussed more in Section VII.

Having identified all the stages of the track formation, we see the final product of an ion impact: an experimentally observable nanometric track. The size of the track may be different for the same ion parameters in different materials. And *vice versa*, changing ion parameters affects the track size in the same material. For each material, we can identify a certain threshold in ion electronic stopping power, at which the radius of a created track in the bulk tends to zero, $S_e^{th}$.

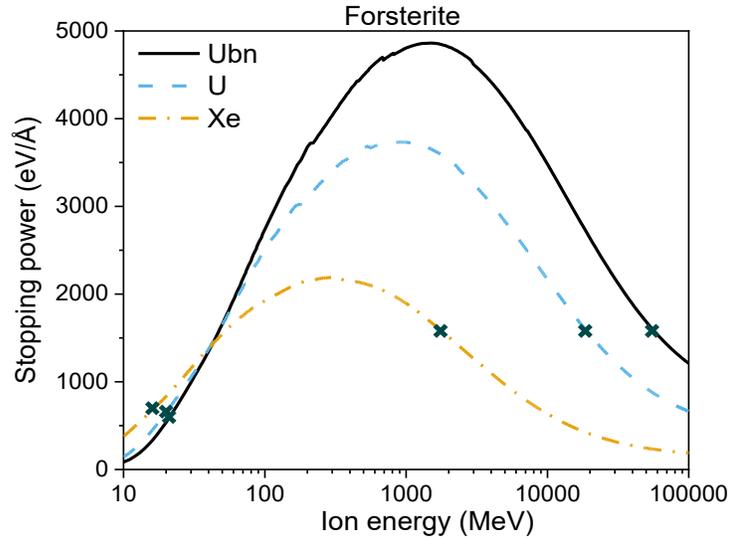

*Figure 29. Energy loss of Xe, U, and Ubn ($Z_{ion}$=120) ions in forsterite. The crosses mark thresholds of track formation at the left and right shoulders of the Bragg curve.*

It is believed that the threshold stopping power is a universal criterion used to evaluate material sensitivity or resistivity to ion irradiation and it does not depend on a particular ion: various SHI seem to have the same threshold stopping power. However, the stopping power is not the sole quantity that defines the threshold of track formation. It is clearly illustrated by the fact that track creation thresholds are different on the left vs. the right shoulder of the Bragg curve.

This manifests the *velocity effect*: ions with the same stopping power but different velocities do not produce identical tracks [302]. The threshold stopping value depends on the ion velocity, $S_e^{th}(v_{ion})$, see an example of the stopping power of Xe, U and Ubn (element 120) ions in forsterite in Figure 29. The threshold values are taken from Ref. [279].

A difference in spectra of electrons excited by an SHI is the reason for the velocity effect: the maximum energy an electron can receive from an ion (see Eq. (17)) is higher for higher incident ion energies (the right shoulder of the Bragg curve vs. the left one). Electrons then travel faster outwards from the track core, bringing more energy farther away. It decreases the energy density in the track core, which reduces energy transferred to the target atoms. Thus, to achieve the same threshold deposited dose, the stopping power of the ion must be higher at the right shoulder of the Bragg curve [279].

The functional dependence of the threshold stopping power on ion velocity $S_e^{th}(v_{ion})$ or energy $S_e^{th}(E_{ion})$ has not yet been studied: the transition between seemingly constant values at the left and the right shoulders of the Bragg curve is still unknown. For example, it is unclear how the thresholds behave for an ion, whose Bragg peak lies above the left threshold but below the right threshold. One may use a set of ions with different atomic numbers or nonequilibrium SHI charges, which would allow to probe the entire plane $S_e(E_{ion})$.

Such a definition of the track formation threshold: $S_e^{th} = \lim_{R \to 0}(S_e(R))$, with $R$ being the radius of a track observed in an experiment, reflects only the track core with a structure clearly distinguishable from the surrounding material. It is not always possible to discern a track core in experiments. In some materials, no track core is formed after an SHI irradiation up to the stopping powers of the heaviest elements used in experiments (U ions) [2]. It includes, for example, LiF, crystalline Si, most metals, and other materials, which are commonly referred to as non-amorphizable materials [201,303].

In conclusion, relaxation of the atomic structure after an SHI impact may be well modeled with the classical MD method applied to small parts of an ion trajectory. Rapid track cooling within sub-nanosecond timescales allows using atomistic simulations with modern days computers. The insights obtained with such simulations allow one to identify the track creation threshold, which is strongly affected by a material capability to recrystallize after a transient disorder. This relaxation stage thus defines the final crystalline or amorphous structure of the formed track, if any. The velocity effect – different thresholds of track formation for ions with the same energy losses but different velocities (the left vs. the right shoulders of the Bragg curve) – may be well reproduced by a simulation that accounts for all stages of track formation, as it is defined by the electron transport and a resulting profile of initial energy deposition into the atomic system.

## VI.B.    Defects kinetics

As was discussed in Section V.B, apart from the severe damage in the track core, a halo containing point defects is also produced by an SHI impact. Point defects are created via atomic cascades due to nuclear stopping, electron-hole recombination, exciton creation and decay, or shock waves.

In the context of point defects, a common and the simplest quantitative measure of damage created in materials is the number of Frenkel pairs formed for given energy transferred to the primary knock-on atom called "displacements per atom" (dpa) [304,305]. This measure allows quantifying the damage produced in terms of the number of point defects, allowing to compare in the "first-order approximation" radiation effects from various kinds of incoming particles (ions, electrons, neutrons) [306].

For a simple evaluation of the dpa produced by an ion in the nuclear stopping regime, the Norgett-Robinson-Torrens (NRT-dpa) model can be used [305] (which is an improved Kinchin-Pease model [304]). In the NRT-dpa model, the number of atomic displacements ($N_d$) can be found as follows:

$$N_d = \begin{cases} 0, & T_d < E_d \\ 1, & E_d < T_d < \dfrac{2E_d}{0.8} \\ \dfrac{0.8T_d}{2E_d}, & \dfrac{2E_d}{0.8} < T_d \end{cases} \tag{36}$$

where $T_d$ is the kinetic energy available for creating atomic displacements, also called the damage energy, which for a single ion is defined as the total ion energy minus the energy lost to electronic ionization. The displacement energy $E_d$ is typically in the range from 20 eV to 100 eV depending on the material [306]. It is a crude but efficient model widely used in radiation physics. More detailed information on the point defects creation in nuclear stopping may be obtained via molecular dynamics simulations, as was already discussed in Section II.B, or *via* Monte-Carlo simulations of atomic cascades [306].

Classical MD simulations do not account for electronic stopping and ensuing atomic heating by excited electrons. So, these effects must be incorporated additionally by means of other methods, such as MC simulations, or at least in the framework of the TTM-MD [282,307]. Such methods, although capable of precise description of the stages of the creation of point defects, are currently too computationally demanding to trace the evolution of the defect ensembles at long timescales (nanoseconds or longer). So far, they found only limited use, although undoubtedly their usage will increase in the future.

Up to now, to trace processes taking place in the defect ensemble, other more efficient methods are used, such as the kinetic Monte Carlo (KMC) simulations [48,58], or rate equations accounting for defect transport [58,290]. The kinetic MC method traces the probabilities of an object (defect such as a vacancy or an interstitial, a dislocation, an impurity, etc.) to transition from one state to another. The states may be different states of a defect (for example, a charged or a neutral defect, a single defect or agglomerate, etc.), or a different position in a lattice (allowing for defect migration). Each transition requires knowledge of the corresponding probabilities or transition rates [308,309]. Then, each possible transition can be sampled with help of random number generation, similarly to the transport MC methods discussed in Section II.C.

Rate equations, or chemical balance equations, are typically written as a set of equations for each type of defect, accounting for their possible conversion into each other, and diffusive transport [58,290]. They require the knowledge of rates (probabilities per time) of the corresponding processes, which define how fast the defect conversions and diffusion take place.

In either of the methods, one of the key parameters is the activation energy of a defect (an energy barrier for defect migration) in a given material. It defines the probability of a defect to move from its potential well to a neighboring site, and correspondingly its transition rate or diffusion coefficient. Reliable activation energies can be obtained by means of *ab-initio* simulations such as DFT *via* evaluating potential energy surfaces of the defects [310].

Activation energies and migration barriers are defined by the collective atomic potential energy surface and are sensitive to the temperature of the material [311]. The probability of a defect migration follows the Arrhenius law $w \sim \exp\left(-E_a/(k_B T)\right)$, with $E_a$ being the activation energy, and thus changes drastically with heating or cooling of the material.

*Ab-initio* methods can be used to calculate defect formation energies (and correlate them with the projectile energy loss) [312–314]. Atomic heating induced by the energy transfer from excited electrons as well as generated shock waves transiently affects the kinetics of defects (preexisting in the target or produced by the SHI impact), creating synergy between the nuclear and electronic stopping effects [59,282,315]. Of particular practical interest is the effect of annealing: heating produced by an SHI impact may be sufficient to trigger significant recombination of preexisting defects, recovering and improving material properties [59].

In conclusion, an SHI impact creates stable point defects with ensuing complex kinetics. The point defects may recombine or aggregate and accumulate, thus changing the properties of the material. The kinetics of defects may be traced by means of kinetic Monte Carlo simulations, or transport rate equations, accounting for possibilities of various defect formation, annihilation, and conversion between different types. It means that the kinetics of SHI-produced point defects can be treated with standard radiation physics models and radiation material science, thereby connecting it to the well-developed field [58].

## VII.    Long timescales: macroscopic relaxation

After the SHI track cools down, its core stays stable. The created halo, however, may continue to be kinetically active and evolve at micro- to macro-scoping timescales. This activity originates from nonequilibrium and spatially inhomogeneous fields of different defects (possibly, chemically active), variations of material density, and induced stresses left by the track formation process. Each of these residual kinetics may be addressed with its own appropriate methods.

In irradiated inorganic solids, a relaxation of macroscopic stresses and inhomogeneous densities may result in the viscous flow of material. At its characteristic timescales of nanoseconds and longer, this flow can be modeled with hydrodynamic models, based on the Navier-Stokes equations [316], or a combination of the continuity equations and those for stresses in the material [317]. Heat transfer, fluid flow, mass transport equations, etc. can be efficiently solved numerically by means of finite element methods (FEM), describing the material response to irradiation at macroscopic scales [33,318,319]. The initial conditions for such modeling may be taken from the MD simulations, which describe the sub-nanosecond timescales very well, as discussed in the previous sections.

Depending on the material properties, namely its viscosity and heat conduction, an amorphous material may freeze in a state with density oscillations, or relax to a nearly unirradiated state (and everything in-between) [317]. Depending on the sensitivity of material properties to temperature, the active stage of density evolution may last only during the cooling time (~100 ps, see Section VI.A), or continue afterward if viscosity at room temperature allows for it. Thus, this process is material specific and requires dedicated modeling for each particular irradiation and target conditions.

The effects of irradiation resulting in macroscopic strains, stresses and swelling are especially important for nuclear power plant components. Modeling of such effects can also be made with multiscale models, exploiting the fact that elasticity equations have no characteristic spatial scale. This fact enables a mathematical treatment based on elastic fields of defects on the nanoscale, readily extending the results to the macro-scale [320]. Such models without characteristic spatial or temporal scales offer a promising perspective for future research.

The track halo is especially important for polymers and biological mater where localized and chemically active defects have drastic effects on material properties: localized bond breaking creates damage fragments, radicals, and chemically active species [172,293]. For the biological targets, three general mechanisms of DNA lesions are suggested: interaction with free radicals and fast hot electrons, dissociative attachment of low-energy electrons to bases, and local atomic heating mechanisms. Realizations of damage channels depend on the parameters of a projectile. E.g., in the case of an energetic carbon ion numerical estimates showed that single strand breaks (SSB) and double-strand breaks (DSB) cannot be neglected [178].

A multiscale simulation tool, MBN Explorer code can be mentioned as an example of a very effective tool for MD simulation of DNA damage [321]. The software is also able to simulate combined systems (e.g., DNA and nanoparticles) and contains a large variety of different-type potentials that makes it a versatile tool. Implemented reactive rCHARMM force field was tested on the DNA and fullerenes [322]. It also supports creation of simple water radicals and their ability to break bonds.

For water, a media surrounding and interacting with biological materials, it is important to trace radiolytic species (such as free electrons, H and OH radicals, $H_2$, etc.) over the time of microseconds to establish the final damage profile after an SHI impact [293]. The kinetics of dissociated molecules leads to complex radiation chemical damage, in which the created free radicals play a crucial role. An additional complication to it is that such effects as transient heating and pressure waves may severely alter the distribution and kinetics of radicals [265]. The indirect damage due to secondary kinetics may be responsible for up to 90% of injuries of cells [323]. Since indirect effects are so damaging, mechanisms of radical creation, distribution, and bond rupture must be considered in adequate models.

This complex problem may be addressed by combined and multiscale approaches, in which various sub-problems are described with their own methods. For example, an SHI and electron kinetics is usually traced with the MC method, which then may be coupled to the chemical dynamics simulated with chemical balance equations [293]. Alternatively, MD simulations of initial stages can be combined with KMC methods for long-timescales dynamics of radicals [265]. A more detailed description of the principles of multiscale methods will be discussed in Section VIII.

Ref. [293] presented modeling of generation and propagation of radicals. The Monte-Carlo code TRAX was modified to describe the production, diffusion, and reactions of radicals with one another and the surrounding medium (TRAX-CHEM). The authors achieved a reasonable agreement of the calculated spatio-temporal radical distributions with available experimental data, albeit experimental data are scarce. This multiscale approach was applied to unveil the role of oxygen in the so-called FLASH effect when normal tissue toxicity is significantly reduced under irradiation with ultra-high dose rates (ion fluxes) [324]. The extension of the TRAX-CHEM code to different rates of oxygenation in a cell made it possible to estimate oxygen contribution and to test the hypothesis of the crucial oxygen role [325].

Direct interaction of reactive species with genetic material was shown, e.g., in [326]. This theoretical study was based on a consistent application of the three tools: (a) GEANT4-DNA Monte-Carlo code was applied for initial physical interactions like ionizations and electronic excitations [327]; (b) *ab-initio* Car-Parrinello Molecular Dynamics was used to figure out which $H_2O$ and $O_2$ molecules were converted into reactive oxygen species (ROS); (c) an MD simulation with reactive ReaxFF potential was utilized up to nanosecond timescales to study reactive species merging into new non-reactive clusters. ROS were shown to form spaghetti-like structures with stranded chains. The harmless complexes are more often formed in healthy cells, while active radicals damage tumor cells stronger [326]. A number of fundamental questions remain open, such as the significance of the damage to cellular components compared to DNA double strands breaks, or the main molecular mechanism of DSB [10].

In conclusion, macroscopic effects at macroscopic times may be modeled with finite element methods or continuum approaches, such as hydrodynamic equations that may be solved numerically. Initial conditions for them require reliable simulation of all preceding stages, starting from a swift ion passage, followed by electronic kinetics, transfer of energy to atoms, atomic response, and cooling. Once they are available, further simulations can be coarse-grained. Applications of well-developed models (viscous flow, diffusion, defect ensemble evolutions) deliver sufficiently precise results. At the same time, the kinetics of point defects may be traced with chemical balance and diffusion equations. The kinetics for chemically active environments, such as biological samples, where the creation of radicals and damage fragments may have dramatic effects, can be followed efficiently by means of chemical balance equations, and rate equations tracing the evolution of each kind of species, however, they often require input from *ab-initio* calculations.

## VIII.    Multiscale model: covering the entire track formation

As we have seen throughout the previous sections, the SHI track formation problem is well separable in time. Moreover, at each timescale of various processes involved in track creation, multiple methods are available to model involved processes with the required precision, which gives a researcher freedom in constructing a hybrid multiscale model.

A hybrid model relies on the well-known theoretical methodology of identifying parameters, with respect to which approximations can be made. This forms the basic idea of creating a hybrid model: finding a parameter allowing for a division of the entire system into subsystems that can be described efficiently with their own models. This reduces the problem to defining proper interconnections between the employed models [33].

Time is one of the most convenient parameters, along which the simulation may be divided. If the processes involved in the problem at hand have different characteristic timescales, each of them may be adequately described with its own method. Models that are divided in time or space are commonly called multiscale models. It is a subclass of hybrid approaches, which, in general, can use other parameters to divide the problem into easily manageable parts (e.g., energy or momentum, particle mass, degeneracy, the strength of coupling, etc. see e.g. Chapter 15 in Ref. [33]).

Having a description of all relevant effects taking place at different timescales during and after an ion impact, we can combine them to provide insights into the entire process of creation of an SHI track. It will be made in the framework of a multiscale model covering many relevant stages of SHI track formation.

To date, multiple variations of combinations have been tried for a description of a track creation in solids, such as MC with MD or TTM-MD in early works [47,51] and in TREKIS+MD model [120], MC with PIC simulations [133], and for damage in biological matter, such as MC and rate equations in TRAX-CHEM code [293], or MD and MC/KMC in MBN-Explorer [82].

Below we will consider in detail a hands-on illustrative example of TREKIS+MD simulations, which uses the Monte Carlo model for the description of the SHI penetration, induced electron kinetics, and energy transfer to the atomic system (up to timescales of ~100 fs), followed by molecular dynamics simulations of atomic response, relaxation, and formation of the observable tracks (up to timescales of ~100 ps). Setting up a reliable MC model will be discussed, with an emphasis on the simple yet efficient connection to the MD simulation. It will be shown that the multiscale model can predict track parameters without adjustable parameters, which is the current state-of-the-art in the SHI track creation problem. A few examples of significant results will be discussed, which could not have been achieved without the application of a multiscale approach.

### VIII.A.    Comprehensive modeling

Here we provide an example of a model based on the combination of MC code TREKIS[d] [5,169] and MD code LAMMPS [328]. During the stage of energy exchange between excited electrons and atoms, the latter can be considered frozen during the simulation of electron kinetics. Thus separated in time, the processes may be treated sequentially, because the short overlapping stage of the relaxing electronic system and starting-to-move exited atoms does not influence the final track parameters [163].

Excitation and evolution of the electronic system by an SHI until timescales ~100 fs after an impact can be efficiently traced with a Monte Carlo code [5,169]. It includes SHI passage triggering ionization of target atoms, transport of primary excited high-energy electrons, generation and transport of secondary electrons, valence band and inner shell holes, photons, and energy transfer to atomic subsystem *via* direct (scattering) channel, utilizing the methodology described throughout Sections II and III; the model for approximating nonthermal contribution (from Section IV.C) will be discussed below.

An SHI is considered to decelerate in the electronic stopping regime, so nuclear losses of the SHI are neglected in the model. In every scattering act, an ion transfers energy to excited electrons according to Eq. (21). A scattering cross section of ions is calculated within the linear response theory through the loss function of a target, see Eqs. (19)-(20). For practical applications, we need to specify the model parameters, which will be done below.

TREKIS models the processes induced by high-energy electrons propagating through a solid outward from the SHI trajectory. Electrons scatter via two dominant channels – elastic and inelastic, whereas Bremsstrahlung photon emission is neglected in this model (considering only nonrelativistic energies).

Inelastic scattering events cause the ionization of secondary electrons, which are calculated in the same manner as for the SHI. Elastic scattering corresponds to a kinetic energy transfer from electrons to atoms, and it dominates for low-energy electrons, calculated either using Mott cross sections Eq. (11) or *via* the phonon part of the loss function Eq. (20) as described in Section III.B.

The probability of a deep-shell hole to decay via Auger/radiative mechanism is chosen according to Poisson distribution depending on their characteristic times (from EPICS2017 database [190]) for an ionized shell. Valence holes produced in direct ionization and/or in Auger decays involving valence band electrons are considered to be mobile. Their transport is simulated similarly to low-energy electrons accounting for the hole effective masses and dispersion relation, Eq. (25), see Section III.C.

---

[d] Available at https://github.com/N-Medvedev/TREKIS-3

Produced in radiative decays of core holes, photons propagate through a target and interact with surrounding atoms. The dominant process in photon transport is photoabsorption (see Section III.C), thus others (Rayleigh and Compton scattering, electron-positron pair production) are neglected in TREKIS. Probabilities of photoabsorption are taken from the EPICS2017 database [190].

If the simulated target is a semi-infinite layer or a thin film, the TREKIS model also accounts for the possible emission of electrons and photons from the target surface (will be discussed in the dedicated Section X.A). The probability of emission is defined via transmission coefficients. In the case of electrons, the transmission coefficient is the solution of the 1D Schrödinger equation for the model surface potential, and in the case of photons, it is derived from Fresnel equations [329].

The key parameter of MC simulations is scattering cross sections, see Eqs. (19)-(20). For practical applications, it requires the inverse complex dielectric function (the energy loss function, ELF, $Im\left(\frac{-1}{\varepsilon(\omega,q)}\right)$) of material for arbitrary values of transferred energy $W=\hbar\omega$ and momentum $\hbar q$. The standard methods to obtain the ELF are based on a reconstruction of the optical limit, $Im\left(\frac{-1}{\varepsilon(\omega,q=0)}\right)$, and then extending it to $\hbar q > 0$.

The optical limit of ELF can be obtained from *ab initio* calculations or from the complex refractive index of a material $n(\omega) + ik(\omega)$ known from experimental data, e.g. [104,105], as follows [100]:

$$Im\left(\frac{-1}{\varepsilon(\omega,0)}\right) = \frac{2n(\omega)k(\omega)}{[n(\omega)^2 - k(\omega)^2]^2 + [2n(\omega)k(\omega)]^2} \qquad (37)$$

For high photon energy involving deep-shell electrons excitations, optical data are often presented in terms of X-ray photon attenuation lengths $\lambda_i(\omega)$ for $i^{th}$ inner shell [190,330]. A high-energy part of the ELF can be calculated via the Fano expression [54,100]:

$$Im\left(\frac{-1}{\varepsilon(\omega,0)}\right) = \sum_i \frac{c}{\omega\lambda_i} \qquad (38)$$

An example of thusly restored ELF for $Al_2O_3$ is shown in Figure 30.

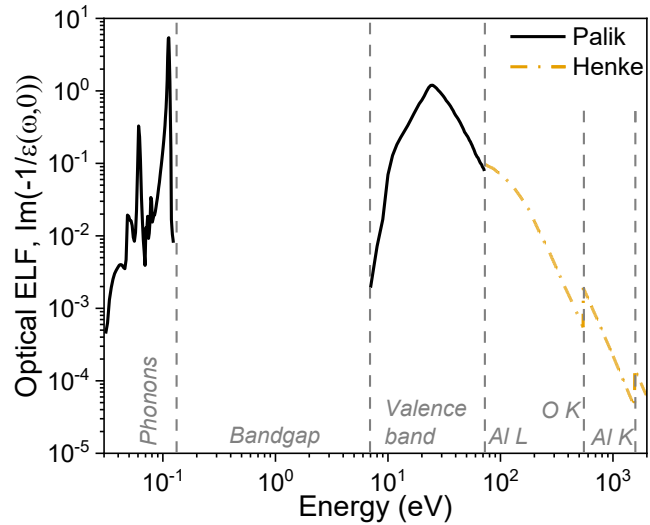

*Figure 30. Optical energy loss function (ELF) of Al$_2$O$_3$. Data for phonon and valence band parts are taken from [104], and deep shells from [330].*

Once optical ELF is reconstructed, it needs to be extended to arbitrary values of transferred momentum $\hbar q > 0$. There are a few standard methods to do so. Penn proposed an algorithm of analytical continuation of optical ELF based on convolution with momentum-dependent weights taken in the form of Lindhard loss functions [107]. Although such a continuation provides an automatic extension to the $\hbar q > 0$ plane, it does not include the finite time of electron excitations and is computationally demanding, which limits its application in MC models. A similar method was developed by Ashley, where a single-pole approximation of Penn's algorithm is used [109]. In this simplified version, the Lindhard loss function is replaced with a delta function with the plasmon dispersion relation for transferred energy. Ashley's model is computationally more efficient, but significantly overestimates the ionization threshold for deep-shell excitations and still does not include finite damping of ELF. Recently, an extension to Penn's algorithm was proposed replacing the Lindhard function with the Mermin model [331–333]. This approach takes into account nonzero damping but additionally increases computational costs, prohibitively so for MC applications.

An alternative to Penn's algorithm and its derivatives, more suitable for MC modeling, is Ritchie and Howie method [106]. This method is widely used in MC simulations due to its relative simplicity. In this method, the optical ELF is approximated with a finite sum of Drude-Lorentz (DL) oscillator functions:

$$Im\left(\frac{-1}{\varepsilon(\omega,0)}\right) \approx \sum_i \frac{A_i \gamma_i \hbar \omega_i}{((\hbar\omega_i)^2 - E_i^2)^2 + (\gamma_i \hbar \omega_i)^2} \tag{39}$$

This simple analytical representation depends on a set of parameters $(A_i, E_i, \gamma_i)$ determined from the fitting procedure of the optical data (Eqs.(37,38)). The parameters represent the amplitude $A_i$, the position $E_i$ and the width $\gamma_i$ of the $i^{th}$ oscillator, and may be interpreted as intensity, energy, and an inverse lifetime of collective excitation (plasmon or phonon) associated with peaks in the optical ELF.

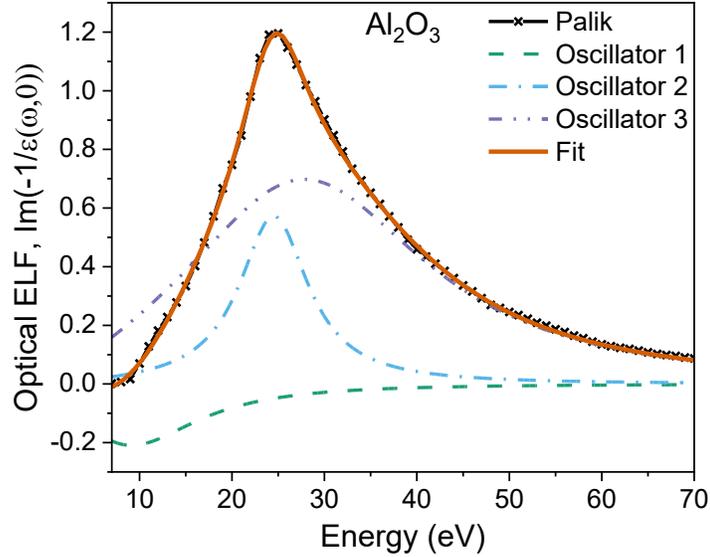

Figure 31. Drude-Lorentz fitting of valence band part of optical ELF in Al₂O₃ compared with the data from Palik [104].

The fitting parameters are additionally constrained by the sum rules for ELF. The *f*-sum rule states that the value $Z_{eff}$ must be equal to the number of electrons on a shell with ionization potential $I_p$ (or in the valence/conduction band) [100]:

$$Z_{eff} = \frac{2}{\pi \Omega_P^2} \int_{I_p}^{\infty} Im\left(\frac{-1}{\varepsilon(\omega,0)}\right)_{shell} \omega d\omega = N_{e,shell}. \tag{40}$$

Here $Im\left(\frac{-1}{\varepsilon(\omega,0)}\right)_{shell}$ is a partial ELF for the considered shell, and $\Omega_P^2 = 4\pi e^2 n_{at}/m_e$ is the corresponding plasma frequency.

The KK-sum rule (the limit of Kramers-Kronig dispersion relations) is [334]:

$$\frac{2}{\pi} \int_0^\infty \left(\frac{-1}{\varepsilon(\omega, 0)}\right) \frac{d\omega}{\omega} = 1 \qquad (41)$$

An example of Drude-Lorentz fitting for the valence band of $Al_2O_3$ is shown in Figure 31. The corresponding values of the parameters and sum rules are shown in Table 1.

Table 1. Fitted coefficients of the loss function in $Al_2O_3$ according to Eq.(39). The f-sum rule (Eq.(40)) are listed in comparison with the number of electrons $N_e$ in the corresponding shells (or the number of atoms $N_a$ in the unit cell, for phonons). Corresponding kk-sum rule (Eq.(41)) is 0.934.

| Shell | A | E | γ | f-sum rule ($N_{e,a}$) |
|---|---|---|---|---|
| Valence band | -42.659 | 12.613 | 19.246 | 23.998 (24.000) |
| | 128.226 | 24.925 | 9.043 | |
| | 694.967 | 32.087 | 33.244 | |
| $L_3$-shell of Al | 282.8 | 109.9 | 104.0 | 3.961 (4.000) |
| $L_2$-shell of Al | 141.4 | 111.1 | 103.3 | 1.989 (2.000) |
| $L_1$-shell of Al | 94.27 | 150.69 | 212.8 | 1.365 (2.000) |
| K-shell of O | 282.8 | 540 | 350 | 2.028 (2.000) |
| K-shell of Al | 117.83 | 1519.3 | 969.77 | 1.790 (2.000) |
| Phonons | 0.003 | 0.1125 | 0.005 | 3.483 (5.000) |
| | 0.000045 | 0.061 | 0.002 | |

The analytical expression for ELF in Eq.(39) does not include an inherent extension to the $q > 0$ plane. Instead, various dispersion relations for the position of the oscillator $E_i$ and its width $\gamma_i$ have been proposed [106]:

1) The simplest and the most common dispersion relation is a free-electron approximation:

$$E_i(q) = E_i(0) + \frac{\hbar^2 q^2}{2m_e} \qquad (42)$$

2) Plasmon-pole approximation (where $v_F$ is the Fermi velocity of electrons in a target):

$$E_i^2(q) = E_i^2(0) + \frac{1}{3} v_F^2 (\hbar q)^2 + \left(\frac{\hbar^2 q^2}{2m_e}\right)^2 \qquad (43)$$

3) Extended Ritchie model:

$$E_i(q) = \left[\left(E_i(0)\right)^p + \left(\frac{\hbar^2 q^2}{2m_e}\right)^p\right]^{1/p}, \qquad p = \frac{2}{3} \qquad (44)$$

$$\gamma_i(q) = \left[ \left( \gamma_i(0) \right)^2 + \left( \frac{\hbar^2 q^2}{2m_e} \right)^2 \right]^{1/2}$$

All of these dispersion relations satisfy the two limits: 1) $q \to 0$ where $E_i(q) = E_i(0)$ recovers the optical limit, and 2) $q \to \infty$ where $E_i(q) = \frac{\hbar^2 q^2}{2m_e}$ corresponds to free electrons, restoring the Bethe ridge [106]. The choice of the particular dispersion relation is rather arbitrary, conditioned only on the agreement of calculated electronic properties – namely, the mean free path and the stopping power – with experimental ones.

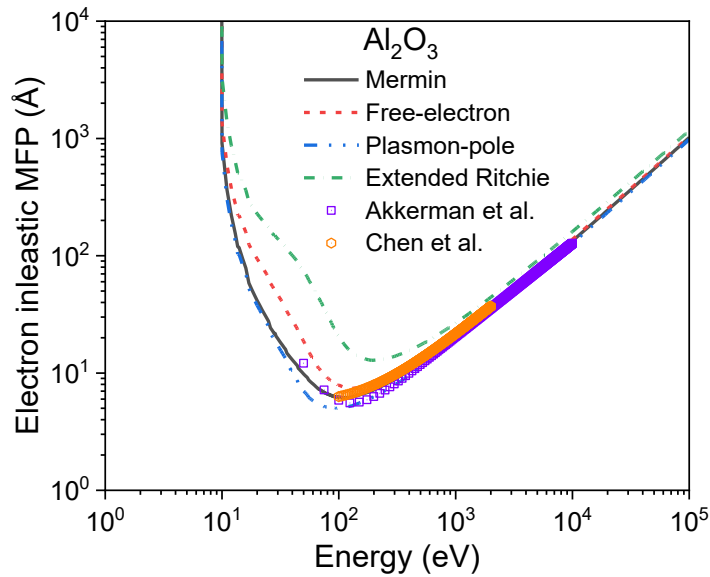

*Figure 32. Total electron inelastic mean free paths in $Al_2O_3$ calculated with TREKIS using Mermin-type ELF as well as dispersion relations (42)-(44). Data of Akkerman et al. and Chen et al. are taken from the NIST database for comparison [335].*

Another approach to the extension of optical oscillators to the ($\omega$,$q$>0) plane was proposed in the works of Abril *et al*. [108]. Instead of the dispersion relations within the oscillators, it uses Mermin dielectric function. Considering oscillators in Eq. (33) as a $q$=0 limit of the Mermin ELF, the extension to the $q$>0 is then done *via* the analytical continuation of the Mermin function. As it was pointed out in Section II.C, the Mermin-based ELF function seems to provide a better agreement of mean free paths (see, e.g., Figure 32 for $Al_2O_3$) and stopping powers with experimental data in the case of protons and electrons. However, it is computationally more demanding than the Richie-Howie approach and thus found only limited use in practical applications in MC models so far.

To avoid numerical integration of cross sections, required within Ritchie-Howie or Mermin models, further approximations can be made. Noticing that at high particle energies the ELF peak width is negligibly small, it can be approximated with a delta-function, as was proposed already by Fano [54], and realized by Liljequist [41,336], and more recently in a rigorous derivation in Refs. [6,337]. At high values of transferred energy and momentum $\{W, Q\} \gg \gamma_i$ the Drude-Lorentz oscillators reduce to delta-functions [6]:

$$\left( \frac{-1}{\varepsilon(\omega, q)} \right) \approx \sum_i \frac{\pi A_i}{2W} \delta(W - (E_i + Q)) \tag{45}$$

In that case, the cross-sections (19) and stopping powers (14), as well as transferred energies in scattering events (21), can be evaluated analytically which significantly reduces the computational demands of MC simulations [6,337].

Once the energy- and momentum-dependent energy loss function $Im\left(\frac{-1}{\varepsilon(\omega,q)}\right)$ is reconstructed e.g. *via* the Ritchie-Howie fitting algorithm, one may perform the Monte Carlo simulation of an SHI impact and ensuing kinetics of the electronic system. The electronic kinetics is traced in TREKIS until the chosen time of ~100 fs, after which the density of excited electrons in the track center drops to negligibly small values. The majority of energy by this time is already redistributed to valence holes, to atoms, or carried far away by the delta-electrons [120].

The MC-calculated energy distributions then serve as the initial conditions for molecular dynamics simulations of the triggered atomic kinetics in an SHI track. The energy transfer to the atomic system at such short timescales is provided *via* three channels: (i) energy transfer in (quasi)-elastic scattering of electrons with target atoms; (ii) scattering of valence holes with atoms (both of these channels are discussed in Sections III.A and III.B); (iii) the ultrafast nonthermal heating of atoms due to modification of the interatomic potential (discussed in Section IV.C or Ref. [139]).

The spatial distributions of the energy transferred to atoms *via* the first two channels are readily available in the MC method. The last one is not accounted for in MC simulations, and neither can it be easily included in MD calculations. In covalent materials, the modification of the potential energy surface resulting from the electronic excitation is accompanied by an ultrafast band gap collapse (see Section IV.C), which releases this energy into the kinetic energy of atoms on the characteristic timescale of ~100 fs. The energy release converted into an increase of the kinetic energy of atoms in such a process is defined by Eq.(34).

As was seen in the example in Figure 24 (Section IV.C), during the nonthermal rearrangement of the band structure, the valence band levels spread somewhat symmetrically from their original values, whereas the conduction band levels lower towards the top of the valence band. Thus, the energy change in this process may be approximated as a release of the energy equal to the band gap for each excited electron/valence hole pair, at the time of electron/hole stopping resolved in space [163]:

$$\Delta E_{kin}(r) = N_h(r) E_{gap} \tag{46}$$

That means, in covalent materials, it is possible to account for this effect very efficiently in a combined MC-MD model [163]: the potential energy stored in electron-hole pairs at the end of MC simulations at ~100 fs can be simply fed to atoms as additional kinetic energy [120]. Such a simple but efficient way of energy transfer between MC and MD models, accounting for the nonthermal heating of atoms in an SHI track, proved sufficiently precise to provide a good agreement with experiments [256].

These three energy distributions (provided by elastic scattering of electrons, of valence holes, and potential energy of electron-hole pairs representing nonthermal contribution) obtained from MC calculations with TREKIS code are then serving as the initial velocity distribution of atoms in the MD part. With the appropriate choice of the interatomic potential or a force field for a system (as discussed in Sections V and VI), MD describes the response of the atomic system to this instantaneous energy increase.

In practice, it is done in the following way: energy density profile is provided to the beforehand equilibrated supercell, introducing normally distributed kinetic energies and uniformly distributed momenta to atoms in cylindrical layers around an SHI trajectory. Then, the LAMMPS MD package is used to simulate the atomic dynamics within up to times of ~100 ps, when the average temperature of the track region is dropping down to $300 - 400$ K so no further structural changes are expected after that in the track core. The choice of the interatomic potential for the initial supercell equilibration and subsequent atomic relaxation depends on the particular target material; the chosen potential has to be tested in extreme conditions similar to those achieved in an SHI track core (see Section V.A). The X- and Y-borders of the supercell (in the plane perpendicular to the ion impact) are kept at 300 K temperature with the Berendsen thermostat [338] with a characteristic time of 0.1 ps, representing heat exchange with the surrounding unirradiated material. In the case of bulk modeling [120], periodic boundary conditions in all directions are applied, while for thin films and surfaces [259,339], periodicity is turned on only in X- and Y-directions.

We emphasize that this hybrid model does not require any *a-posteriori* fitting to the track data to work. Once the cross sections of the processes for the MC model are known, and an interatomic potential for the MD model is provided, the combined model is capable of delivering information on the track formation for an arbitrary combination of an SHI and a target. Below we will consider a few examples of the results obtained with this model, which were not possible to achieve without a hybrid approach tracing all stages of SHI-induced electronic and atomic kinetics.

## VIII.B.   Examples of results

The hybrid approach allows e.g. to describe structural changes for different ion species and energies, or at different points along the ion trajectory covering almost the entire ion path. In the recent works [256,279], the model described above in Section VIII.A was used to study an interplay of the ion energy loss and velocity during the formation of a damaged region around the trajectories of Xe (30-1650 MeV), U (45-15000 MeV), and hypothetical $^{304}_{120}$Ubn (50-25000 MeV) ions in olivine ($Mg_2SiO_4$, forsterite allotrope).

Figure 33 shows the MD snapshots of amorphous tracks for different U ion energies in olivine (or, equivalently, at different depths along the ion trajectory). Figure 33b also demonstrates the corresponding dependence of the track radius on the electronic energy loss of Xe, U, and Ubn ions in $Mg_2SiO_4$. It has a loop-like form with two distinct thresholds of track formation – for slow and fast ions. Extrapolation of MD data points with polynomial functions gives the threshold value of *dE/dx$_{th}$* ~6.3 keV/nm for slow ions and ~ 15.8 keV/nm for fast ions. Ref. [340] reported the experimental lower limit for the track formation threshold in $Mg_2SiO_4$ around 6.67 keV/nm and the estimated value of ~7.5 keV/nm, which is in reasonable agreement with the calculated one.

A similar loop-like dependence of the track radius on electronic energy loss was observed experimentally in $Y_3Fe_5O_{12}$ in Refs. [341,342], suggesting that it is a universal effect.

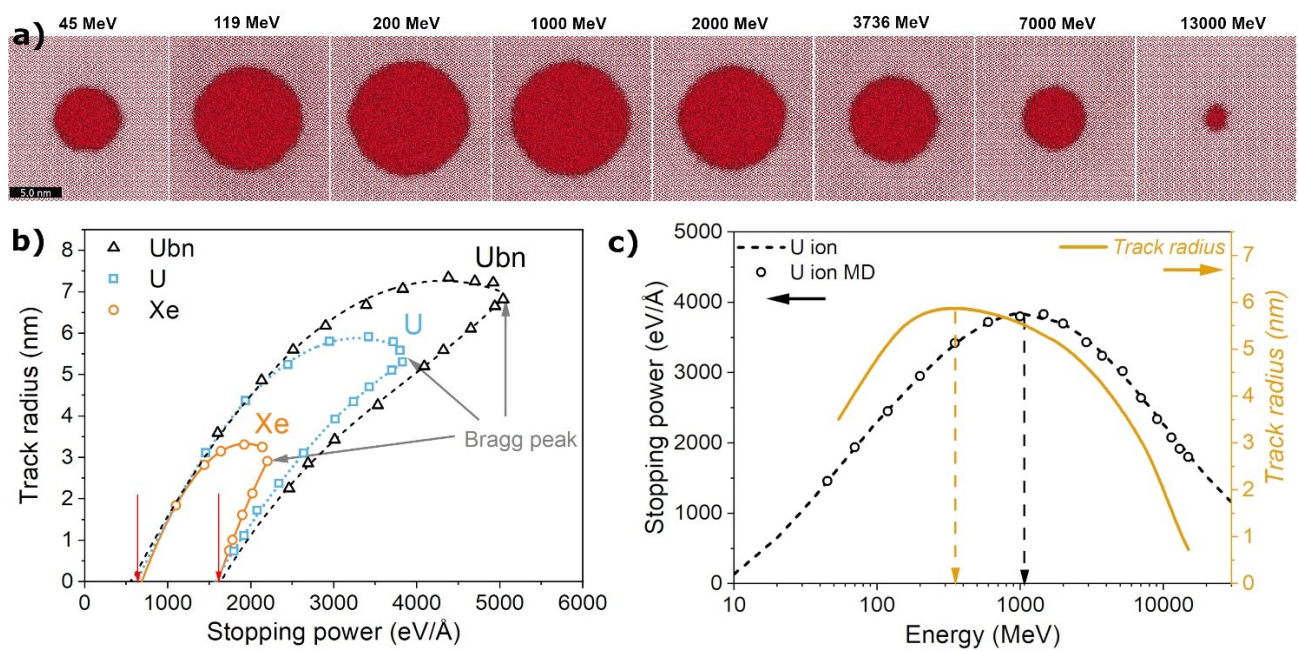

*Figure 33. (a) Results of MD simulations of U impacts in olivine with different energies. (b) Dependence of the track radius on the energy loss of Xe, U ions, and hypothetical Ubn ion in forsterite. Lines represent polynomial fits of the data points for slow and fast ions. Blue arrows mark calculated track formation thresholds. Grey arrows point at the Bragg peaks, the maximal energy loss (note their mismatch with the maximal track radii). (c) Track sizes and stopping power of U ion vs. SHI kinetic energy. Dashed arrows indicate the positions of maxima of the curves.*

Figure 33b also shows the energy loss corresponding to the Bragg peak (pointed out with grey arrows). The peak energy loss value does not produce a damaged region with the maximal size. This effect is elucidated in Figure 33c where the maximal damage produced by U ions lies at much lower energies (~250 MeV) than the Bragg peak position (~1000 MeV). This effect also manifests in terms of ion projected ranges: the maximal damage production along the ion trajectory appears deeper than the Bragg peak position (see Ref. [279] for details).

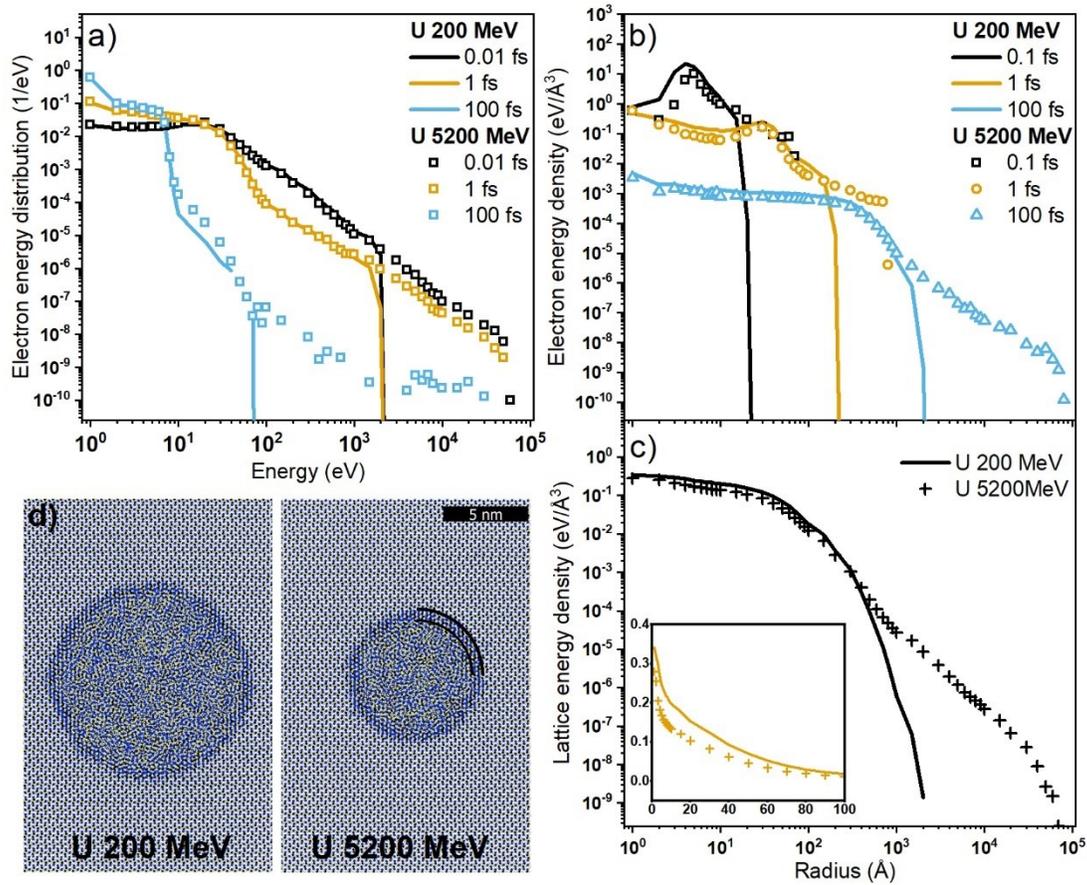

*Figure 34. The velocity effect in tracks of 200 MeV and 5200 MeV U ions in Mg₂SiO₄: (a) spectra of generated electrons and (b) radial electron energy density at different times after the ion impact; (c) calculated lattice energy at 100 fs with inset zooming at the region of radii < 100 Å; (d) MD snapshots of U ion tracks. Black arcs show the core-shell structure of the track.*

Figure 34 shows that the origin of the velocity effect lies in a difference of δ-electrons spectra created in Mg₂SiO₄ by fast (200 MeV, 0.84 MeV/u) vs. slow (5200 MeV, (21.8 MeV/u) uranium ions with the close values of stopping power of ~30 keV/nm. The faster ion (Figure 34a) produces a much larger yield of high-energy electrons bringing a part of deposited energy out of the track core, which significantly influences further electronic cascading (Figure 34b) and finally resulting in lower heating of a central region of a track (Figure 34c and inset). This, in turn, produces a smaller observable track despite the same SHI stopping power (Figure 34d). This figure also demonstrates that a track in olivine has a structure typical for amorphizable solids: an amorphous core surrounded by a narrow shell of an intermediate structure. In the case of $Mg_2SiO_4$, the width of the shell is almost the same for all ion species (~0.5 nm) [279]. The same effect was found experimentally in $SiO_2$ [226].

To conclude, a hybrid or multiscale model allows for tracing nonequilibrium kinetics of electrons and holes with the transport Monte Carlo approach and the response of target atoms to various channels of energy deposition with molecular dynamics. It predicts fundamental effects of the SHI irradiation: a mismatch between the Bragg peak and the maximal track radius as well as provides a quantitative explanation of the velocity effect. Both of the phenomena result from the difference in the electronic transport triggered by the impact of ions with energies at the left vs. the right shoulders of the Bragg curve. The model does not employ *a posteriori* fitting parameters, thereby showing that predictive models of SHI track creation are possible to build with the current state-of-the-art knowledge in the field.

## IX.   High fluences: track overlap

A large number of ions in an accelerated bunch may lead to tracks overlap. Its possibility is characterized by the ion fluence – the number of impinging ions per unit area: $\Phi = N_{ion}/A$. Depending on the fluence, even multiple track overlap may occur.

With a gradually increasing fluence, a larger and larger area of a sample will be covered with tracks, so eventually, track halos will start to overlap. At higher fluences, track cores will start to overlap with a region of transient disorder, which may be larger than the final size of the track as discussed in Section VI.A. At even higher fluences, overlapping of track cores may lead to macroscopically observable changes in material properties [2,343]. Each of the overlapping regimes will produce its own kind of synergy and induce phenomena not observable in an individual track.

Apart from the fluence, another important factor is the ion flux: a fluence per unit of time, $\varphi = d\Phi/dt$. The flux is closely related to the dose rate: a dose deposited per unit of time, a concept widely used in radiation chemistry and biology [8,344]. At a low ion flux, a successive ion impacts an already formed and cold track created by a prior ion. At a high flux of ions, an ion may hit a still hot and forming track, whose properties are different from that of a cold matter. Accumulation of the deposited dose at hot stages during high-flux irradiation may lead to extreme heating of an entire area of interaction in the sample to a temperature sufficient to form a plasma within the ion bunch duration [345]. Below, we will consider the effects of track overlap in both regimes.

IX.A.    Low flux: successive overlap

Low flux irradiation of matter realizes the regime at which SHIs arrive at the target after significantly long-time intervals as to meet already relaxed and cold fully formed tracks created by prior ions. Various effects take place in this scenario with increasing fluence.

At a low fluence regime, individual isolated tracks occur in a target. Each track can be considered independent, and thus the effect of damage accumulation is linear: each SHI in the bunch adds an identical damaged region to the target.

With the increase of the fluence, track halos start to overlap. This effect can be most clearly demonstrated in materials, which do not form a track core, for example, alkali halide crystals. Measuring the number of point defects allows tracing damage accumulation vs. fluence. An example of LiF irradiation with Pb ions (1600 MeV energy) is shown in Figure 35 [346]. At fluences below ~$10^{10}$ ions/cm$^2$, the area density of created $F$-centers is indeed linear. With an increase in ion fluence, deviation from the linear behavior is seen at fluences approaching $10^{11}$ ions/cm$^2$. A saturation of the defects concentration occurs at ion fluences ~$10^{11} - 10^{12}$ ions/cm$^2$ [174,291]. It allows estimating the halo radius, which gives the values of ~$5 - 50$ nm.

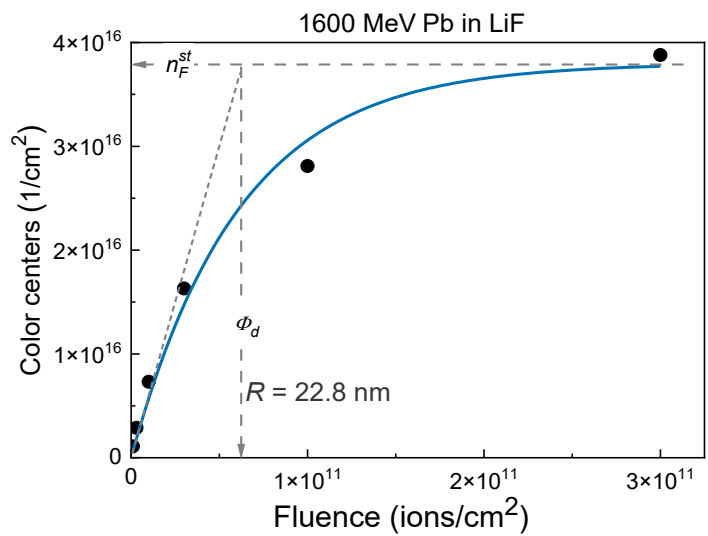

Figure 35. The fluence dependence of the number of F-centers per surface area detected in LiF crystals (circles) irradiated at room temperature with 1600 MeV Pb ions, and its interpolation (red line). Data taken from Ref. [346]. Saturation level $n_F^{st}$, corresponding linear saturation fluence $\Phi_d$ and the halo radius are shown with grey arrows.

When the track halos start to overlap, complex defects kinetics start to play a role. Electronic excitation, local heating, and formation of new defects in the regions where a defect halo was already formed lead to such competing effects as defect aggregation and annealing [194,282,347]. As discussed in Section VI.B, changes in the target temperature, as well as the addition of newly formed defects, significantly affect preexisting defect ensembles.

At sufficiently high fluences, each successive ion arrives in an already damaged region. Transient heating at the impact leads to annealing of prior defects and the formation of agglomerates ($F_n$ centers, etc.), which results in the saturation behavior of $F$-center concentration seen in Figure 35 [290,348]:

$$n_F = \mathrm{n}_F^{\mathrm{st}}(1 - \exp{(-\Phi/\Phi_d)}) \qquad (47)$$

where $n_F$ is the area density of $F$-centers, $\Phi$ is the SHI fluence, and the saturation level $n_F^{st}$ and corresponding linear saturation fluence $\Phi_d$ are empirical parameters defined from experiments. This saturation law is typical for point defect accumulation not limited to $F$-centers in alkali halides [349].

At even higher fluences, track cores start to overlap (in materials that form track cores). This leads to the accumulation of damaged regions, which may be amorphous or disordered, as discussed in Section VI.A. The initial linear accumulation of individual independent tracks starts to deviate from the linear behavior, similarly to deviations in the number of defects in the halo but at higher fluences. These effects may be modeled straightforwardly with an MD simulation, in which the second ion may be sent into the simulation box after it cooled after the first ion impact.

As we have seen above in Sections V.A and VI.A, transiently disordered areas around an SHI trajectory may be larger than the final track radius in materials with efficient recrystallization. The overlap of tracks starts with an overlap of the transiently hot disordered area with the formed prior track. For example, $Al_2O_3$ under successive irradiation with two Bi ions (700 MeV energy each) at different distances, calculated with a multiscale MC-MD model (see Section VIII), is shown in Figure 36 [256]. At the distance of 8 nm, where the two tracks do not overlap, the recrystallization of the second track is almost not affected by the presence of lattice damage induced by the first ion. This results in the formation of two isolated tracks. In contrast, in the case of 6.5 nm distance, the transient hot disordered track overlaps with the pre-existing damaged track from the previous ion. Recrystallization of the track starts from the periphery of this highly excited disordered region. The existence of the pre-damaged first ion track precludes the periphery of the second track from perfect recrystallization. Instead, a new damaged region in-between the two ion trajectories forms during solidification. In this case, the damage in the $Al_2O_3$ crystal irradiated with 700 MeV Bi ions increases nonlinearly with increase of the fluence.

At higher fluences, corresponding to an even shorter distance between the impact points of two ions, the transiently disordered region created by the second ion covers completely the first pre-existing track, see a comparison in Figure 37. In this case, it anneals the pre-existing track, and only the second track is observed (see Figure 37f, and experimental confirmation in Figure 37b) [256].

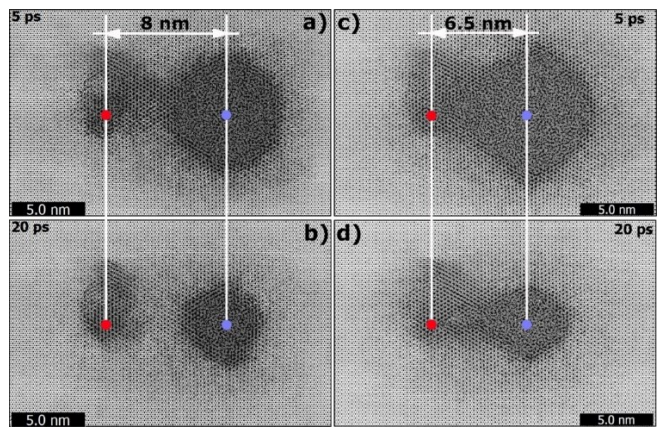

*Figure 36 Snapshots of an $Al_2O_3$ supercell at (a) 5 ps and (b) 20 ps after passage of a 700 MeV Bi ion in the crystal at 8 nm from a pre-existing 700 MeV Bi track; at (c) 5 ps and (d) 20 ps after passage of a Bi ion at 6.5 nm from a pre-existing track. Red and blue dots indicate trajectories of first and second ions correspondingly [350].*

In materials, where recrystallization does not shrink transient disorder and the final track has the same radius, the annealing does not take place [256]. There, track overlap has a minor effect, and linear accumulation of defects proceeds up until the fluence where track cores start to overlap and saturate.

Saturation of damage typically takes place at ion fluences ~$10^{12} - 10^{13}$ ions/cm$^2$ which corresponds to the average radius of the damaged regions of ~$0.5 - 5$ nm [2]. These values, indeed, correspond to the sizes of track cores, see Section VI.A.

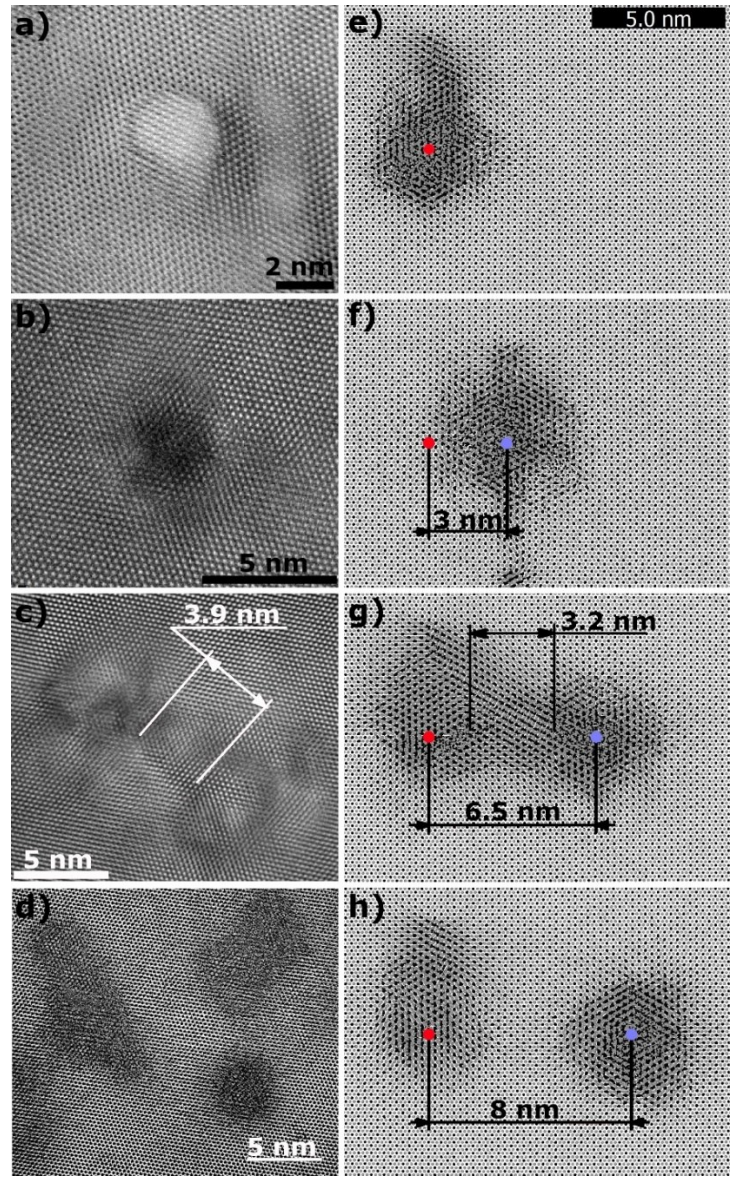

*Figure 37. Bi 700 MeV ion tracks in Al₂O₃. (a)-(b) HRTEM images of single ion track; HRTEM images of tracks in samples irradiated to fluence of (c) 5×10¹¹ cm⁻²; (d) 6.5×10¹² cm⁻². Image plane is perpendicular to the ion trajectories in all cases. (e)-(h) Results of a MD simulation of two subsequent impacts of Bi ions: (e) single track; the second ion at (f) ~3 nm, (g) ~6.5 nm, (h) ~8 nm from the first track. Red dot in (e)-(h) images indicates trajectory of first ion, while blue dot shows path of the second projectile* [256].

Amorphous materials under high-fluence ion irradiation exhibit plastic flow, leading to them changing their shape macroscopically. This effect is known as "ion hammering", as amorphous layers under irradiation typically exhibit anisotropic growth getting thinner and wider, similar to an effect of a hammer blow [351]. The ion hammering effect is used in practice for material shaping with carefully controlled ion beams [352]. It can be modeled with two-temperature plastic flow of a material, accounting for the evolution of viscosity and stresses in the target [353,354]. Ion-induced anisotropic growth of amorphous targets is a result of a viscosity reduction instigated by the accumulation of atomic displacements, simultaneous with local material heating in an ion track [354].

To summarize, at low fluxes, ion tracks start to overlap their defects halos at fluences ~$10^{11}$ ions/cm². Accumulation of defects starts to deviate from the linear behavior due to complex defect kinetics, such as annealing of preexisting defects *via* heating by the successive ions, and aggregation of defects into defect clusters. These effects may be traced with defect kinetics methods (chemical balance rate equations), and for estimation of saturation levels, empirical relations exist. At higher fluences at ~$10^{12}$ ions/cm², a transient disorder in hot tracks may create defected regions between two track cores, thereby synergistically enhancing material damage. At even higher fluences, track cores start to overlap, with possible annealing of preexisting tracks, in materials with efficient recrystallization. Damage saturates at high fluences when the entire material is covered with tracks ($\Phi$ ~ $10^{13}$ ions/cm²). Those effects may be modeled with the same kind of molecular-dynamics-based multiscale models (including electronic kinetics), as used for modeling individual tracks.

### IX.B. High flux: overlap of hot tracks

At high fluxes, when a few ions hit the same place nearly simultaneously, it leads to the accumulation of the deposited energy. The irradiation spot in the matter does not have sufficient time to relax before a successive ion arrives. Properties of matter at such conditions are different from the cold matter state, altering all the stages of SHI interaction, electron kinetics, and atomic response. This regime of track overlap is much less studied than the low flux regime.

The electronic stopping of an SHI depends on the properties of the target, including the target temperature [345]. It especially clearly manifests at high temperatures, corresponding to plasma formation, which may be achieved at ultrahigh fluxes. The nuclear stopping of ions is also affected by the target temperature at extremely high levels of excitation, in the state of warm and hot dense matter. The nuclear stopping may be significantly higher than its cold-target values in such a case, this effect becomes pronounced at plasma temperatures on the order of above some 100 eV [355]. Such temperatures may only be achieved in extremely high fluxes of ions due to multiple tracks overlapping.

At modern accelerators, a typical ion bunch duration is on the order of hundreds of picoseconds to tens of nanoseconds [356,357]. These timescales match characteristic timescales of hydrodynamical behavior of plasma formation and relaxation. At such long timescales, the problem connects to the theory of plasma physics, which may be described with standard tools such as plasma hydrodynamics [358,359]. They become applicable after the created plasma has had sufficient time to expand and relax.

Interaction of swift heavy ions with rare plasma can be modeled with the linear response theory (see Section II.C) if the plasma state parameters (temperature, ionization, density) are accounted for in the complex dielectric function [345,360–362]. Evolution of the plasma state needs to be included in such simulations, e.g. by means of hydrodynamic simulations [363,364], PIC simulations [49,365], or, for non-equilibrium states, rate equations using such codes as FLYCHK [366].

Thus, we have a situation in which the early stages can well be modeled with standard solid-state technics, late stages of the behavior of formed plasma may be treated with plasma-physics approaches, but the intermediate states lie in-between the solid and plasma. The transitional region between a cold solid and a hot plasma is called a warm dense matter (WDM) state.

The WDM state is neither a solid, where the average potential energy of atoms $U$ is larger than their average kinetic energy $K$ ($U \gg K$), nor a plasma, where $U \ll K$. In the WDM, the kinetic energy of particles is comparable to their potential energy ($U \sim K$), rendering standard technics based on perturbation theory inapplicable [151,250]. Nonperturbative approaches are required to describe WDM. This is still a poorly understood state of matter, although it is abundant in the Universe: e.g., it is present in cores of giant gas planets [224,250,367].

WDM properties may be studied with precise *ab-initio* methods like path integral Monte Carlo (PIMC), which follows the evolution of quantum electronic ensembles over thermodynamic states [367–369]. PIMC is currently limited to thermodynamic equilibrium, and cannot describe out-of-equilibrium states relevant to SHI irradiation scenarios. Simplified *ab-initio* approaches are also used with such methods as orbital-free DFT mentioned in Section III.A [370–372]. Standard simulation technics like finite-temperature DFT may be applied to the low-energy border of WDM (temperatures of a few eV), at which sufficiently large basis sets can capture the essential effects [367,373]. However, it is important to keep in mind that most of the exchange-correlation functionals used in DFT calculations are constructed for ground states (low temperature) of matter, and their extensions are required to reliably treat WDM states, e.g., with help of Green's function formalism [249]. Methods based on the Hartree-Fock theory are also recently finding their applications in WDM physics, including two-temperature states (nonequilibrium between the electronic and the atomic systems) [374]. It is still a formidable task for future research to develop a precise and efficient methodology to treat SHI irradiation in a rapidly evolving non-equilibrium WDM state.

It is especially challenging to treat the WDM in far-from-equilibrium conditions, as produced in ultrafast irradiation scenarios [375–377]. The kinetics of WDM formation may be very intricate, due to a complex interplay of electronic nonequilibrium effects, ultrafast nonthermal phase transitions and associated changes in the electronic structure, and enhanced nonadiabatic coupling in the rapidly evolving interatomic potential [378].

To conclude, a high flux of swift heavy ions leads to multiple SHI impacts overlap. In an SHI bunch so intensive that ions impact on a still hot track, it leads to an accumulation of excited electrons and heat in the target. It eventually leads to the creation of hot plasma, which may be described with such methods as plasma hydrodynamics, or kinetic equations for non-equilibrium effects. The transition from a cold solid to a hot plasma goes through the states of warm and hot dense matter (highly degenerate plasma). Theory of these states of matter is an actively developing field of research in both directions: fundamental theory and numerical methods. Appropriate methods, when developed, may be added into a framework of a multiscale approach, tracing the evolution of the target under multiple ions impacts, and modifying the material parameters on-the-fly.

# X. Surface effects

A presence of a free surface influences the dynamical response of the irradiated material. Several important effects are not present in the bulk: emission of particles from the surface, changes in the geometry of the problem (surface breaks the symmetry), and modification of the interatomic potential near the surface. They all contribute to the final SHI track formation, as will be discussed below.

## X.A. Electron emission

Photons and electrons may cross a boundary between two media or get reflected, whereas valence holes being quasiparticles cannot leave the material into a vacuum. For photons, reflection coefficients can be obtained from the scattering angle and refractive indices of the corresponding media [379]. For electrons, there are a number of models for a transmission coefficient evaluation [329,380], based on an empirically defined barrier width and the work function of the material.

The emission of particles enables experimental measurements of various material properties and thereby serves as validation of models. Spectra of emitted electrons are one of the standard quantities used to cross-check simulations of electron kinetics. An example of a transmission coefficient through a surface of germanium is shown in Figure 38a. In a typical case, electrons with energies below the work function of the material cannot be emitted, then the probability of emission increases for a few eV of energy, after which the probability is nearly unity and faster electrons may cross the boundary almost without noticing it, except for the cases when crossing into a material with a higher density, which may induce a sudden Bremsstrahlung-type emission due to instantaneous slowing down known as transition radiation [381]. Spectra of electrons emitted from germanium after irradiation are shown in Figure 38b.

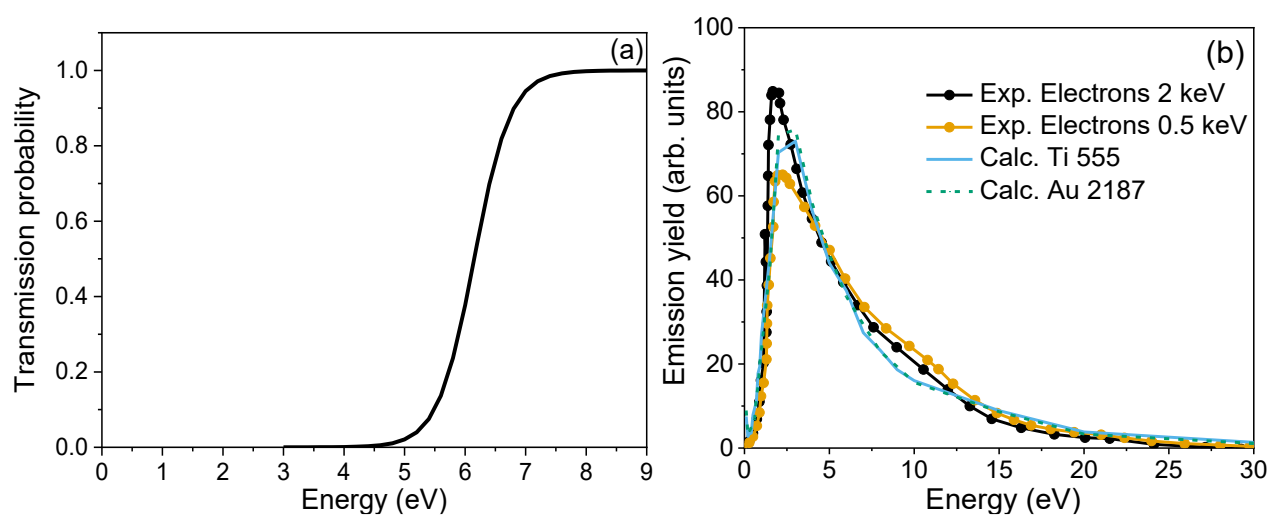

Figure 38. (a) Transmission coefficient through the surface of germanium [382]. (b) Spectra of emitted electrons from germanium after irradiation with Ti (555 MeV) or Au (2187 MeV) ions calculated [382], compared with those after irradiation with 0.5 keV or 2 keV electrons in experiments [383].

It is important to keep in mind that the presence of a surface itself affects the spectra of emitted particles. Thus, surface effects need to be taken into account in modeling. Not only the fact that a fraction of electrons with low energies will not be transmitted, but additionally the surface properties are different from the bulk which affects the kinetics of such electrons in a near-surface layer. Under intense irradiation, if many electrons are emitted, a positive charge accumulates near the surface [384]. It may slow down and attract back slow emitted electrons, especially in finite-size samples (such as ultrathin layers or nanoclusters) [133].

An analysis of photon emission (photon spectroscopy) allows finding ionization potentials of an ion, whose deep-shell holes emit photons *via* radiative decay [385]. Ionization potentials, or more generally energy level structure, allow to unambiguously define the ion state [386,387]. At the same time, each ion configuration has its own characteristic Auger- and radiative decay times, different from neutral atoms [119,388]. This effect potentially enables to reconstruct the time-resolved kinetics in SHI tracks by time-sorting of the photon or Auger spectra emitted from a target [67,119]. Measured spectra of photons or Auger electrons allow connecting the energy position of each peak with its own characteristic decay time. Thus, in measured spectra, time-resolved information of the irradiated system is encoded. Auger-decays spectra enabled to extract information on the times of charge neutralization in the track core, discussed in Section III.A [67]. The same scheme for constructing a femto-clock on the basis of photon spectra was proposed theoretically but has not yet been implemented experimentally [119].

The emission of particles from the surface carries out some fraction of energy, thereby reducing the dose deposited by an SHI. It is especially noticeable in the first few nanometers of the material, see an example of 200 MeV Au ion irradiation of a layer of $CaF_2$ in Figure 39 [259]. After the depth of a few nm, the distribution of the energy deposited to the target atoms turns cylindrical. For normal incidence SHI, only a few electrons are emitted from the surface, whereas the majority of them remain inside because backscattering electrons are rare.

Under a grazing incidence, however, many electrons, traveling perpendicular to the SHI trajectory, reach the surface and are then emitted. In this case, the emission of electrons and a corresponding energy sink are very important to account for, as they may bring out up to half of the deposited energy [339].

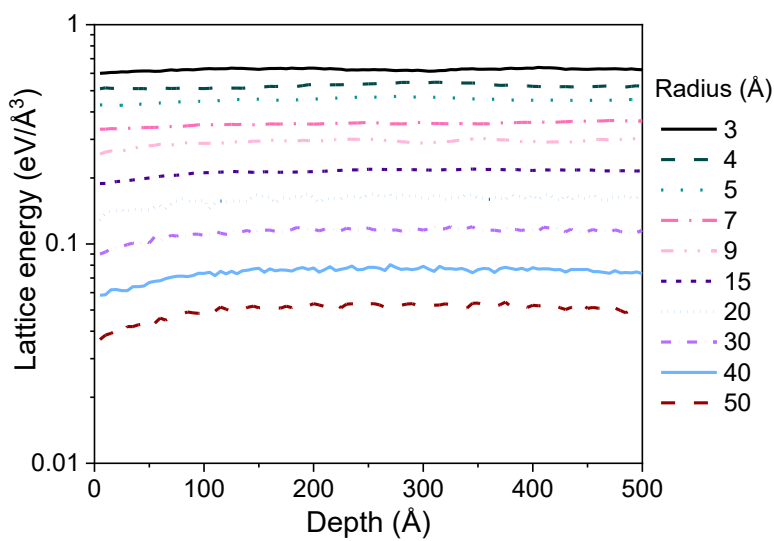

*Figure 39. Dependence of excess lattice energy deposited by an ion on layer thickness, considered for different radii. Layer thickness is 500 Å.*

To conclude, the presence of a surface induces multiple effects in electron kinetics, the most important of which are the emission of particles bringing a part of the deposited energy out of the target. This can be modeled within the MC method, accounting for a surface transmission probability. It leads to reduced energy deposition to the atomic system, which then will affect the observable track formation. It is especially important for grazing incidence ions.

### X.B. Atomic response: sputtering and nanohillocks

In this section we consider specifics of the atomic response to SHI irradiation at material surfaces and in thin layers. A presence of a free surface breaks the symmetry of the material, and correspondingly the local cylindrical symmetry of a forming track.

The effect of emission of electrons reducing the deposited energy is, in a sense, compensated by a softer atomic potential in a near-surface region [389]. A disordered track region near the surface is then larger than in the bulk [390]. This specific under-surface damage region is usually about a few to few tens nanometers deep [390]. It forms a cone-like track from the surface towards a cylinder-like bulk shape of a track, see an example of yttrium aluminum garnet ($Y_3Al_5O_{12}$, YAG) irradiated with 700 MeV Bi ion in Figure 40 [390,391]. This cone-like connection between the surface damage and the bulk track was also observed experimentally [391,392].

The behavior of atoms at this stage can be described well with classical molecular dynamics simulations [393]. As discussed above, such simulations need to ensure a capability to reproduce the surface properties of materials at extreme conditions (high temperatures and pressures).

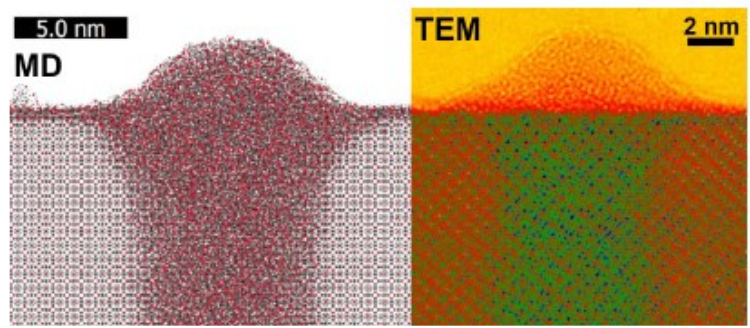

*Figure 40. The surface of YAG irradiated with a 700 MeV Bi ion, simulated with MD (left), and experimentally measured with TEM (right) [390].*

The formation of a subsurface region is accompanied by the emission of atoms and fragments from the surface of the material – the effect known as inelastic sputtering, or sputtering in the electronic stopping regime [79,393,394]. It is in contrast to elastic or nuclear sputtering, which is a result of ion-ion collisions, knocking atoms out of the surface. Inelastic sputtering is a result of atomic heating due to relaxation of excited electrons. It is a two-step process, consisting of the direct atomic emission with high kinetic energies, and a later hydrodynamic emission of atomic ensembles (nanoparticles or nanoclusters) [259,394].

Increased atomic kinetic energy overcomes the cohesive energy of atoms, allowing them to escape from the surface during the first ten picoseconds after an ion impact. An example of it is seen in Figure 41, which shows $CaF_2$ irradiated with 200 MeV Au ion [259]. Atoms and small fragments from the center of a track with the highest kinetic energy are emitted.

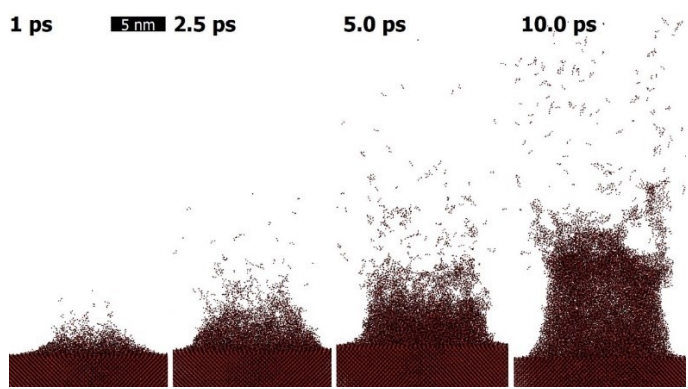

*Figure 41. Snapshots of emission of atoms and nanoclusters from CaF₂ after the impact of 200 MeV gold ion* [259].

Melted material protruding from the surface due to increased pressure may result in the emission of nano-droplets, see the example of CaF₂ irradiated with 200 MeV gold ion in Figure 42. The melt hydrodynamically protrudes from the surface at the timescales of tens of picoseconds and may break apart if its velocity is sufficiently high [259]. Otherwise, it is attracted back by the surface tension of the melted material. The part that is attracted back then cools down at the characteristic time of about 100-200 ps, forming a hillock.

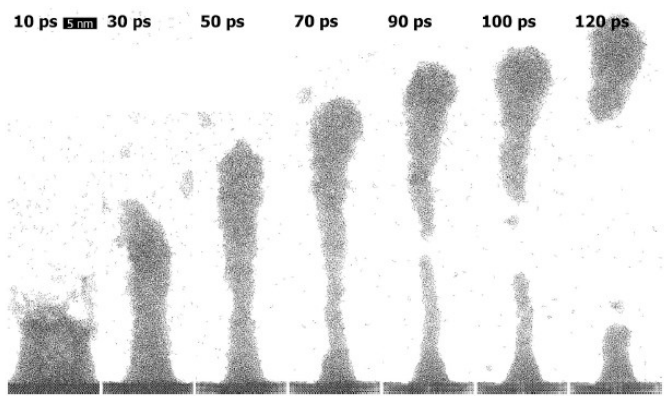

*Figure 42. Snapshots of emission of atoms and nanoclusters from CaF₂ after the impact of 200 MeV gold ion* [259] *(extended timescales cf. Figure 41).*

Thus, there are three different mechanisms of sputtering at play after an SHI impact at a surface of a solid. (i) Elastic sputtering due to atomic cascades near the surface, (ii) inelastic sputtering due to ultrafast heating of atoms by electrons, and (iii) hydrodynamic expansion due to increased pressure leading to ablation-like emission of nanodroplets. It seems, the presence of the different mechanisms of sputtering is reflected in the angular distribution of emitted fragments in experiments, forming a central jet on top of a wider distribution [79,395].

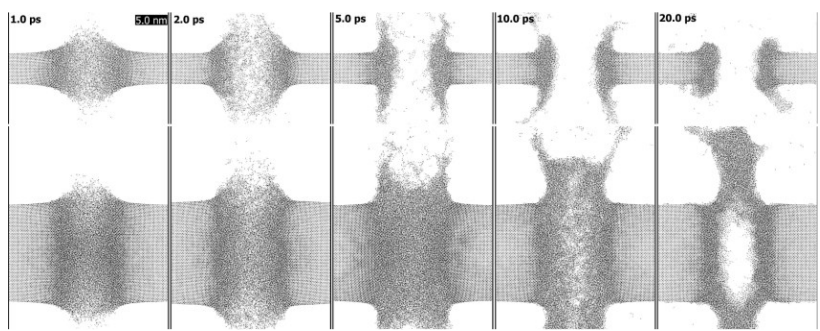

Figure 43. Initial kinetics of damage formation in 5 nm (top panel) and 15 nm (bottom panel) films of CaF$_2$ irradiated with 200 MeV Au ions. 5 nm slice of the central part of the box is shown [259].

In ultrathin layers of material under SHI irradiation, the high pressure due to heating leads to the formation of a through-hole, see the example in Figure 43 for 5 nm and 15 nm films of CaF$_2$ irradiated with 200 MeV Au ions [259]. In this figure, the 5 nm layer material emits all the atoms from the track core already within about ~2 ps. In contrast, increasing the layer thickness to even 15 nm does not allow emission of all atoms from the track core, forming only an undersurface void instead of a through-hole [259]. The process of void formation takes longer, as the closing of the top (and bottom) parts of the track is a hydrodynamic process. The liquid melt needs to be attracted back to the surface due to surface tension and recrystallize afterward. Thicker layers than a few tens of nm behave very similar to bulk under SHI irradiation, discussed in Section VI.A.

In the case of an ion impact under a grazing incidence, the picture looks different. Interestingly, in some materials, the tracks observed at the surface consist of a series of separated nanosized bumps [396], while in others an SHI creates a groove [397,398]. The particular form of created surface defects also depends on SHI energy loss and velocity. The dependence of the formed tracks on the material properties still requires dedicated studies to be performed, both experimental and theoretical.

The formation of a series of nanohillocks at a surface of a material under grazing incidence of SHI was observed experimentally [399]. A series of nanobumps or hillocks created by an SHI grazing impact was first hypothesized to reflect an ion crossing the interplanar distance between atomic layers [396]. It was assumed that due to inhomogeneous electronic density in the target, SHI energy loss is also inhomogeneous along the path, which leads to discontinuous surface damage after the energy relaxation [400]. This suggestion was based on the TTM modeling with the initial excitation profile taken from the profile of the local electron density around atoms of materials calculated with DFT. Further research has shown that such effects of variation of electron density in the material when an SHI crosses planar distance quickly smear out due to electron transport and result in a more or less uniform atomic heating. The series of hillocks form at a later stage as a result of hydrodynamic instabilities in the melt protruding from the surface of material along the SHI trajectory [339]. This insight was achieved with a detailed simulation combining MC modeling of electron transport with MD modeling of atomic dynamics.

At a grazing incidence, an ion excites material locally nearly parallel to the surface, thereby creating a track partly reaching an open surface [339]. In such a case, melted material may protrude forming a "wall", which then experiences an interplay of the hydrodynamic expansion and surface tension attracting it back. For some materials and ions, such as, e.g., 23 MeV iodine in MgO shown in Figure 44, it leads to the emission of nanoclusters, accompanied by the formation of nanohillocks along a groove following the SHI trajectory. In others, such as in $Al_2O_3$, it forms an amorphous bump with a height of ~1.5 nm [339].

A systematic study of material and ion parameters to identify the conditions defining the formation of various surface damages is yet to be performed. As of now, it is only clear that material properties such as surface tension and its ability to recrystallize are important factors in defining surface nano-features formation after an SHI impact [339]. Recrystallization also affects the formation of surface defects, defining the size and structure of formed hillocks, which creates a variety of different structures as dome-shaped, semi-spherical, or spherical nano-features, with amorphous or crystalline atomic structures [259,390].

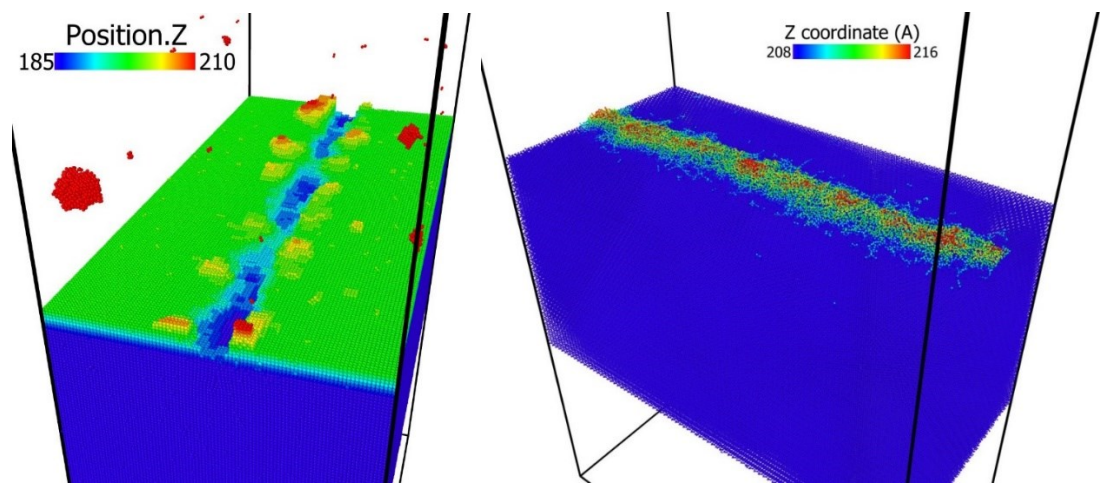

*Figure 44. Snapshots of MD simulation boxes at 100 ps after 23 MeV iodine ion passage at a depth of 1 nm parallel to the surface in MgO (left panel) and $Al_2O_3$ (right panel)* [339]*.*

To summarize, surface modifications in SHI irradiation may be noticeably different from the tracks in the bulk due to the emission of electrons, atoms, and fragments, and a difference in material properties at the surface vs. the bulk. Provided that electron kinetics is faithfully traced with such methods as Monte Carlo simulations, the response of the atomic system can then be modeled with classical molecular dynamics simulations, the same as bulk simulations. This kind of simulation should ensure that surface properties such as surface tension and properties of melted material are reproduced well with the employed interatomic potential. It allows then to study the formation of conical undersurface damage in the transition region between the surface and the bulk; the formation of surface defects such as hillocks or craters; sputtering effects due to emission of atoms and nanoparticles from the surface; and the formation of long grooves or series of nanobumps at the surface under grazing ion incidence.

## XI.     Track etching modeling

Exposure of SHI-irradiated samples to chemically active reactants causes the etching of SHI tracks. It results in the formation of pores with diameters ranging from several nanometers to micrometers and lengths up to hundreds of microns [2,401]. SHI track etching has a long history and wide area of applications. It is used for the production of microdiaphragms, polymer filters, nanowires and nanotubes, nano- and microstructured films, and promising microelectronic devices [15]. Specific applications of nanopores include biological analysis such as DNA/RNA sequencing and protein profiling, polymer and chemical molecule analysis, gas separation, water desalination, ion selective filtering, power generation, ion and nanoparticle detection (see review [402] for details).

Due to its practical importance, understanding and theoretical description of SHI track etching are of interest. It is a challenging task, since not only all initial stages of track formation, discussed throughout the previous sections, should be described correctly, but also chemical and physical processes driving the interaction of a damaged target with an etchant should be modeled in sufficient detail.

Modeling of SHI tracks etching remains mostly phenomenological or semi-phenomenological (see, e.g., reviews and monographs [15,403,404]). More detailed and physically justified numerical approaches for track etching, rather than simplistic phenomenological ones, are in high demand.

For that, at least the following problems should be solved: (1) development of a model describing changes in chemical structure and chemical reactivity of a material in the vicinity of trajectories of various ions; (2) description of chemical interaction of a damaged target with an etchant, which may include a wide number of channels, especially in case of polymers; (3) a model of diffusion of etchant molecules and reaction products to and from the etching front; (4) description of mechanisms, governing etching of spatially anisotropic systems (e.g. crystals) which may result in different shapes of polygonal sections of etched pores [405].

Modeling of SHI track etching is generally based on the chemical master (the mass action law) and Arrhenius equations (transition state theory). If two substances X and Y with stoichiometric coefficients $\alpha_{sc}$ and $\beta_{sc}$ interact chemically to form a product $P$ with the coefficient $\kappa_{sc}$, the reaction equation is:

$$\alpha_{sc}X + \beta_{sc}Y = \kappa_{sc}P. \tag{48}$$

The rate of this reaction, $\chi_{react,}$ is proportional to molar concentrations of the reactants, $[C_X]$ and $[C_Y]$, in the powers of their stoichiometric coefficients, and is independent of other concentrations and reactions:

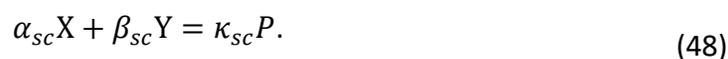

$$\chi_{react} = -\frac{\partial [C_X]}{\partial t} = K_{react}(T)[C_X]^{\alpha_{sc}}[C_Y]^{\beta_{sc}} \tag{49}$$

The rate constant of a chemical reaction $K(T)$ is described by the Arrhenius equation [406]:

$$K_{react}(T) = A_C \, exp\left(-\frac{E_a}{k_B T}\right)$$

(50)

here $A_C$ is the fitting constant depending on specific substances, $E_a$ is the activation energy, and $T$ is the temperature of the chemical reaction (in the case of track etching, equal to the temperature of etchant). In the transition state theory, activation energy is replaced by the Gibbs free energy barrier of the chemical reaction [406]. This exponential in the reaction rate $K_{react}(T)$ inspires various authors to construct adjustable functions of the etching rate in the form of exponentials, as will be discussed below.

### XI.A.    Lengthwise track etching rate

The schematics of the track etching problem is shown in Figure 45. In Ref. [407], the dependence of the lengthwise track etching rate $v_T$ (see Figure 45) on the charge and mass of an SHI and its residual range (recall Figure 1) was discovered. However, for various combinations of targets, ions, and etching conditions, it is still impossible to calculate parameters that may connect the ion charge and energy with lengthwise or radial etching rates of tracks without adjustable parameters, obtained from track etching experiments [408,409]. Various authors use various parameters with various variables to fit lengthwise etching rate in a semi-phenomenological way.

For example, the authors of [407] suggested that a chemical reactivity increase depends on the ionization rate $J$ of a target, i.e. a number of ions formed per unit distance along the ion path [410] (the parameter of the Coulomb explosion model of track formation they assumed). The resulting lengthwise track etching rate $v_T$ was approximated by a fitting function of $J$:

$$v_T = v_B + v_{cc}(1 - e^{a_{j1}\Delta J} - e^{a_{j2}\Delta J})$$

(51)

here $v_B$ is the bulk etching rate (see Figure 45), $a_{j1}$, $a_{j2}$, and $v_{cc}$ are fitting parameters, and $\Delta J = J - J_C$ is a difference between the ionization rate $J$ in a point on the ion trajectory and a threshold ionization rate $J_C$ switching on an enhanced etching, which is also fitted for each material.

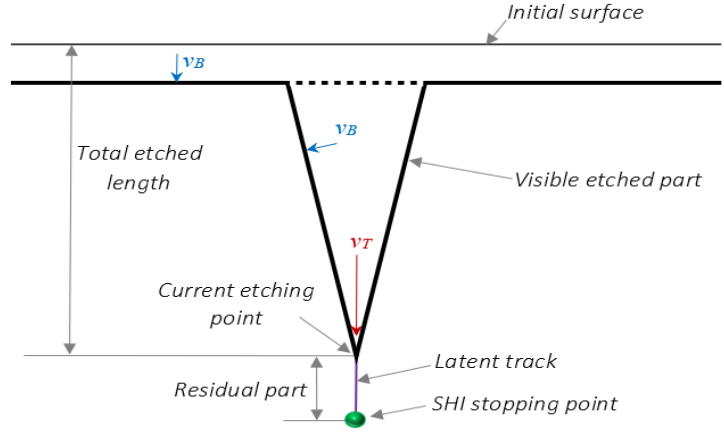

*Figure 45. The scheme of track etching.*

Nowadays, several fitting functions $v_T$ depending on the residual range are found in the literature, e.g.:

$$v_T(Z_{ion}, L) = \frac{A(Z_{ion}) \cdot (1 + E(Z_{ion}) \cdot L^2)}{1 + B(Z_{ion}) \cdot \exp\left[(L - C(Z_{ion}))/D(Z_{ion})\right]} \tag{52}$$

here $Z_{ion}$ is the atomic number of the SHI, $L$ is its residual range, and *A, B, C, D, and E* are fitting parameters [411], or

$$v_T(L) = 1 + \exp\left(-a_1 L + a_2 - a_3/L + a_4/L^{a_5}\right) \tag{53}$$

here $a_1$-$a_5$ are fitting parameters [412]; as well as many other models [413].

Since the parameters in these models are fitted to experimental data on track etching, they must be changed when changing a target or an etchant (conditions of etching). That means, applications of the models need a large calibration dataset for different targets and irradiation conditions.

### XI.B. Radial track etching rate

As discussed in the previous sections, electrons may travel far from the SHI trajectory (up to micrometers), and free radicals created can migrate during track relaxation. This may lead to changes in chemical structure at a distance up to micrometers from the ion trajectory, and, consequently, to variation of track etching rates not only in axial but also in radial direction [414].

Refs. [415–417] developed an approach describing this phenomenon in polymers assuming that a polymer around an SHI trajectory has a complicated structure after track relaxation: a mixture containing low mass fragments, cross-linked structures, and intact macromolecules. This model defines the radial etching rate as follows:

$$v_R(r) = v_f c_f(r) + v_s c_s(r) + v_m c_m(r)$$

(54)

here $v_f$, $v_s$, and $v_m$ are the etching rates for the low-mass fragment fraction, cross-linked fraction, and intact macromolecules, respectively, and $c_f(r)$, $c_s(r)$, and $c_m(r)$ are their concentrations.

The model assumes that the consistence of this mixture is connected to the concentration of transient states (transients, $n_t(r)$), such as free radicals or molecular ions [416], forming during track relaxation, by the system of equations:

$$c_f(r) = k_f n_t(r) n_t(r) \tau_0^{-1}(r), \quad c_s(r) = k_s n_t(r)\big(1 - n_t(r)\big)\tau_0^{-1}(r)$$

$$c_m(r) = \big[1 - n_t(r) - k_s n_t(r)\big(1 - n_t(r)\big)\big]\tau_0^{-1}(r),$$

$$\tau_0(r) = 1 - n_t(r) + k_f n_t(r) n_t(r)$$

(55)

here $k_f$, $k_s$ are fitting parameters, and $\tau_0(r)$ is the normalization factor.

According to Ref. [415], low-mass fragments are formed mostly in the track core, cross-linked structures appear mostly in the track halo, and intact molecules are beyond the track area ( recall Sections V.B and VI.B), which, we guess, leads to such forms of these functions.

The concentration of the transients was described there as:

$$n_t(r) = 1 - exp\big(-G_t \cdot e_{LR}(r)\big)$$

(56)

here $G_t$ is the fitting constant, and $e_{LR}(r)$ is the local radiation dose assumed as:

$$e_{LR}(r) = \frac{\gamma_f}{2\pi R_{tr}^{\gamma_f}} \frac{dE}{dx} r^{\gamma_f - 2}$$

(57)

here $\gamma_f$ is a fitting parameter, and $R_{tr}$ is a maximum radial range of the secondary electrons, also used as an adjustable parameter [415]. For example, in Ref. [416], the authors of the model defined $R_{tr}$=84$E/\rho_t$ (nm), where $E$ is the ion energy in MeV/u and $\rho_t$ is the polymer density (g/cm$^3$).

The model uses many fitting parameters, *e.g.* $\gamma_f$, $R_{tr}$, and etching rates. It does not take into account the velocity effect, since only the value of $dE/dx$ is used. It also cannot describe effects caused by anisotropy of the crystalline structure. Despite that, the model proved useful in describing the chemical structure of the latent SHI tracks and their etching in amorphous polymers [417].

## XI.C. Etching and pre-etching conditions

As seen from the law of mass action and the transition state theory, a change of concentration of an etchant or its temperature affects the track etching dynamics. Several works attempted to describe the effects of the etchant concentration and temperature, mostly in a phenomenological way. Refs. [418,419] proposed a model of CR-39 etching based on the transition state theory with a number of fitting parameters to describe an increase in the lengthwise track etching rate:

$$v_T = K_{TG} \frac{d_b^e}{d_t^e} \frac{\lambda_b}{\lambda_t} \frac{N_e^t}{N_0^t} \frac{N_0^b}{N_e^b} e^{\frac{E_t - E_b}{k_B T}} v_B \tag{58}$$

here $v_T$ and $v_B$ are the lengthwise track and bulk etching rates $N_0^t$ and $N_0^b$ are the numbers of atoms in a bulk volume $d_b^e a$ and track volume $d_t^e a$, where $a$ is the cross-sectional area of a latent damaged trail of an incident particle, and $d_b^e$ and $d_t^e$ are a bulk and track layer thicknesses; $\lambda_t$ and $\lambda_b$ are molecules collision frequencies; $E_t$ and $E_b$ are the activation energies in the track and the bulk, correspondingly; and $K_{TG}$ is a fitting parameter.

Ref. [419] suggested studying the dependence of tracks etching dynamics on the concentration and the temperature of the etchant using the parameters $S_{T0}$ and $S_0$:

$$S_{T0} = S_0 e^{\frac{E_b - E_t}{k_B T}} = K_{TG} \frac{d_b^e}{d_t^e} \frac{\lambda_b}{\lambda_t} \frac{N_e^t}{N_0^t} \frac{N_0^b}{N_e^b}; \quad S_0 = \frac{v_T}{v_B} \tag{59}$$

In the range from 60°C to 80°C of the etching solution (NaOH of varying concentrations: 3 *mol/L* to 7 *mol/L*), the experiment demonstrated a much weaker temperature dependence of $S_{T0}$ compared to that of $S_0$. Thus, $S_{T0}$ allows for describing the dependence on other etching conditions, such as types of targets and etchant or etchant concentration.

Track annealing is another phenomenon affecting the etching. Annealing is usually used to remove the background noise in heavy nuclei detectors [420,421]. Changing a defect ensemble that appeared in an SHI track before etching, annealing may considerably affect the etching kinetics. Annealing can be applied either after irradiation of a target to change latent track parameters, or before the irradiation to change the parameters of the target itself [422].

For example, attempting to connect the lengths of etchable tracks before and after the annealing of irradiated target ($l(t, T_{ann})$ and $l_0$, respectively), Ref. [423] assumed that a difference between these lengths may be described as:

$$\frac{l(t, T_{ann})}{l_0} = \left(1 - \frac{t}{t_0} e^{\frac{E_{act}}{K_B T_{ann}}}\right)^{\beta_{ann}} \tag{60}$$

here $\beta_{ann}$ is a fitting parameter; $t$ is annealing time; $T_{ann}$ is the annealing temperature initiating the atomic movement to a relaxed position over the annealing activation energy $E_{act}$, and $t_0$ is a fitting constant [423].

The above-mentioned models investigating track annealing are capable to describe track shrinking. However, all these models use fitting parameters, which are suitable only for specific materials and conditions of irradiations and annealing.

The application of surfactant may also affect track etching, but this effect is usually described in a phenomenological way. For example, Ref. [421] describes an effect of surfactant on the shape of the etched pore, explaining cigar-like shapes of etched tracks by a decrease in diffusivity of the etchant molecules to the sample surface through the surfactant.

### XI.D.    Etching of anisotropic crystalline targets

The studies mentioned above are mostly related to the etching of tracks in isotropic systems, where etched pores have cylindrical symmetry. The etching of tracks in anisotropic crystalline targets may result in a polygonal shape of the section of etched pores. Ref. [424] studied this effect in apatite. It considered various crystal orientations relative to SHI trajectory resulting in an anisotropy of radial etching. Polygonal shapes of produced etched channels were described taking into account the crystal symmetry and assuming sequential removal of material layers, starting from instantaneous removal of the track core. No relationship between an incident ion charge and energy with the etching kinetics has been investigated, even though these parameters may be crucial for the anisotropic etching of crystalline solids. Even the same level of the linear energy losses *dE/dx* of an ion may result in various shapes of polygonal pore when ions from different Bragg curve shoulders are used (a result of the velocity effect, see Sections VI.A and VIII.B) [405].

A numerical iterative algorithm with the sequential and selective removing of atoms with broken chemical bonds was developed in Ref. [425] describing track etching in crystalline mica. It is one of the most advanced models of chemical etching of SHI tracks based on molecular dynamics simulations. However, the model uses experimental data to set the initial radius of a track and its shape to reproduce the experimental diameter and shape of the track core, and does not account for peculiarities of interatomic potential, and, thus, is not a predictive one.

### XI.E.    Multiscale model of track etching

An analysis of the above-mentioned studies shows that focusing only on one of the problems of track etching while ignoring microscopic mechanisms of material excitation in a track gives no possibility to describe the track etching kinetics without fitting parameters. Alternative approaches are required.

A successful attempt to convert initial material excitation into track etching parameters in olivine (($Mg_{0.89}Fe_{0.11})_2SiO_4$) starting from the impact of a projectile was realized in Refs. [426–428]. A multiscale approach described in Section VIII provided initial conditions for this model of wet chemical etching, forming its basis. The transition state theory (see above) was used to determine a change of track etching rate with respect to the bulk one [426]:

$$\frac{K_{Track}(r,L)}{K_{Bulk}} = exp\left(-\frac{\Delta G^{++}_{Bulk} - \Delta G^{++}_{Track}(r,L)}{k_B T}\right) \tag{61}$$

here $K_{Track}(r,L)$ is the reaction rate inside the track, which depends on $r$, the distance from the ion trajectory to the current point, and $L$, the residual range of an ion; $K_{Bulk}$ is the reaction rate in the unirradiated material; $\Delta G^{++}_{Bulk}$ and $\Delta G^{++}_{Track}(r,L)$, are Gibbs activation barriers for chemical reactions of the etchant with the undamaged material and with the material in the track – both can be calculated with *ab-initio* methods such as DFT or even classical MD potentials [426–428].

The chemical reaction rate of the track core with a given etchant in olivine is several orders of magnitude higher than that for an undamaged crystal [426]. In this case, the velocity of the etching front in a track core is limited not by the reaction rate, but by the rate of diffusion of etchant molecules to this front (see Figure 46). This effect was simulated using a set of reaction-diffusion equations for the transfer of etchant molecules and reaction products along an appearing track pore, which were solved numerically (see Ref. [427] for details). A variation of generated damage along the ion trajectory was also taken into account [428]. The model demonstrated a good agreement with experimental data within error bars, see Figure 47.

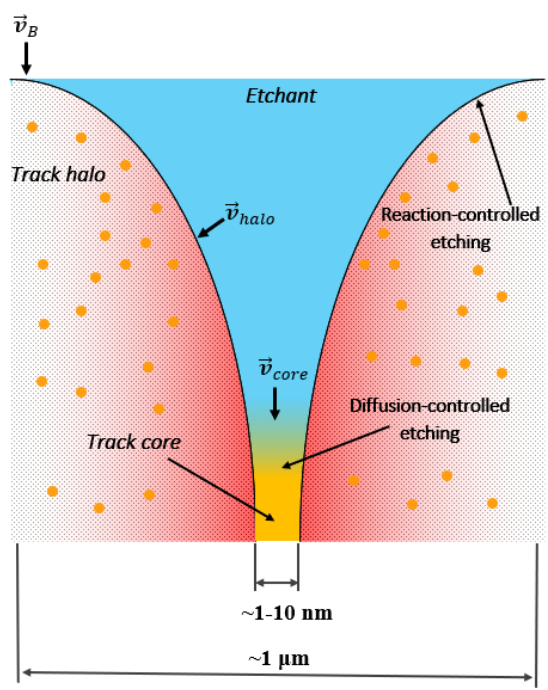

*Figure 46. Scheme of the reaction-diffusion SHI track etching.*

A significant advantage of this approach is an absence of fitting parameters (such as the track radius or the lengthwise etching rate). In contrast to the simple geometrical approach (cf. Figure 45), it decomposes $v_T$ to the lengthwise track-core etching velocity $v_{core}$ and the radial and lengthwise track-halo etching $v_{halo}$ and describes automatically a transition between these areas. However, it still requires the target bulk etching rate as an input parameter, which may be obtained from *ab-initio* simulations. In the presented form, the model was not applied yet to anisotropic crystals, annealing, or surfactant effects.

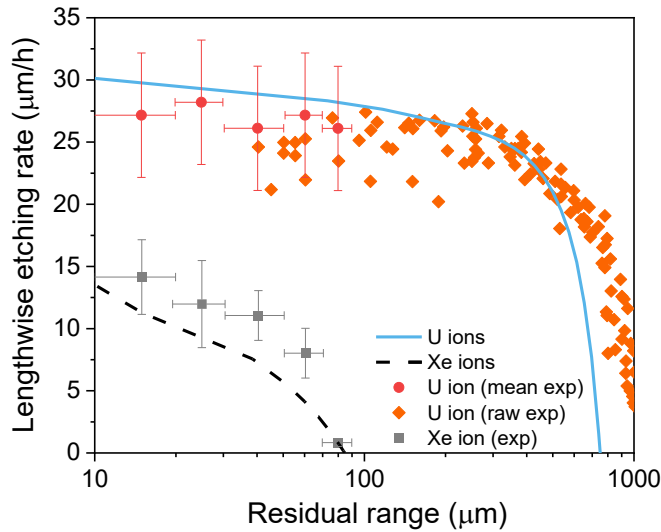

*Figure 47. Calculated lengthwise etching rates as functions of the residual ranges of 16 GeV U (red solid line) and 1.7 GeV Xe (dashed black line) ions in olivine calculated with a multiscale model and Eq.(61) [428]. Experimental values for corresponding ions are shown with symbols, taken from [22].*

Summarizing, the modern state of the theory of SHI track etching remains mainly semi-phenomenological. Effects of a target anisotropy on shapes of etched tracks as well as effects of annealing on track kinetic need the development of microscopic self-consistent models. So far, a complicated numerical technique using a multiscale numerical approach was only used for chemically simple material (olivine in [426–428]) but demonstrated inspiring capabilities of such a methodology.

## XII.    Summary and outlook

The swift heavy ion impacts on matter trigger a variety of processes across multiple timescales. A traversing ion excites in a target an ultrashort-lived nonequilibrium plasma column embedded in cold dense matter. Intertwined effects from solid-state physics, plasma physics, nano-science, nonequilibrium kinetics, atomic and molecular physics, and femtochemistry, naturally combine those various fields into a unified research topic in the SHI track creation problem. This review is by no means exhaustive but aims to provide the reader with a solid background on the current understanding of the SHI track formation science and to direct them to appropriate references forming an up-to-date starting point for further studies.

In this review, we gave a glimpse of the multitude of processes instigated by an ion impact spanning over ten orders of magnitude in time. Each characteristic timescale requires its own theoretical and computational technics giving a chance to build up hybrid multiscale models, which combine a few models within a unified approach describing various relevant aspects of track creation.

The insights obtained with multiscale simulation tools within the past decade brought major advances in the broad field of SHI research. It allowed identifying the most important effects in a track formation. After over half a century, finally, swift heavy ion track formation can now be described from start to end without adjustable parameters, in good agreement with experiments.

The following effects are now understood to play the major role in an SHI track formation: nonequilibrium electron transport at femtosecond timescales; scattering of valence holes on target atoms and phonons providing them with additional kinetic energy; nonthermal modification of the interatomic potential, which converts the potential energy of atoms into the kinetic one within ~100 fs; atomic relaxation, cooling, and recrystallization of a transiently disordered region around an SHI trajectory. These effects ultimately define the observable track formation in the matter. We described appropriate simulation tools that capture all these stages of a track formation, which the reader may implement and apply.

Multiscale modeling, we believe, will make the new gold standard in the field. Multiscale and hybrid codes naturally have a modular structure, which makes them easy to develop. An advantage of hybrid models is that each individual module may later be replaced with a more precise approach if required. It paves the way for a systematic improvement of the model of the SHI track formation. The limitations and approximations used in the current multiscale models, which we discussed throughout this review, may gradually be improved in future research without an overhaul of the entire simulation framework.

### XIII.    Acknowledgments

We are indebted to many colleagues for helpful discussions throughout the years of collaboration on various aspects of science relevant to swift heavy ion irradiation. NM gratefully acknowledges financial support from the Czech Ministry of Education, Youth and Sports (Grants No. LTT17015 and No. EF16_013/0001552). This work benefited from networking activities carried out within the EU-funded COST Action CA17126 (TUMIEE) and represents a contribution to it. AV, SG, RV, and PB acknowledge support from the Russian Science Foundation (grant number №22-22-00676, https://rscf.ru/en/project/22-22-00676/).